\documentclass{article}
\newcommand{\undertitle}{}
\usepackage{arxiv}

\usepackage[utf8]{inputenc} 
\usepackage[T1]{fontenc}    
\usepackage{hyperref}       
\usepackage{url}            
\usepackage{booktabs}       
\usepackage{amsfonts}       
\usepackage{nicefrac}       
\usepackage{microtype}      
\usepackage{amsmath}        
\usepackage{subcaption}     
\usepackage{cleveref}       
\usepackage{lipsum}         
\usepackage{graphicx}
\usepackage{natbib}
\usepackage{doi}
\usepackage{textcomp}
\usepackage{array}
\usepackage{fancyhdr}
\usepackage{float}
\newcommand{\headeright}{}
\title{Weather Jiu-Jitsu: Exploring the Feasibility of Control Paradigms in Weather Foundation Models}

\newif\ifuniqueAffiliation
\uniqueAffiliationtrue

\ifuniqueAffiliation 
\author{ {\hspace{1mm}Prakriti Biswas} \\
	School of Electrical, Computer and Energy Engineering\\
	Arizona State University\\
	\texttt{pbiswa12@asu.edu} \\
	\And
	\hspace{1mm}{Kobi Abayomi} \\
    The GAME School\\
	School of Mathematical and Statistical Sciences\\
	Arizona State University\\
	\texttt{kabayomi@asu.edu} \\
	\AND
	\hspace{1mm}{Upmanu Lall} \\
	School of Complex Adaptive Systems\\
	Arizona State University\\
	\texttt{ulall@asu.edu} \\
}

\hypersetup{
pdftitle={Weather Jiu-Jitsu: Exploring the Feasibility of Control Paradigms in Weather Foundation Models},
pdfsubject={},
pdfauthor={Prakriti Biswas, Kobi Abayomi, Upmanu Lall},
pdfkeywords={weather foundation models, Aurora, atmospheric control, perturbation theory, latent space analysis, chaotic systems, feasibility study},
}

\begin{document}
\maketitle
\begin{abstract}
	Weather Jiu-Jitsu is a control paradigm for extreme climatological events, inspired by chaos theory. As a proposition, small, precise, targeted, and cost-inexpensive perturbations can redirect trajectories of a large dynamical system. This strategy has been demonstrated analytically in the Lorenz-63 system, where a naturally chaotic trajectory switching between two attractors can be confined to a single attractor, indefinitely, via arbitrarily small perturbations.\\
    This paper examines the feasibility of Microsoft's Aurora -- a 1.3 billion parameter global atmospheric model -- as a test bed for this strategy. This paper explores three questions: (1) Is Aurora a reliable enough simulation environment to serve as a meaningful testbed? (2) Are the perturbations required to redirect its trajectories small enough to be physically plausible? (3) Does Aurora's learned latent space (the parametric estimators on climatological attributes) yield any apparent, structured, and/or perhaps interpretable features that can convey a geo/atmospheric response to initial conditions?\\
    We find evidence consistent with all three: Aurora's modeled trajectories respond to perturbations beyond measurement drift, the perturbation magnitudes required are small relative to the model's own forecast uncertainty, and its latent representations exhibit directional structure that responds to Jiu-Jitsu-type interventions, even though that structure does not separate extreme from normal states outright. These results should be read as feasibility diagnostics rather than a demonstration of control: we do not implement or test an actual steering intervention on Aurora, and several of our findings, particularly around the model's latent-space geometry, are explanatory.
\end{abstract}


\section{Introduction}
The Weather Jiu-Jitsu paradigm \citep{10.1371/journal.pwat.0000562} originates from the classical problem of controlling chaos. The Lorenz-63 system, a three-dimensional ordinary differential equation exhibiting sensitive dependence on initial conditions and an attractor with two wings, can, counterintuitively, be controlled with arbitrarily small inputs if those inputs are applied at the right moment and in the right direction. The key insight is that chaotic trajectories, precisely because of their sensitivity, can be redirected using less energy than one might expect: the system cooperates with the controller. This idea, formalized by Ott, Grebogi, and Yorke (OGY) in 1990 \citep{PhysRevLett.64.1196} showed that small perturbations applied near unstable periodic orbits embedded in the attractor can stabilize desired behaviors indefinitely.

This paper asks whether this paradigm translates to a weather foundation model. A new generation of large-scale neural weather models, including Pangu-Weather \citep{Bi2023}, GraphCast \citep{doi:10.1126/science.adi2336}, FourCastNet \newline\citep{pathak2022fourcastnetglobaldatadrivenhighresolution}, and Microsoft's Aurora \citep{Bodnar2025}, has demonstrated that data-driven models can match or exceed operational numerical weather prediction systems at a fraction of the computational cost. These models are differentiable, globally realistic, and capable of high-resolution simulation, offering an end-to-end differentiable environment in which gradients with respect to initial conditions can be computed efficiently, enabling gradient-based perturbation strategies.

The Lorenz-63 system is three-dimensional, exactly known, and operates near well-characterized unstable periodic orbits. Aurora operates in a space of roughly $10^8$ degrees of freedom, its dynamics are learned from data rather than derived from first principles, and the geometric structure of its attractor is unknown. It is unclear whether the control strategy that works in three dimensions transfers to this setting.

This work conducts a systematic feasibility study. We characterize Aurora's forecasting skill and generalization properties, develop and apply a concrete metric, the \textbf{Jiu-Jitsu Feasibility Ratio}, that tests whether trajectory-redirecting perturbations can be small compared to the model's own chaotic drift, analyze Aurora's latent representation space to ask whether it encodes the kind of structured atmospheric geometry a control strategy would exploit, and carefully assess what our results do and do not imply about the broader program.

The contribution of this paper is a rigorous diagnosis of the conditions that must hold for Weather Jiu-Jitsu to work inside a foundation model, an evaluation of which of those conditions currently hold, and a map of the open problems that remain.

\subsection{Related Work}
\begin{description}
    \item[Weather Foundation Models] The modern era of neural weather prediction began with FourCastNet \citep{pathak2022fourcastnetglobaldatadrivenhighresolution}, followed by Pangu-Weather \citep{Bi2023}, GraphCast \citep{doi:10.1126/science.adi2336}, and most recently Aurora \citep{Bodnar2025}. Aurora is distinguished by its use of a Swin transformer backbone \citep{9710580} with 3D attention over pressure levels, trained on ERA5 reanalysis data from 1979-2021. It achieves competitive performance on the WeatherBench2 \citep{https://doi.org/10.1029/2023MS004019} benchmark across most variables. \citet{huang2026evaluating} evaluate Aurora against 16 high-impact extreme weather events and find it provides reliable guidance for extreme prediction out to roughly a week, establishing Aurora's relevance as a testbed for extreme-event-focused control work, such as ours. Prithvi WxC \citep{schmude2024prithviwxcfoundationmodel} is a related foundation model from IBM and NASA that we use as a secondary baseline.

    \item[Chaos Control] The OGY method established the theoretical foundation for controlling chaos via small perturbations. Subsequent work generalized the approach to continuous-time systems \citep{PYRAGAS1992421} and adapted it to spatially extended systems. Applications to atmospheric science have been largely theoretical, with Lorenz-63 serving as the canonical testbed. The Weather Jiu-Jitsu control paradigm \cite{10.1371/journal.pwat.0000562} is a proposal to nudge atmospheric circulation to redirect or defuse extreme weather events by leveraging the sensitivity of chaotic atmospheric dynamics to initial conditions. Follow-up work has moved beyond the classical OGY prescription: \citet{egusphere-2025-3997} develop an optimal-control formulation for the Lorenz-63 and Lorenz-84 systems that minimizes total control energy over a full event rather than targeting a single unstable periodic orbit, and characterize when such control is and isn't feasible, showing it becomes unreliable when local Lyapunov exponents are large or intervention is delayed. \citet{liu2026instabilityawaresteeringextremeatmospheric} extend full-event, optimization-based steering directly to an AI weather foundation model, demonstrating instability-aware control of an extreme atmospheric river. Beyond this steering demonstration, the direct application of chaos control ideas to data-driven atmospheric models is, to our knowledge, unexplored. 

    \item[Predictability and the Butterfly Effect] A separate strand of work asks whether AI weather foundation models exhibit a butterfly effect at all. \citet{selz2023} report that initial-condition uncertainty in an early AI weather model grows far more slowly than in physics-based NWP, showing "no signs of a butterfly effect". \citet{shen2025butterfly} contest this, arguing the result reflects the coarse grid spacing and narrow definition of "butterfly effect" used rather than a fundamental property of AI models, and show comparable finite-predictability error-growth slopes between AI and physics-based models at longer lead times. \citet{hassanzadeh2026missingbutterfly} trace the missing butterfly effect to the coarse-graining inherent in training data, showing that reducing coarse-graining restores butterfly-like divergence at the cost of forecast accuracy, directly implicating the space-time discretization of the map, not just the model architecture, in whether chaotic sensitivity is observed. This debate is a reminder that Lyapunov-exponent estimates are properties of the discretized map itself -- its resolution, time step, and degree of smoothing -- rather than some fixed ground truth about atmospheric dynamics: a 6-hour-step and a 1-hour-step model of the same underlying system can yield materially different divergence estimates even when both are physically grounded. Our approach, therefore, sidesteps the question of whether Aurora reproduces the literal atmospheric butterfly effect at PDE resolution, and instead asks the narrower, operational question of whether Aurora (at its own 6-hour operating scale) exhibits local Lyapunov exponents that vary meaningfully across space, season, and flow regime, which is the quantity the Jiu-Jitsu framework actually requires.
    \item[Latent Space Analysis of Weather Models] Prior work has analyzed the representation spaces of weather foundation models primarily in terms of variable reconstruction quality. The work in \cite{TheRiseofDataDrivenWeatherForecastingAFirstStatisticalAssessmentofMachineLearningBasedWeatherForecastsinanOperationalLikeContext} surveys the field and frames all progress in terms of skill metrics, which are useful supporting evidence that the field as a whole thinks in reconstruction-quality terms. The geometric structure of these latent spaces: clustering, separability, and the encoding of meteorological regimes, has received less attention, with \cite{hu2026mechanisticinterpretabilityaifoundation} representing a recent exception focused on atmospheric chemistry fine-tuning. Our analysis contributes to this gap.

    \item[Direction-based Analysis of Latent Space] Standard clustering methods identify group structure through geometric separation in projection space, but fail when groups differ in the distribution of tail activations rather than cluster centers. We draw on a richer toolkit from statistics and atmospheric science.
    We scan individual principal components of Aurora's learned representations for preferential activation on extreme-event dates, asking which directions in the encoder's feature space carry extreme-event information rather than seasonal variance. Independent Component Analysis (ICA) \citep{HYVARINEN2000411} has been used to extract statistically independent, physically interpretable modes for climate data~\citep{https://doi.org/10.1029/2000JD900152}; we apply it to Aurora's internal representations to identify directions that are both non-Gaussian and sensitive to extreme events. Archetypal analysis~\citep{alcacer2025surveyarchetypalanalysis} identifies the most extreme prototype configurations on the convex hull of the data distribution, making meteorological heterogeneity explicit in a way that centroid-based methods cannot. Principal Nested Spheres~\citep{monem2025principalnestedsphereshighdimensional} generalizes PCA to non-Euclidean manifolds via a greedy backward-reduction sequence, capturing intrinsically curved latent structure that linear methods miss. Tail-dependence PCA constructs a covariance matrix weighted toward the upper quantile of joint activations, following the framework of \citet{10.1093/biomet/83.1.169} for characterizing joint extreme behavior. None of these methods has previously been applied to probe the learned representation space of a weather foundation model. We apply all five in a unified comparison.

    \item[Gradient-based Input Diagnostics] Gradient-based saliency methods compute $\partial(\text{output})/\partial(\text{input})$ via backpropagation to identify which input features drive a model's output. Introduced by \citet{simonyan2014deepinsideconvolutionalnetworks} for image classifiers, the technique has been adapted for geoscientific models to identify physical drivers of learned predictions \citep{https://doi.org/10.1029/2019MS002002,AMachineLearningExplainabilityTutorialforAtmosphericSciences}. These methods produce a spatial map over input variables and pressure levels, directly interpretable as a perturbation target.
    Layer-wise attribution, involving computing gradients with respect to intermediate activations rather than inputs, traces where inside the network a signal is encoded. This idea appears in mechanistic interpretability work on language models \citep{gurnee2026verbalizablerepresentationsformglobal}. We adapt this into a "Jacobian Lens" (J-lens) that evaluates $\partial s_{\mathrm{IC3}}/\partial\mathbf{h}^{(k)}$ at each Swin3D backbone stage, revealing the depth at which extreme-event sensitivity freezes. To our knowledge, this is the first application of layer-wise gradient-based analysis to a weather foundation model's internal representation.
\end{description}

\section{Background and Problem Formulation}\label{sec:Background and Problem Formulation}
\subsection{The Weather Jiu-Jitsu Paradigm}\label{subsec:The Weather Jiu-Jitsu Paradigm}
Consider a deterministic dynamical system $\mathbf{x}_{t+1} = f(\mathbf{x}_t)$ evolving on a chaotic attractor. Two trajectories initialized at nearby points $\mathbf{x}_0$ and $\mathbf{x}_0 + \delta$ will diverge exponentially:
\begin{equation}
    \|\mathbf{x}_t - \tilde{\mathbf{x}}_t\| \approx e^{\lambda t} \|\delta\|
\end{equation}
where $\lambda > 0$ is the leading global Lyapunov exponent: an infinite-time average of the local divergence rate, taken over the whole attractor and independent of where on it a trajectory currently sits. 
The global exponent characterizes a chaotic system in the abstract, but it is the wrong quantity for control. An extreme weather event is a finite-time excursion through a specific, evolving region of state space, and what determines whether a small perturbation can redirect it is the local, or finite-time Lyapunov exponent (FTLE) in the neighborhood of that trajectory:
\begin{equation}
    \lambda(\mathbf{x}_t,T)=\frac{1}{T}\ln\frac{\|Df^T(\mathbf{x}_t)\delta\|}{\|\delta\|}
\end{equation} the finite-horizon state- and time-dependent stretching rate at $\mathbf{x}_t$ evaluated over control-relevant horizon $T$, rather than in the $t\rightarrow\infty$ limit. Unlike $\lambda$, $\lambda(\mathbf{x}_t,T)$ varies across the attractor and across time, and can be large in the neighborhood of an evolving extreme even when the attractor's long-run average is unremarkable. \citet{egusphere-2025-3997} make exactly this substitution operational: their optimal-control scheme for the Lorenz-63 and Lorenz-84 systems is triggered by locally estimated Lyapunov exponents rather than the global exponent, and control becomes unreliable precisely where and when the local exponent is large, independent of the system's long-run average. \citet{liu2026instabilityawaresteeringextremeatmospheric} apply the same local, state-dependent logic to an actual atmospheric river event, using FTLE fields computed from the evolving flow to identify where and when it is locally unstable enough to redirect. We adopt this local view throughout: every stretching-rate quantity we estimate from Aurora, including the "effective Lyapunov exponent" of \S\ref{sec: Aurora Characterization} and the feasibility ratio below, is local and finite-time, not a claim about a global exponent that a data-driven model of this kind does not straightforwardly possess. 
In the Lorenz-63 setting, this local view enables "wing pinning": a trajectory naturally switching between the two lobes of the butterfly attractor can be confined to one wing indefinitely using perturbations whose magnitude vanishes as the target time horizon grows. The control cost is not zero, but it is far smaller than one might expect, and it exploits the trajectory's local instability rather than requiring a global model of the system.

\subsection{Aurora as a Simulation Environment}
\label{subsec:Aurora as a Simulation Environment}
Aurora is an autoregressive neural model that maps an atmospheric state $\mathbf{x}_t \in \mathbb{R}^D$ to a predicted state $\hat{\mathbf{x}}_{t+1} = f_\theta(\mathbf{x}_t)$ at a 6-hour time step\footnote{Divergence and Lyapunov-exponent-style quantities estimated from an autoregressive map are properties of that map's own space-time discretization; a 6-hour-step model need not agree with a 1-hour-step model of the same underlying system (see Related Work).}, where $\theta$ is the fixed, full collection of learned parameters of Aurora. The model architecture consists of three learned components: an encoder $E_\theta$ that maps the high-dimensional atmospheric state to a latent representation $\mathbf{z}_t = E_\theta(\mathbf{x}_t)$, a Swin 3D Transformer backbone $B_\theta$ that evolves the latent state forward in time, and a decoder $D_\theta$ that maps the evolved latent state back to physical space.
 
For the purposes of this study, Aurora acts as a surrogate dynamical system. We apply perturbations to its inputs, roll the model forward autoregressively, and measure how the perturbed and unperturbed trajectories diverge relative to the model's overall forecast error. The key question is: do these trajectories diverge by more than the model's own uncertainty, or by less?
 
\subsection{The Jiu-Jitsu Feasibility Ratio}\label{subsec:The Jiu-Jitsu Feasibility Ratio}
We formalize the feasibility question with the following metric. Let $\mathbf{x}_t^{(0)}$ denote the unperturbed trajectory initialized from ERA5 analysis at time $t=0$, and let $\mathbf{x}_t^{(\delta)}$ denote the trajectory initialized with a perturbation $\delta$ added to a single variable. Define:
\begin{description}
    \item[Error Drift:] The forecast error of the unperturbed run against ERA5 truth is formulated as:
    \begin{equation}
        \epsilon_t = \text{RMSE}(\mathbf{x}_t^{(0)}, \mathbf{x}_t^{\text{ERA5}})
    \end{equation}
    \item[Chaotic Drift:] The difference between the perturbed and unperturbed runs is formulated as:
    \begin{equation}
        \gamma_t = \text{RMSE}(\mathbf{x}_t^{(0)}, \mathbf{x}_t^{(\delta)})
    \end{equation}
\end{description}
The \textbf{Jiu-Jitsu Feasibility Ratio} is defined as:
\begin{equation}
    \mathcal{R}_t = \frac{\gamma_t}{\epsilon_t}
\end{equation}
 
Intuitively, $\mathcal{R}_t < 1$ means the perturbation-induced chaotic drift is smaller than the model's own forecast uncertainty. This is a necessary condition for feasibility: a perturbation is only useful as a control input if it produces displacements that are non-negligible relative to the model's error floor yet still small enough to be physically realizable. A ratio well below 1 indicates that the perturbation is absorbed by the model, i.e., trajectories are not yet diverging as fast as errors are growing, which is consistent with the Jiu-Jitsu regime. A ratio above 1 would indicate that the perturbation is destabilizing the model beyond its forecast horizon, which would be inconsistent with the control paradigm.
Framed against \S\ref{subsec:The Weather Jiu-Jitsu Paradigm}, $\gamma_t$ is precisely a finite-time, empirically estimated measure of local divergence at the specific state $\mathbf{x}_t^{(0)}$ and horizon $t$, which is a variable-specific, data-driven stand-in for $\lambda(\mathbf{x}_t,t)$ that requires no explicit Jacobian or tangent-linear model. The Feasibility Ratio $\mathcal{R}_t$ therefore tests whether this local, FTLE-like divergence rate is small relative to the model's own local error growth $\epsilon_t$ and not whether either quantity is small relative to any global, attractor-wide average.
\section{Experimental Setup}\label{sec:Experimental Setup}
\subsection{Data and Model}\label{subsec:Data and Model}
All experiments use ERA5 reanalysis data \citep{https://doi.org/10.1002/qj.3803} as ground truth. Aurora is run in autoregressive mode, starting from ERA5 initial conditions. In this work, we use two initialization dates: 2019-01-01 for in-sample evaluations and 2023-01-01 for out-of-sample evaluations. The 6-hour autoregressive step is used throughout.
The variables evaluated span both surface and atmospheric levels:
\begin{description}
    \item[Surface:] 2-meter temperature (2t, K), 10-meter u-wind (10u, m/s), 10-meter v-wind (10v, m/s), mean sea-level pressure (msl, Pa)
    \item[Atmospheric (700hPa unless stated):] geopotential height (z, m²/s²), u-wind (u, m/s), v-wind (v, m/s), temperature (t, K), specific humidity (q, kg/kg)
\end{description}
The reference baseline is persistence (predicting that the atmospheric state at time $t+\Delta t$ equals the state at time $t$) and a secondary comparison is made against Prithvi WxC.

\subsection{Perturbation Experiments}\label{subsec:Perturbation Experiments}
We apply a spatially localized perturbation to a single variable at initialization: a Gaussian bump of amplitude $\Delta$ centered at the middle of the domain, with standard deviation $\sigma$ set to $5\%$ of the grid dimension in each spatial direction, so that the perturbation decays to negligible amplitude away from the center rather than being spread uniformly across the globe. Aurora is then rolled forward for two autoregressive steps (0-12h), and the Jiu-Jitsu Feasibility Ratio $\mathcal{R}$ is computed for each experiment at the final step (12h) across all output variables.

\subsection{Latent Space Analysis}\label{subsec: Latent Space Analysis}
To assess whether Aurora's internal representation encodes structure amenable to targeted control, we extract encoder output representations $\mathbf{z} = E_\theta(\mathbf{x})$ for a set of 48 dates spanning 2022–2023, covering all four seasons. We also extract backbone (post-processing) representations. Dimensionality reduction is performed using PCA, t-SNE, and UMAP, and cluster quality is assessed using the silhouette score. We label each sample by: (1) meteorological season (DJF/MAM/JJA/SON), (2) extreme-event intensity (normal vs. extreme, defined by an area-weighted RMS standardized-anomaly score on the combined mean-sea-level pressure/2-meter temperature field exceeding a fixed threshold, per the criterion detailed in \S\ref{subsec:extreme event encoding}), and (3) the joint season-by-intensity label.

\subsection{Direction-Based Analysis}\label{subsec:Direction-based Analysis}
All direction-based methods operate on the Stage-2 bottleneck representations $\mathbf{h}^{(2)}(t)$ extracted from a larger set of 608 dates spanning 2022 to 2023, which provides the statistical depth needed to resolve non-Gaussian structure in the latent distribution. The representations are centered by subtracting the sample mean $\boldsymbol{\mu}$ before each analysis.
The higher-PC analysis scans the top PCA directions by projecting each date onto each PC and computing a sensitivity score that measures separation between dates of extreme events and normal dates. FastICA \citep{HYVARINEN2000411} is then applied under the generative model $\mathbf{h}^{(2)}(t)-\boldsymbol{\mu}=\mathbf{A}\mathbf{s}(t)$, yielding unmixing vectors $\mathbf{g}_c$ and per-date IC scores $s_c(t)=\mathbf{g}_c\cdot(\mathbf{h}^{(2)}(t)-\boldsymbol{\mu})$, which we examine across seasons and extreme-event labels.
Archetypal analysis finds $K$ extreme convex-hull points that reconstruct the data as convex combinations, with each date characterized by its archetype-mixture coefficient vector. PNS provides a non-Euclidean analogue of PCA where we $\ell_2$-normalize the centered representations and fit a sequence of nested subspheres of decreasing dimension.
Finally, tail-dependence PCA is applied to the full 608-date set, targeting co-exceedance structure in the tails of the latent distribution that standard PCA hides.

\subsection{Gradient-Based Input Diagnostics}\label{subsec:Gradient-Based Input Diagnostics}
Gradient-based analyses trace the information flow from Aurora's input fields to its latent representations by backpropagating through the frozen model. All gradients are computed with respect to the aforementioned 608-date set used in the direction-based analyses.
For a target IC score $s_c(t)$, computed in the direction-based analysis, we compute the input Jacobian $\mathbf{J}(t)=\partial s_c/\partial\mathbf{x}$ via a single backward pass through Aurora's encoder. To identify the input fields most consistently associated with high $s_c$ across extreme-event dates, we form the sign-normalized aggregate shown as,
\begin{equation}
    \mathbf{J}^{-}=\frac{1}{N}\sum_{i=1}^N\operatorname{sgn}\!\bigl(s_c(\mathbf{x}_i)\bigr)\, \frac{\partial s_c}{\partial\mathbf{x}_i},
\end{equation}
which cancels gradient noise while preserving directionally consistent sensitivity. An anomaly Jacobian $\mathbf{J}'(t)=\mathbf{J}(t)-\bar{\mathbf{J}}$, where $\bar{\mathbf{J}}$ is the sample mean of the gradient, isolates the extreme-specific sensitivity pattern above the climatological baseline.
The Jacobian lens experiment extends the saliency analysis to intermediate backbone representations. For each backbone stage $k\in\{0,1,2\}$ we compute the stage-wise gradient $\mathbf{J}^{(k)}(t)=\partial s_c/\partial\mathbf{h}^{(k)}$, resulting in a layer-wise view of where in the network extreme-event information concentrates. We quantify the amplification of gradients across the stages by the RMS norm ratio $\|\mathbf{J}^{(k)}\|_\mathrm{rms}$, and assess spatial organization of the Stage-2 gradient field using Moran's I \citep{af54b142-01be-3f96-af47-1730365d8376} and Local Indicators of Spatial Association (LISA; \citep{https://doi.org/10.1111/j.1538-4632.1995.tb00338.x}), hence classifying each grid cell as high-high, high-low, low-high or low-low relative to its spatial neighborhood.

\section{Aurora Characterization}\label{sec: Aurora Characterization}
Before applying perturbations, we establish a clear baseline picture of Aurora's behavior as a dynamical surrogate.

\subsection{Forecast Skill and Generalization}\label{subsec:Forecast Skill and Generalization}
\begin{figure}[h]
  \centering
  \begin{subfigure}[b]{0.49\textwidth}
    \centering
    \includegraphics[width=\textwidth]{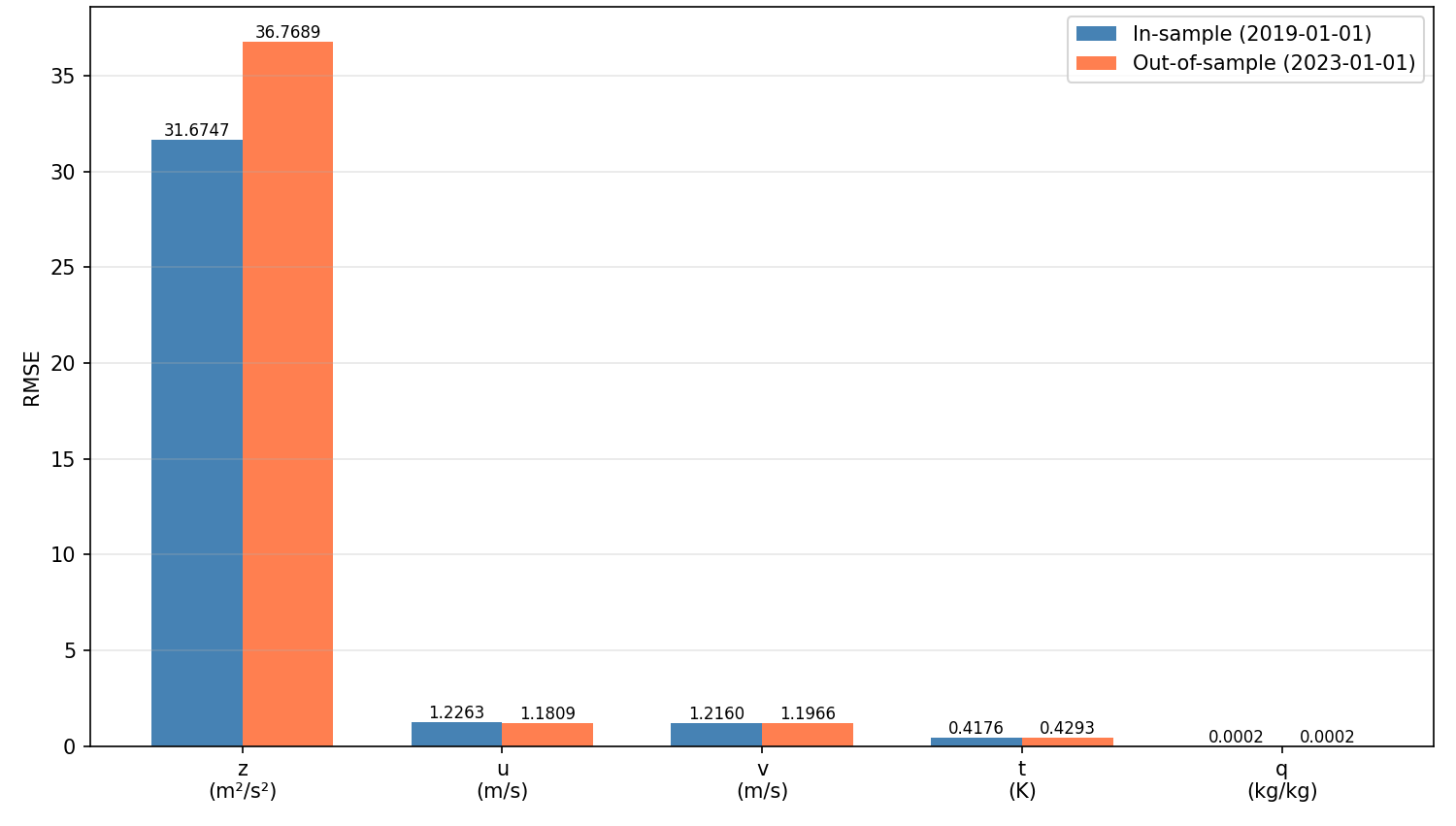}
    \subcaption{Atmospheric Variable RMSE (mean across levels): In-Sample vs. Out-of-Sample}
    \label{fig:atmos_rmse}
  \end{subfigure}
  \hfill
  \begin{subfigure}[b]{0.49\textwidth}
    \centering
    \includegraphics[width=\textwidth]{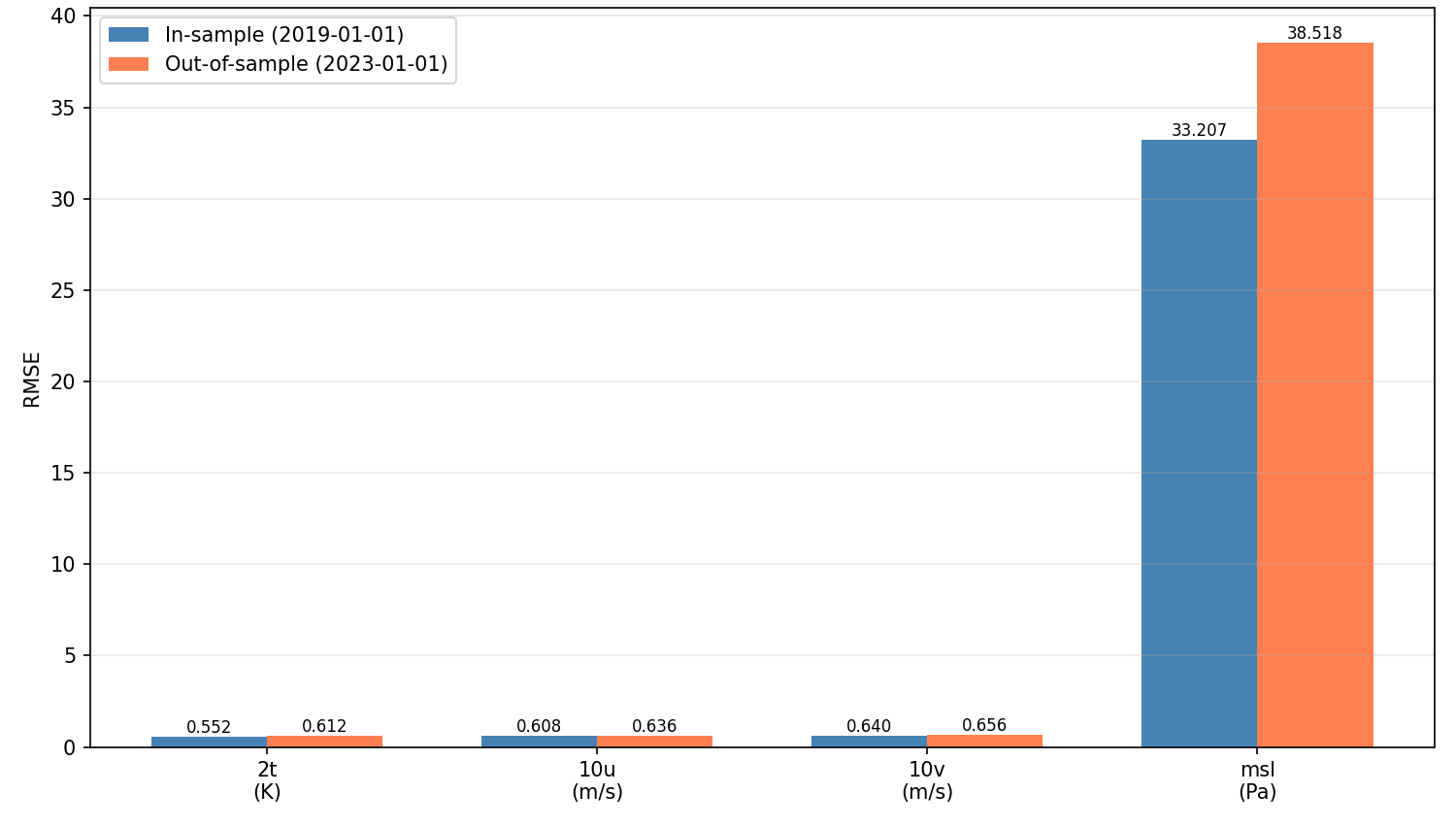}
    \caption{Surface Variable RMSE: In-Sample vs. Out-of-Sample}
    \label{fig:surface_rmse}
  \end{subfigure}
  \caption{In-sample (2019-01-01) vs.\ out-of-sample (2023-01-01) RMSE for atmospheric (left) and surface (right) variables.}
  \label{fig:rmse_comparison}
\end{figure}

\begin{figure}[h]
  \centering
  \includegraphics[width=0.5\textwidth]{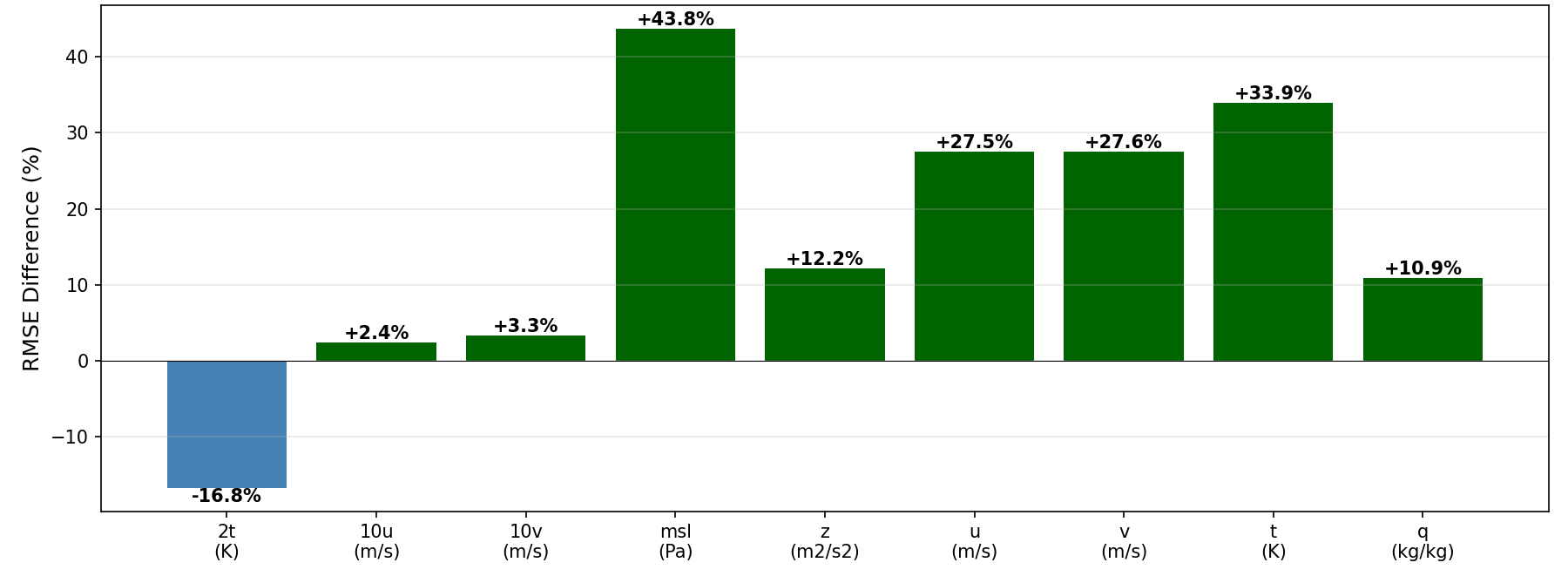}
  \caption{Prithvi WxC RMSE relative to Aurora across all variables (positive = Aurora better, negative = Prithvi better).}
  \label{fig:pct_difference}
\end{figure}

Unless otherwise noted, RMSE throughout this section is the area-weighted (cosine-latitude weighted) global mean, following the standard WeatherBench2 evaluation protocol \citep{https://doi.org/10.1029/2023MS004019}, so that polar grid cells do not disproportionately influence the result. All values are computed from a single representative initialization date rather than averaged across multiple dates. Figure \ref{fig:surface_rmse} shows surface RMSE for in-sample (2019-01-01) versus out-of-sample (2023-01-01) initialization dates, chosen to span Aurora's training/test boundary. Aurora generalizes well for temperature and wind variables: 2t RMSE increases from 0.552 K to 0.612 K (+10.9\%), 10u from 0.608 to 0.636 m/s (+4.8\%), and 10v from 0.640 to 0.656 m/s (+2.4\%). Mean sea-level pressure shows a larger increase (33.2 to 38.5 Pa, +16.0\%), consistent with known difficulties in pressure field prediction at longer lead times.
For atmospheric variables (Figure \ref{fig:atmos_rmse}), geopotential height shows the largest absolute RMSE ($31.7$ in-sample, $36.8$ out-of-sample), while wind components show slight \textbf{improvements} out of sample (u: $-3.7\%$, v: $-1.6\%$), which may reflect favorable initialization conditions on 2023-01-01 rather than improved model generalization.
\begin{figure}[h]
  \centering
  \begin{subfigure}[b]{0.49\textwidth}
    \centering
    \includegraphics[width=\textwidth]{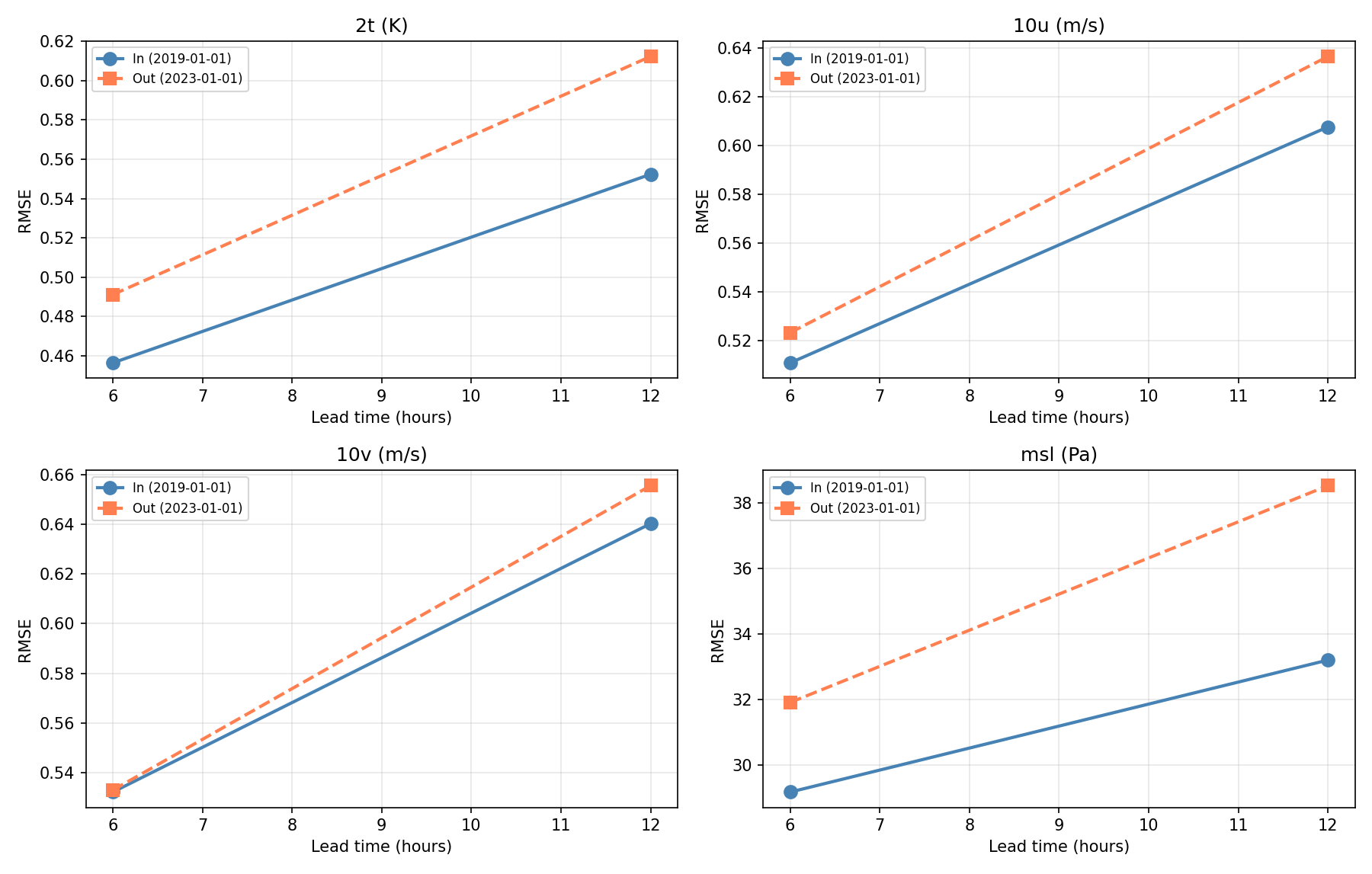}
    \caption{Error Growth Comparison}
    \label{fig:error_growth}
  \end{subfigure}
  \hfill
  \begin{subfigure}[b]{0.5\textwidth}
    \centering
    \includegraphics[width=\textwidth]{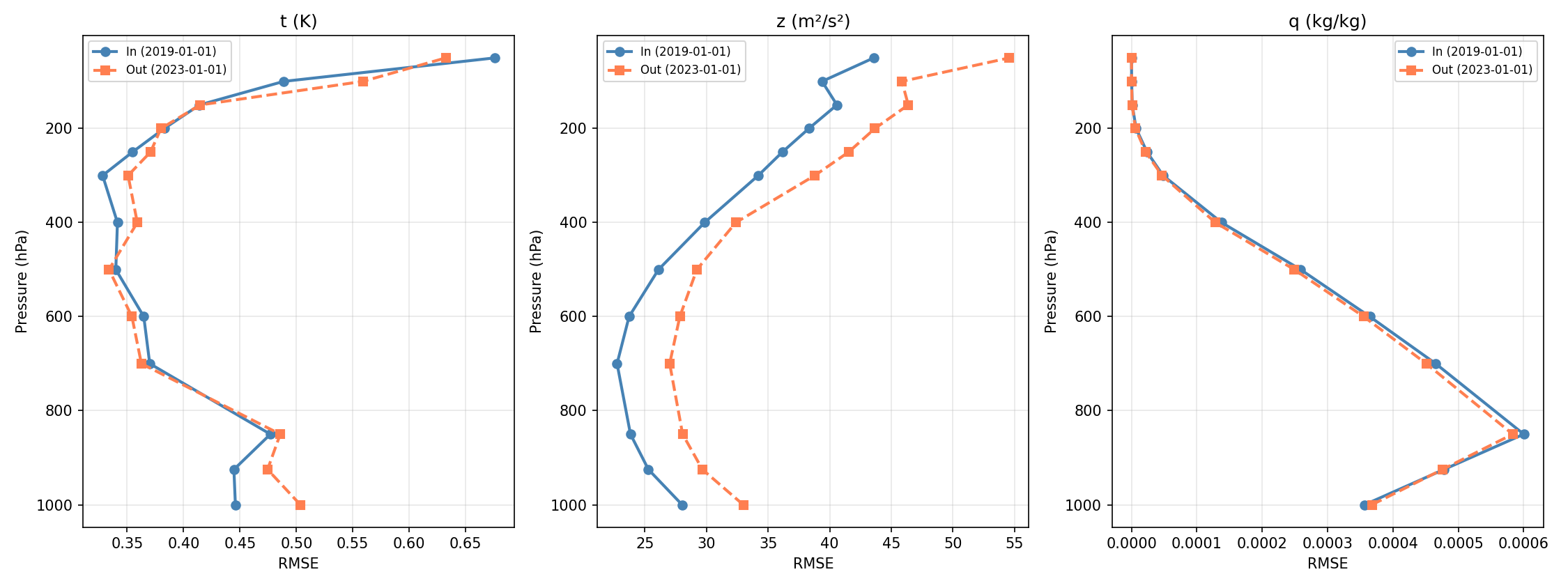}
    \caption{Vertical Profile Comparison}
    \label{fig:vertical_profiles}
  \end{subfigure}
  \caption{Error growth curves (left) and vertical RMSE profiles (right) comparing Aurora against baseline models.}
  \label{fig:error_vertical}
\end{figure}
On the out-of-sample initialization date (2023-01-01), compared to Prithvi WxC (Figure \ref{fig:pct_difference}), Aurora outperforms on most variables: msl ($+43.8\%$ Prithvi relative RMSE), atmospheric u ($+27.5\%$), v ($+27.6\%$), and t ($+33.9\%$). Prithvi outperforms Aurora only on 2t ($-16.8\%$), suggesting Aurora's surface temperature prediction is relatively weaker than its atmospheric-level performance.

These results establish that Aurora is a reliable and generalizing forecast system on the out-of-sample date we use for our perturbation experiments. Aurora achieves competitive RMSE across all lead times and variables. These results are consistent with the published benchmarks in \citet{Bodnar2025}. Competitive RMSE, however, does not itself distinguish genuine dynamic forecasting from a model that exploits the atmosphere's own autocorrelation structure. A forecast that lags or leads the truth by one or two steps ($\pm6$-$12$h) may still produce low RMSE simply because atmospheric fields change slowly. The minimum error need not occur at exact time alignment.

\subsection{Time Registration Analysis}\label{subsec:time registration analysis}
\begin{figure}[h]
  \centering
  \includegraphics[width=0.49\textwidth]{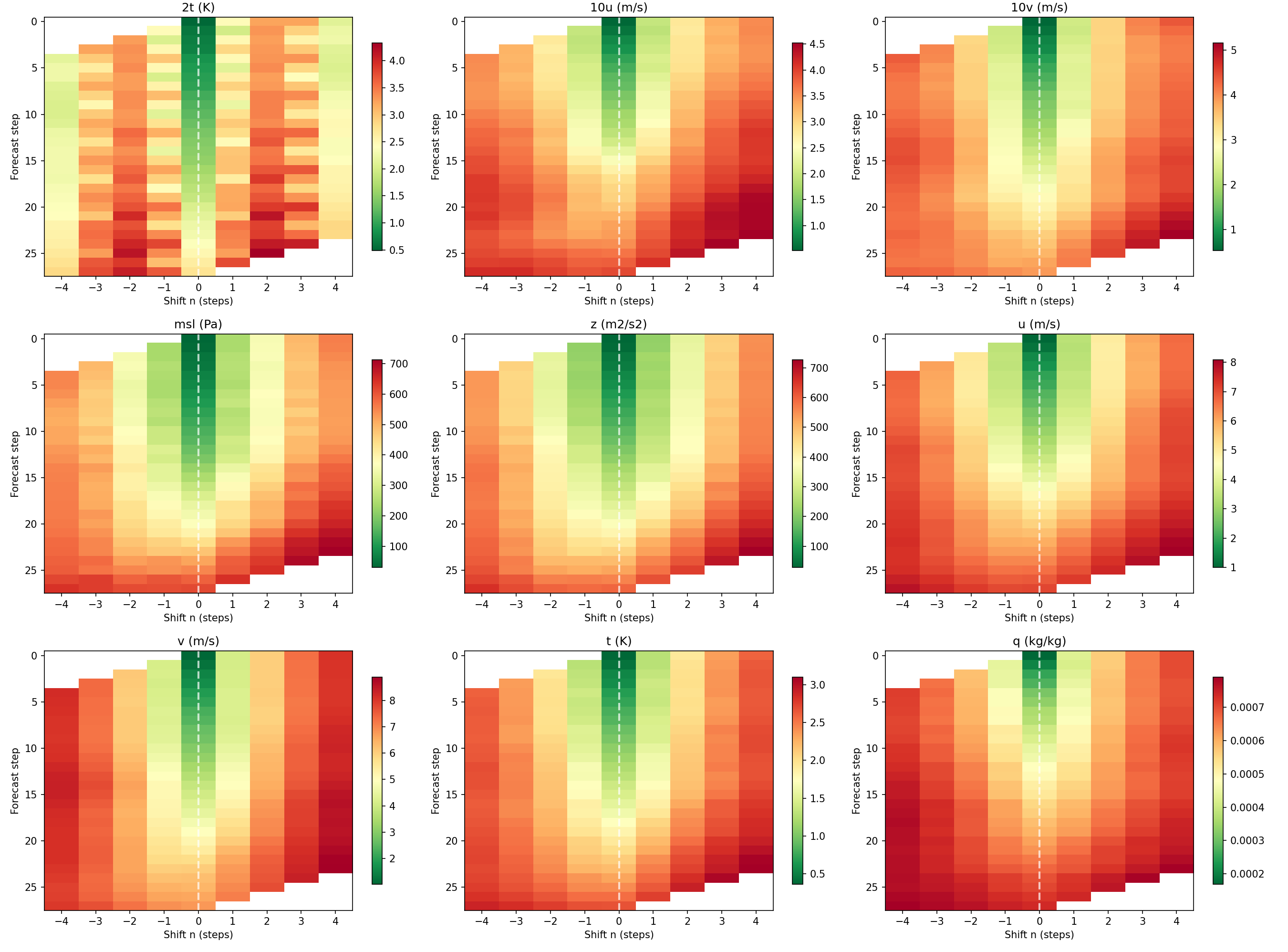}
  \caption{Time-shift analysis for Aurora (2023-01-01). Each cell shows $\text{RMSE}(\mathbf{x}_{t+m}^{\text{ERA5}}, \hat{\mathbf{x}}_{t+m+n}^{\text{Aurora}})$ for forecast step $m$ (y-axis) and time shift $n$ (x-axis). The white dashed line marks $n=0$; minimum RMSE along each row at $n=0$ indicates correct time registration. Aurora is well-registered at short lead times, with minor phase drift toward $n=-1$ emerging at longer forecast steps for pressure and wind variables.}
  \label{fig:time_shift_heatmap}
\end{figure}

A well-calibrated forecast model should achieve its minimum RMSE when compared against the truth at the correct lead time, that is, when the time shift $n=0$. A systematic offset in the minimum would indicate that the model's predictions are temporarily misregistered, either running ahead of or lagging behind the true atmospheric evolution. To determine whether Aurora's predictions are temporally specific, we conduct a time registration analysis across forecast lead times $m=1,\ldots,28$ steps.
\\Figure \ref{fig:time_shift_heatmap} shows the time-shift analysis on Aurora on 2023-01-01. For each forecast step $m$ and each time step $n\in\{-4,\ldots,+4\}$ steps, we compute the $\text{RMSE}(\mathbf{x}^{\text{ERA5}}_{t+m},\hat{\mathbf{x}}^{\text{Aurora}}_{t+m+n})$, that is, the error when Aurora's prediction at step $m+n$ is compared against the truth at step $m$. Across all nine variables, the minimum RMSE consistently falls at or very close to $n=0$ at short forecast steps, confirming that Aurora is correctly time-registered in its early forecast hours. At longer lead times (forecast steps beyond 15), the minimum shifts slightly toward $n=-1$ for several variables, including $2t$, $10u$, and $10v$, suggesting a modest tendency for Aurora's forecast to run slightly ahead of the true atmospheric state at extended ranges. The effect is most pronounced for geopotential height ($z$) and mean sea-level-pressure (msl), where the RMSE gradient across shifts is steeper, indicating these variables are more sensitive to temporal misregistration than temperature or humidity. Overall, the time-shift structure is well-behaved and consistent with a model that faithfully tracks atmospheric evolution at short-to-medium range, with only minor phase drift emerging at longer leads. This rules out step-level persistence as a dominant mechanism. What remains to be quantified is the size of the gap, i.e., how much better Aurora is against persistence, and how that advantage evolves over the forecast horizon.

\subsection{Forecast Skill Against Persistence}\label{subsec:forecast skill against persistence}
\begin{figure}[h]
  \centering
  \includegraphics[width=0.48\textwidth]{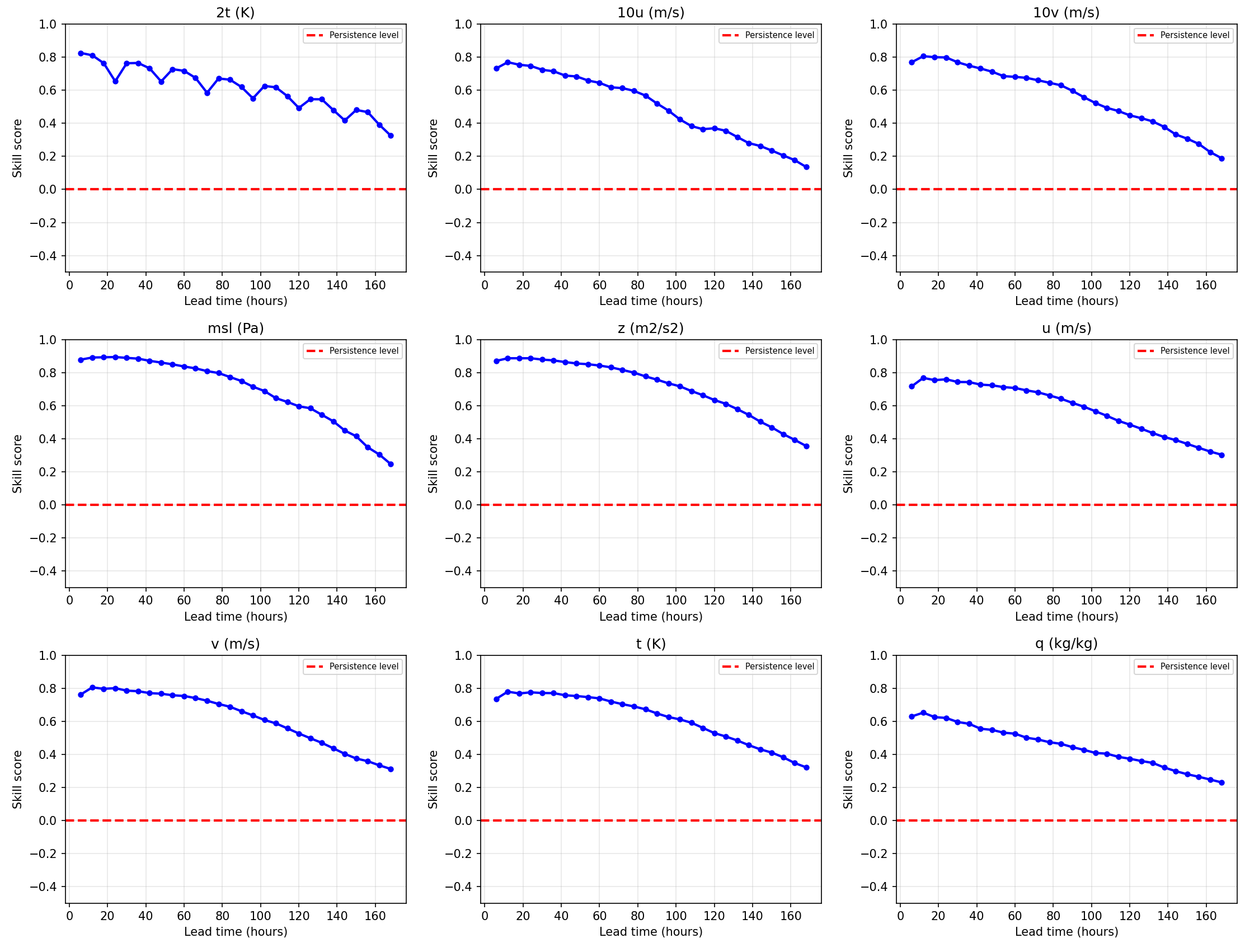}
  \caption{Forecast skill score ($S = 1 - \text{RMSE}_{\text{Aurora}} / \text{RMSE}_{\text{persistence}}$) across lead times from 6 to 168 hours for all surface and atmospheric variables (2023-01-01). Values above zero indicate Aurora outperforms persistence; Aurora maintains positive skill across all variables out to 7 days.}
  \label{fig:skill_score}
\end{figure}

We now compare Aurora's RMSE against a persistence baseline, where the persistence baseline is the trivial forecast that holds the most recent observation constant, to characterize forecast skill as a function of lead time. The persistence errors reported here are identical to those used as the null in \S \ref{subsec:time registration analysis}.\\
Figure \ref{fig:skill_score} shows the forecast skill score $S = 1 - \text{RMSE}_\text{Aurora} / \text{RMSE}_\text{persistence}$ across lead times from 6 to 168 hours (7 days). Aurora consistently and substantially outperforms persistence across all variables out to at least 7 days. At 6-hour lead time, skill scores exceed 0.7 for all variables. At 72 hours, scores remain above 0.5 for msl and z. The skill degrades gracefully, with no variable showing a sudden collapse.
Aurora's global skill score exceeds persistence across all lead times examined, with skill degrading smoothly rather than collapsing abruptly. These results are consistent with a model that tracks genuine atmospheric dynamics rather than statistical regularities. However, the area-weighted global RMSE aggregates over regions with substantially different error characteristics: the tropics, mid-latitudes, and polar regions evolve on different timescales and respond differently to initial perturbations. Therefore, a single global metric obscures the spatial structure of forecast degradation. The following section examines error growth at a finer resolution, decomposing skill decay by latitude band and characterizing the rate of divergence from truth in the first 6-12 hours (the window most relevant to the perturbation experiments that follow).

\subsection{Error Growth Dynamics}\label{subsec:Error Growth Dynamics}
\begin{figure}[h]
  \centering
  \begin{subfigure}[b]{0.49\textwidth}
    \includegraphics[width=\textwidth]{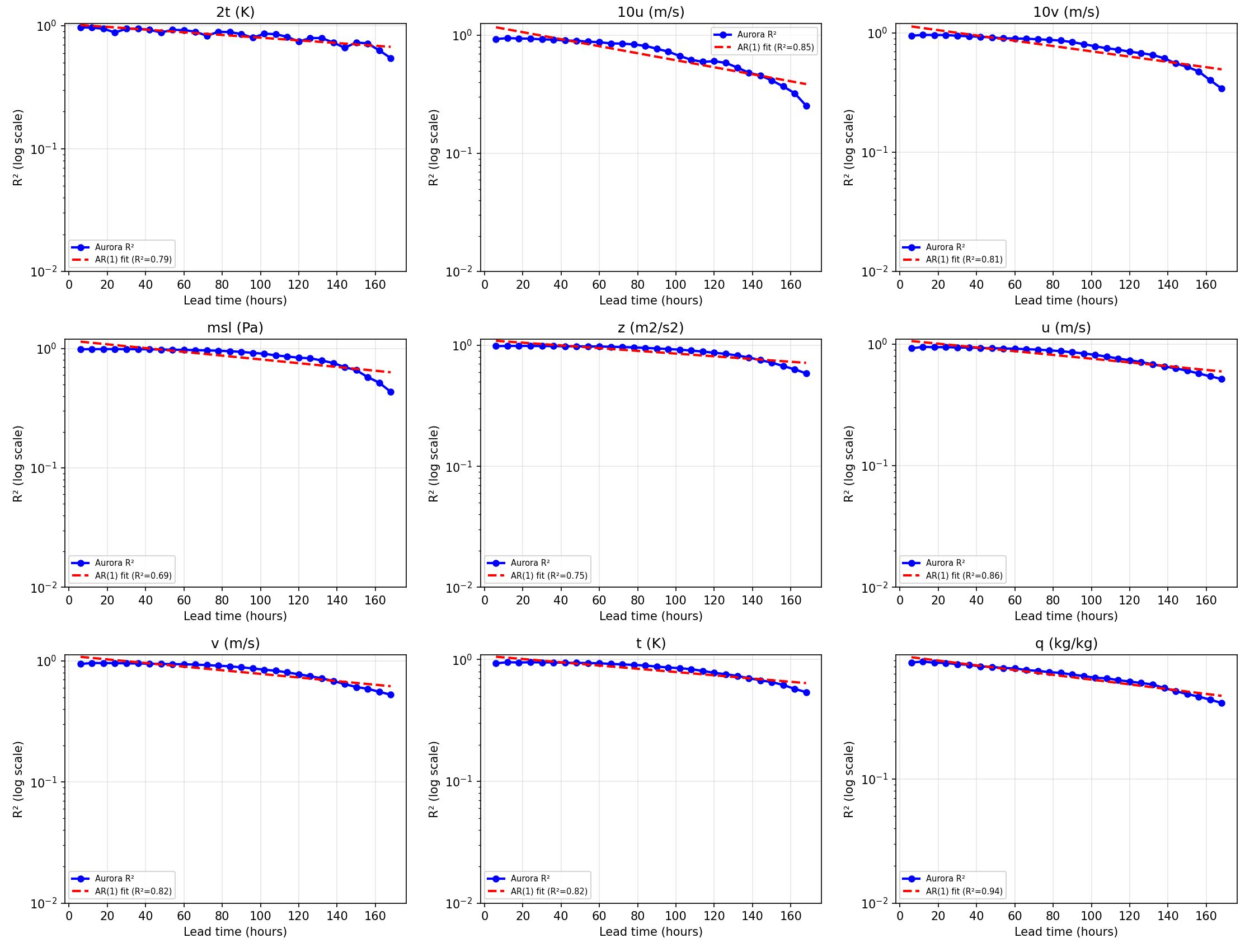}
    \caption{R$^2$ decay compared against AR(1) model.}
    \label{fig:r2_decay}
  \end{subfigure}
  \hfill
  \begin{subfigure}[b]{0.49\textwidth}
    \includegraphics[width=\textwidth]{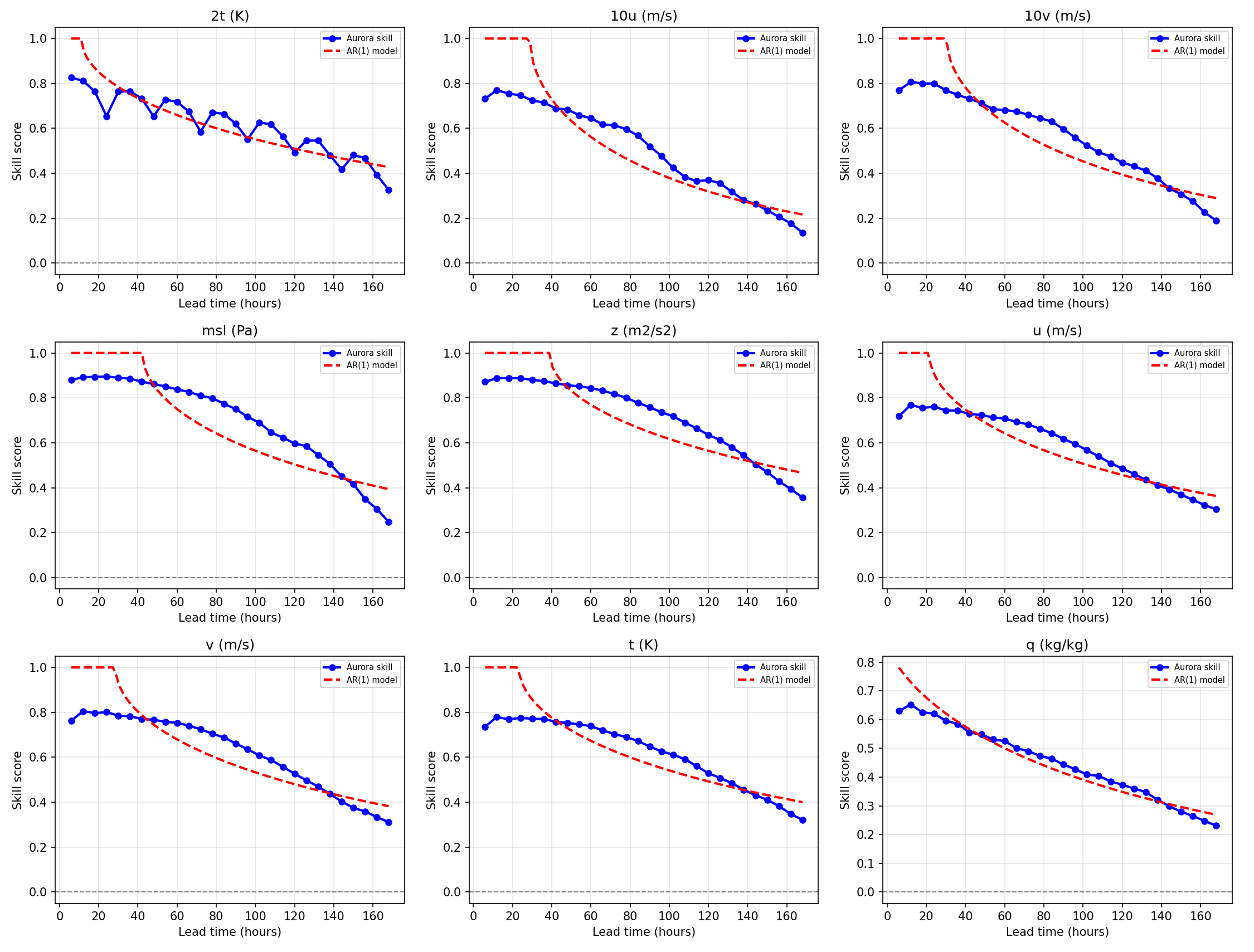}
    \caption{Forecast skill score vs.\ AR(1) theoretical decay.}
    \label{fig:skill_ar1}
  \end{subfigure}
  \caption{Error growth dynamics for Aurora (2023-01-01). Near-linear R$^2$ decay in log space is consistent with exponential error growth at a roughly constant effective Lyapunov exponent. Skill scores converge toward the AR(1) prediction at longer lead times but fall below it in the first 40 hours for wind variables, reflecting an elevated early-hour growth rate.}
  \label{fig:error_growth}
\end{figure}

\begin{figure}[h]
  \centering
  \includegraphics[width=\textwidth]{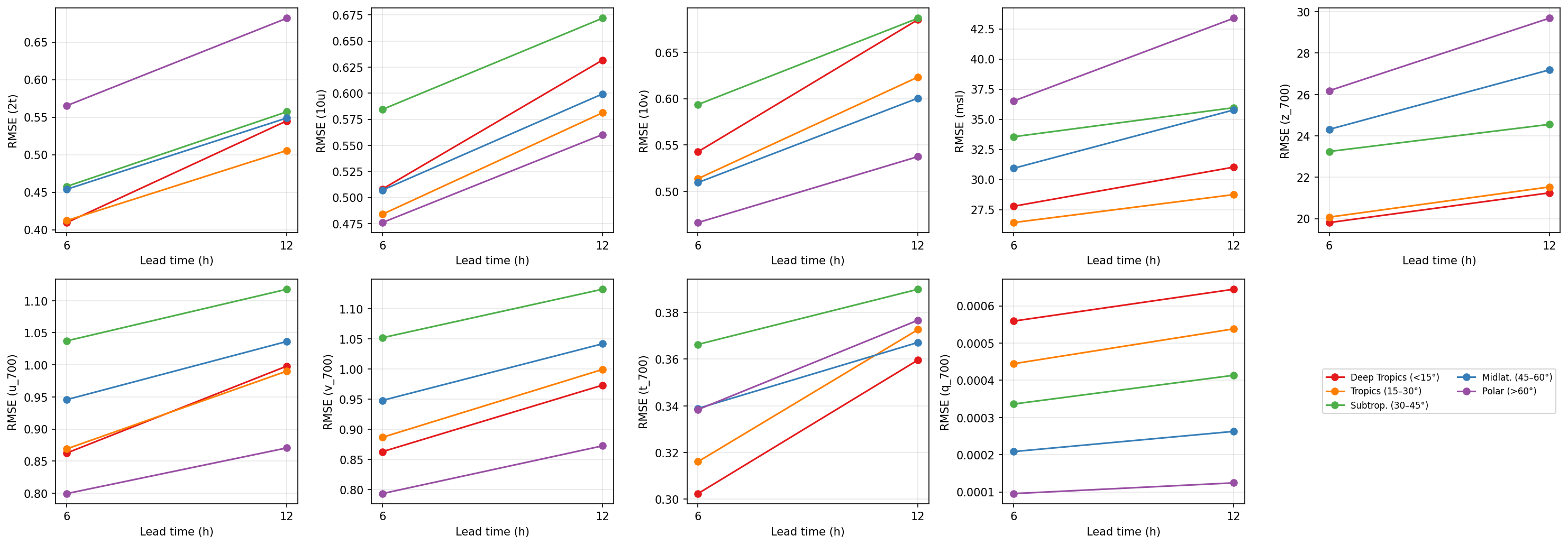}
  \caption{Error growth curves from 6 to 12 hours across latitude bands (Deep Tropics, Tropics, Subtropics, Midlatitudes, Polar) for all surface and atmospheric variables.}
  \label{fig:error_growth_curves}
\end{figure}

\begin{figure}[h]
  \centering
  \begin{subfigure}[b]{0.49\textwidth}
    \includegraphics[width=\textwidth]{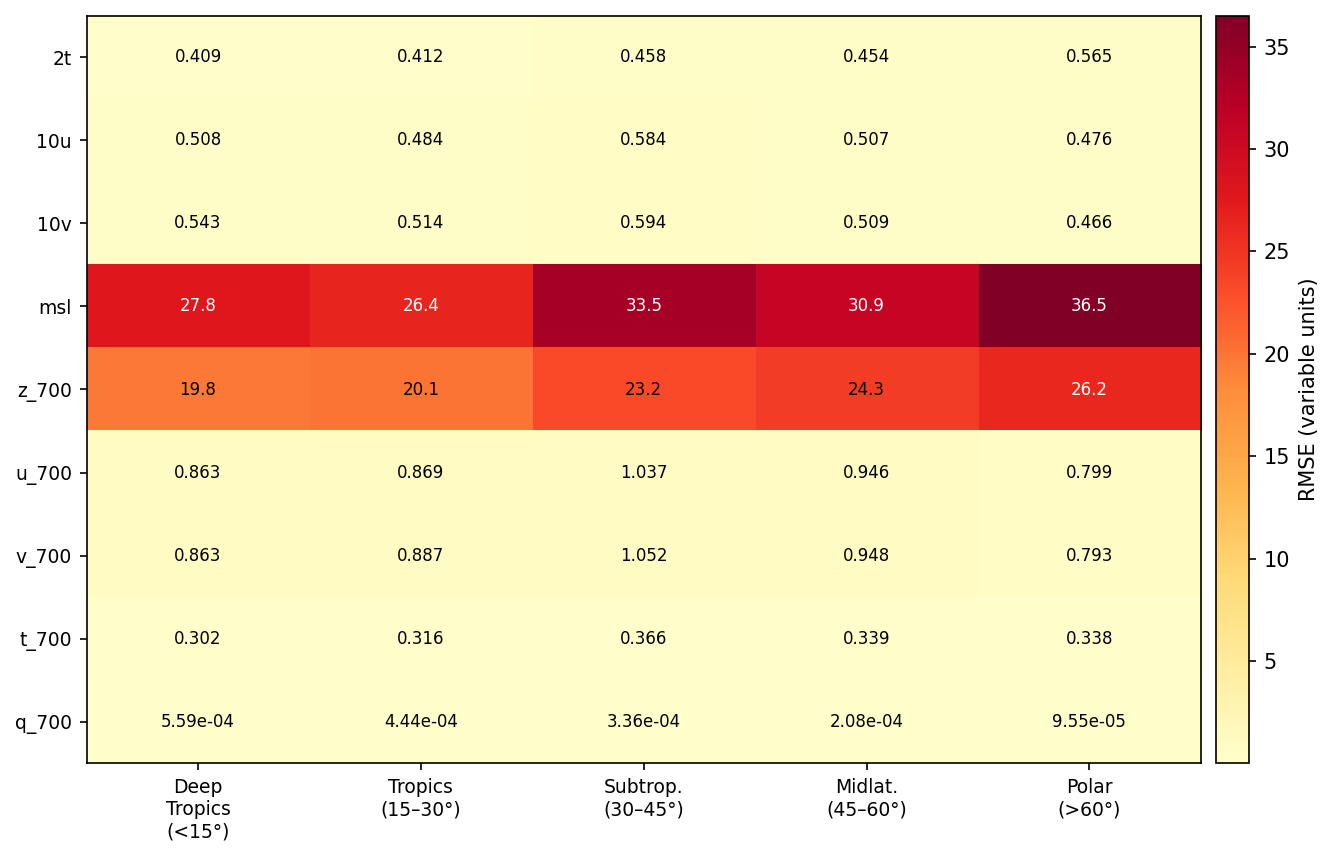}
    \caption{Absolute RMSE at 6-hour lead time (mean over dates).}
    \label{fig:heatmap_rmse_6h}
  \end{subfigure}
  \hfill
  \begin{subfigure}[b]{0.49\textwidth}
    \includegraphics[width=\textwidth]{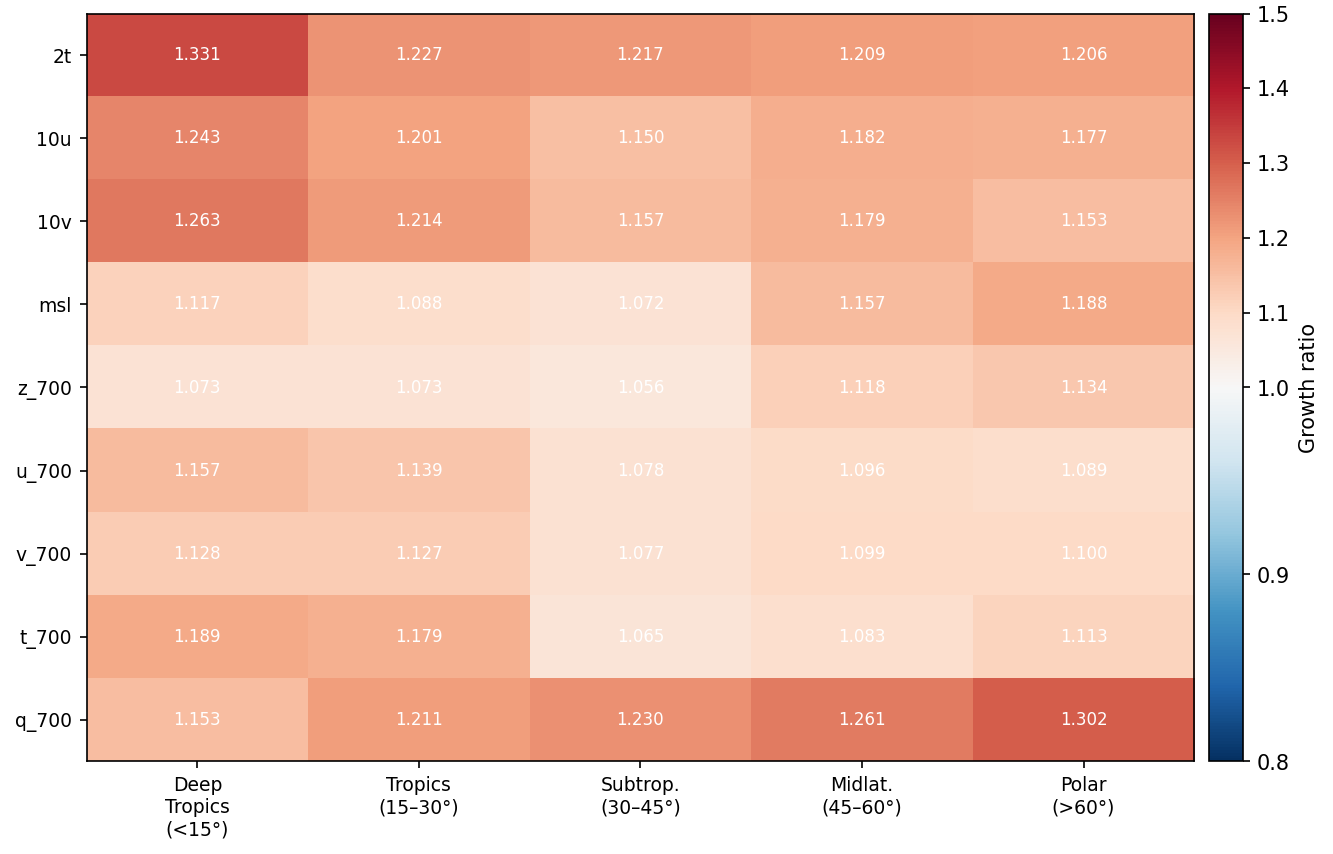}
    \caption{Error growth ratio (12h RMSE / 6h RMSE).}
    \label{fig:heatmap_growth_ratio}
  \end{subfigure}
  \caption{Regional error structure by variable and latitude band (2023-01-01). Left: absolute RMSE at 6 hours. Right: growth ratio between 6 and 12 hours; values above 1 indicate growing errors.}
  \label{fig:regional_heatmap}
\end{figure}

The global skill metrics of the preceding section \S\ref{subsec:forecast skill against persistence} treat Aurora's forecast error as a spatially uniform quantity. We now relax that assumption in two ways: by decomposing error growth across latitude bands, and by characterizing the rate of decay through comparison with an AR(1) process sharing Aurora's empirical autocorrelation. As in the preceding sections, this analysis uses the single out-of-sample initialization date (2023-01-01); it is a per-date diagnostic, not a multi-date average. The AR(1) reference is not fit to the historical ERA5 record for the corresponding season: for each variable we estimate the lag-1 temporal autocorrelation $\rho$ directly from Aurora's own forecast output for this date, and compute the theoretical decay curve as R\textsuperscript{2}$(t)=\rho^{2t}$, the expected squared correlation for a stationary AR(1) process with that persistence. The comparison, therefore, asks whether Aurora's error growth is consistent with a stationary process matched to its \textit{own} autocorrelation structure, not whether it matches climatological variability.\\
Figures \ref{fig:r2_decay} and \ref{fig:skill_ar1} examine error growth from two complementary perspectives. Figure \ref{fig:r2_decay} shows the R\textsuperscript{2} decay of Aurora's forecast error over time, compared against an AR(1) process with matching autocorrelation. For most variables, Aurora's R\textsuperscript{2} decay is well-approximated by the AR(1) model with fits between R\textsuperscript{2}=0.69 (msl) and R\textsuperscript{2}=0.94 (q). This near-linear decay in log-space is consistent with exponential error growth at a roughly constant effective Lyapunov exponent. This behavior is expected of a system operating in the chaotic regime.\\
Figure \ref{fig:skill_ar1} provides a finer view of the same phenomenon by plotting Aurora's skill score directly against the AR(1) theoretical decay curve for each variable. Two features are noteworthy. First, at longer lead times Aurora's skill converges toward and in some variables crosses the AR(1) curve, confirming that the AR(1) is an appropriate asymptotic model for Aurora's error structure. Second, at short lead times (roughly 0-40 hours), Aurora's skill falls systematically below the AR(1) prediction for wind variables (10u, 10v, u, v). This early-hour deficit is not captured by the AR(1) and likely reflects initial condition uncertainty or a brief model spinup (a period during which error growth is faster and more irregular than the constant-exponent assumption implies). This is precisely the window most relevant to the perturbation experiments that follow, and it suggests that the effective Lyapunov exponent is not strictly constant but is elevated in the first several forecast hours.
Taken together, the two figures are consistent with Aurora behaving as a physically reasonable chaotic surrogate: its error growth is exponential in the mean, well-described by a simple AR(1) model at longer lead times, and most sensitive to perturbation in the early forecast window where growth rates are highest.\\
Figures \ref{fig:error_growth_curves} and \ref{fig:regional_heatmap} decompose this error growth spatially, showing growth from 6 to 12 hours across latitude bands (deep tropics, tropics, subtropics, midlatitudes, polar). Polar regions show consistently higher errors for temperature (2t) and wind (10u,10v), while subtropical errors are elevated for wind components, consistent with known patterns in ERA5-based model evaluation. 
Figure \ref{fig:heatmap_growth_ratio} reveals a complementary and in some cases reversed picture when examining growth rates rather than absolute levels. Temperature (2t) errors grow fastest in the Deep Tropics (ratio 1.331), consistent with the convective instability of that region. By contrast, $q_{700}$ shows the opposite gradient: growth is slowest in the Deep Tropics (1.153) and increases monotonically poleward, reaching its global maximum in the Polar band (1.302). This reversal suggests a structurally different error mechanism in the extratropics, likely linked to the advection of moisture anomalies by polar jet dynamics rather than local convective processes. Together, the two heatmaps establish that Aurora's error structure is neither geographically uniform nor variable-uniform: the most dynamically active growth occurs in the Deep Tropics for temperature and winds, and in the Polar band for moisture, a pattern that directly informs the geographic targeting strategy in the perturbation experiments that follow.
These characterizations are consistent with Aurora behaving as a physically reasonable chaotic surrogate, with error growth patterns that are interpretable and consistent with atmospheric dynamics. \S\ref{sec:discussion-conclusion} discusses why we treat this as suggestive of, rather than proof of, genuine chaotic sensitivity at Aurora's operating resolution.\\

\section{Jiu-Jitsu Feasibility Analysis}\label{sec:jiu-jitsu feasibility analysis}
\subsection{Drift Comparison}\label{subsec:drift comparison}

The central question of the Weather Jiu-Jitsu framework is whether a small, targeted perturbation can grow faster than Aurora's own forecast error. To test this, we define two quantities. The \textit{error drift} measures how far Aurora's unperturbed forecast deviates from the ERA5 truth, representing the irreducible uncertainty already present in the model. The \textit{chaotic drift} measures how far a perturbed forecast deviates from the unperturbed forecast, representing the additional divergence induced by the perturbation. Formally, the feasibility ratio is:

\begin{equation}
  \mathcal{F} = \frac{\text{chaotic drift}}{\text{error drift}} = \frac{\text{RMSE}(\hat{\mathbf{x}}_\text{truth},\, \hat{\mathbf{x}}_\text{ctrl})}{\text{RMSE}(\hat{\mathbf{x}}_\text{ctrl},\, \mathbf{x}_\text{pert})}
  \label{eq:feasibility}
\end{equation}

A ratio $\mathcal{F} > 1$ indicates that the perturbation is amplifying faster than the model's background error and hence the regime in which deliberate chaos exploitation is plausible. A ratio $\mathcal{F} \ll 1$ indicates that the perturbation is being damped relative to existing forecast uncertainty, and the intervention has no meaningful leverage. We begin with the simplest perturbation: a spatially localized Gaussian bump (centered at the middle of the domain, $\sigma=5\%$ of the grid dimension) added to surface temperature, to establish a dynamically uninformed baseline before introducing geographic and physics-based targeting strategies that place the perturbation deliberately at flow-sensitive locations.

\begin{figure}[h]
  \centering
  \begin{subfigure}[b]{0.48\textwidth}
    \centering
    \includegraphics[width=\textwidth]{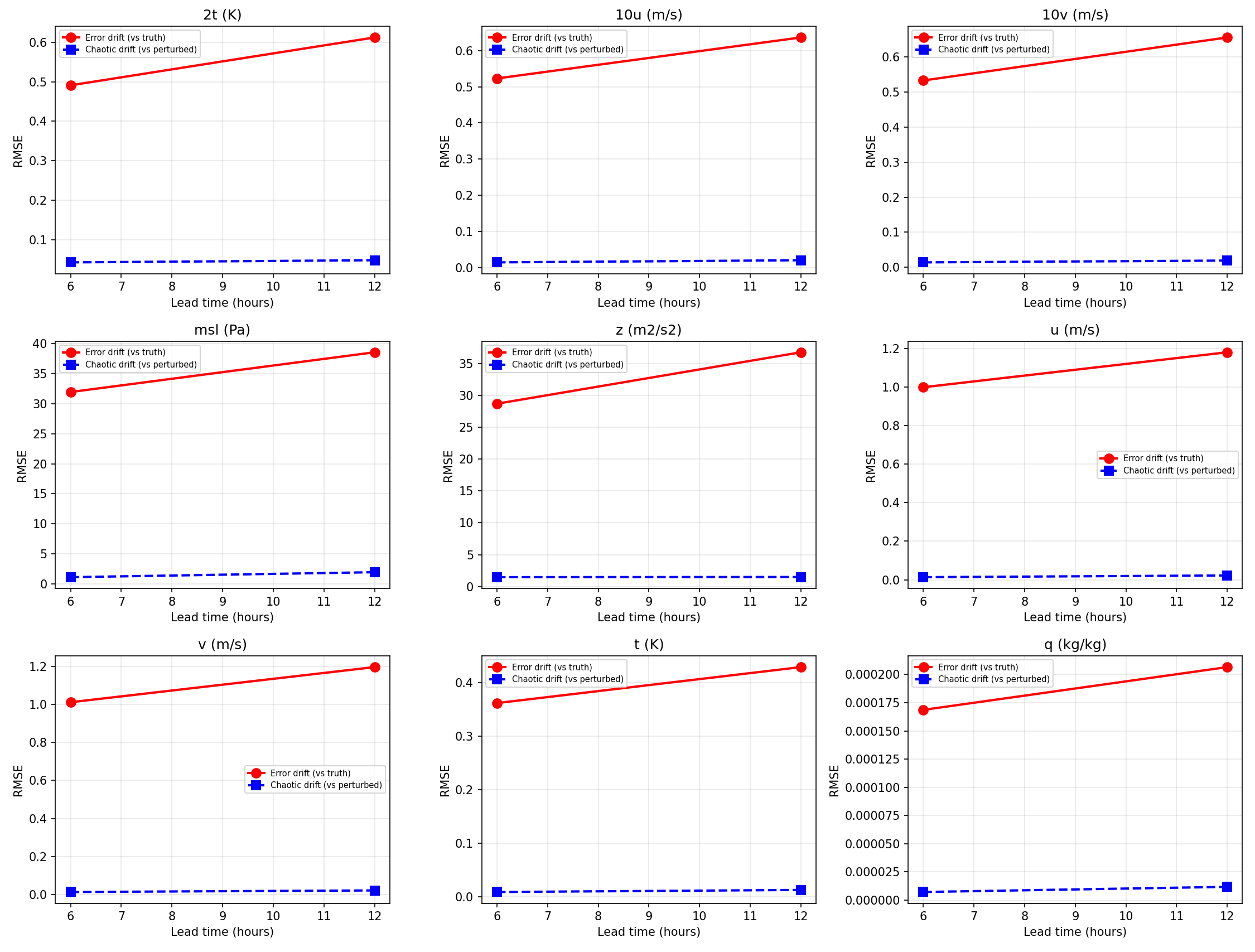}
    \caption{Perturbation: 2t $+= 1.0$ K}
    \label{fig:drift_1K}
  \end{subfigure}
  \hfill
  \begin{subfigure}[b]{0.48\textwidth}
    \centering
    \includegraphics[width=\textwidth]{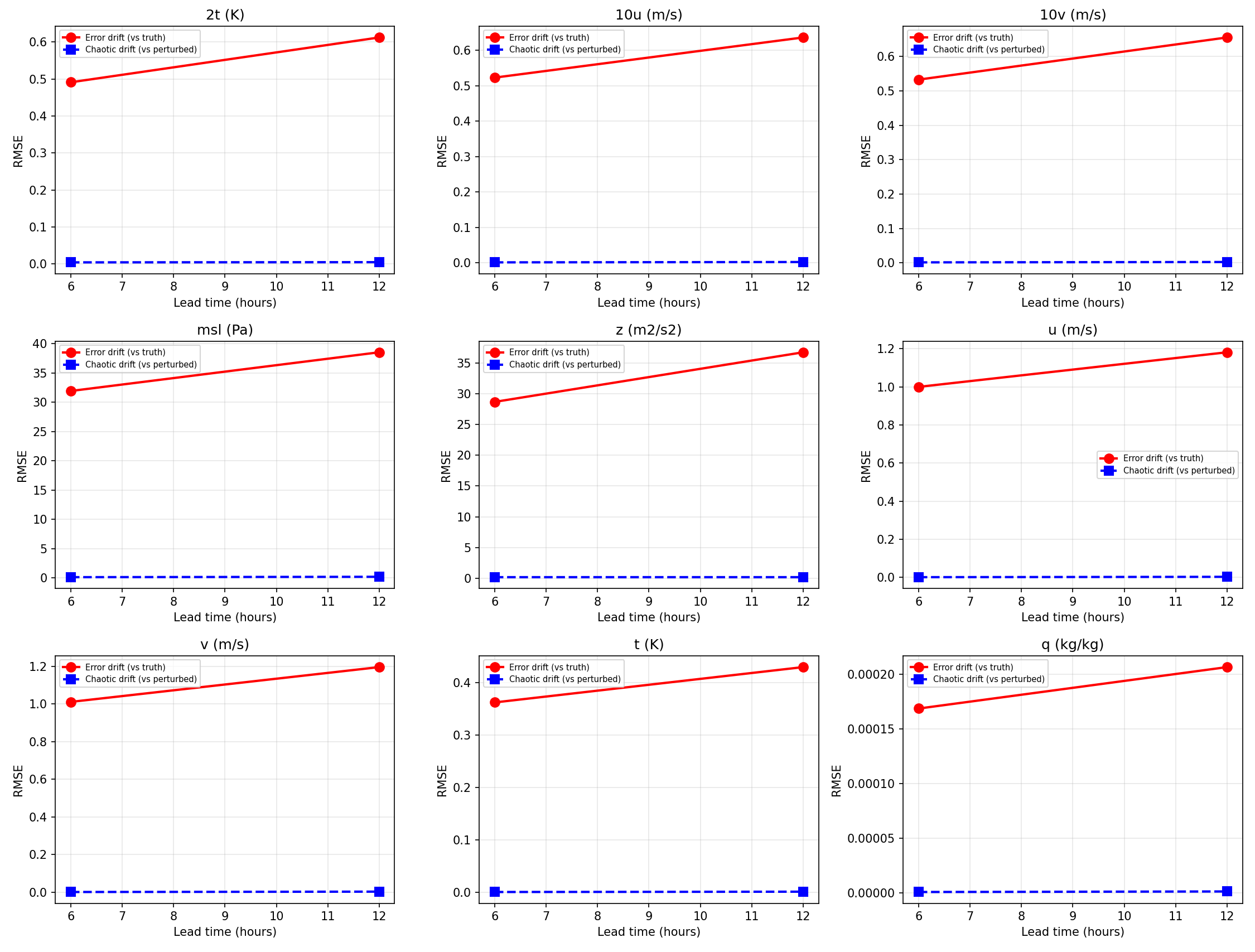}
    \caption{Perturbation: 2t $+= 0.1$ K}
    \label{fig:drift_01K}
  \end{subfigure}
  \caption{Error drift (red, Aurora vs.\ ERA5 truth) and chaotic drift (blue, perturbed vs.\ unperturbed run) at 6 h and 12 h lead times for surface temperature perturbations of 1.0 K (left) and 0.1 K (right). In both cases, chaotic drift remains far below error drift across all variables, yielding feasibility ratios well below 1.}
  \label{fig:drift_comparison}
\end{figure}

Figure \ref{fig:drift_comparison} shows the error drift (red, Aurora vs.\ ERA5 truth) and chaotic drift (blue, perturbed vs.\ unperturbed Aurora) at 6 h and 12 h lead times for perturbations of $\Delta T = 1.0$ K (Figure \ref{fig:drift_1K}) and $\Delta T = 0.1$ K (Figure \ref{fig:drift_01K}). The separation between the two curves is dramatic and consistent across all variables and both magnitudes. At 12 h with the 1.0 K perturbation, error drift for 2t reaches 0.612 K while chaotic drift is only 0.048 K (a feasibility ratio of approximately 0.08). For msl, error drift is 38.5 Pa versus chaotic drift of 1.98 Pa. Reducing the perturbation to 0.1 K does not qualitatively change the picture: chaotic drift shrinks by roughly an order of magnitude while error drift is unchanged, indicating that the perturbation is operating in the linear regime where divergence scales proportionally with magnitude rather than amplifying exponentially. The chaotic drift curves remain far below the error drift curves throughout the 6 to 12h window in both cases (the perturbation has not overtaken the model's own forecast uncertainty at these short lead times), but they are not flat: the chaotic drift grows measurably from 6h to 12h for every variable (Figures ~\ref{fig:drift_01K} and \ref{fig:drift_1K}), a point we return to in\S\ref{subsec:ftle-check} alongside a direct check of the local flow instability at the perturbation site. This absence of rapid, exponential divergence under an untargeted perturbation is consistent with the broader empirical picture in the AI-weather-model literature, where several studies report an apparent lack of a butterfly effect in early AI forecast models \citep{selz2023,shen2025butterfly,hassanzadeh2026missingbutterfly} (see \S\ref{sec:discussion-conclusion} for the fuller discussion); the targeted perturbation strategies that follow are, in part, an attempt to locate the regions of state space where such divergence is nevertheless present. Taken together, these results establish that a spatially fixed, dynamically uninformed temperature perturbation, regardless of magnitude, falls short of the feasibility threshold of 1, motivating the targeted geographic and gradient-based strategies examined in the sections that follow, which place perturbations at locations selected by the flow's own sensitivity structure rather than at a single fixed point.

\subsection{Local Instability Check: Forward FTLE at the Perturbation Site}\label{subsec:ftle-check}
\begin{figure}[h]
  \centering
  \includegraphics[width=0.75\linewidth]{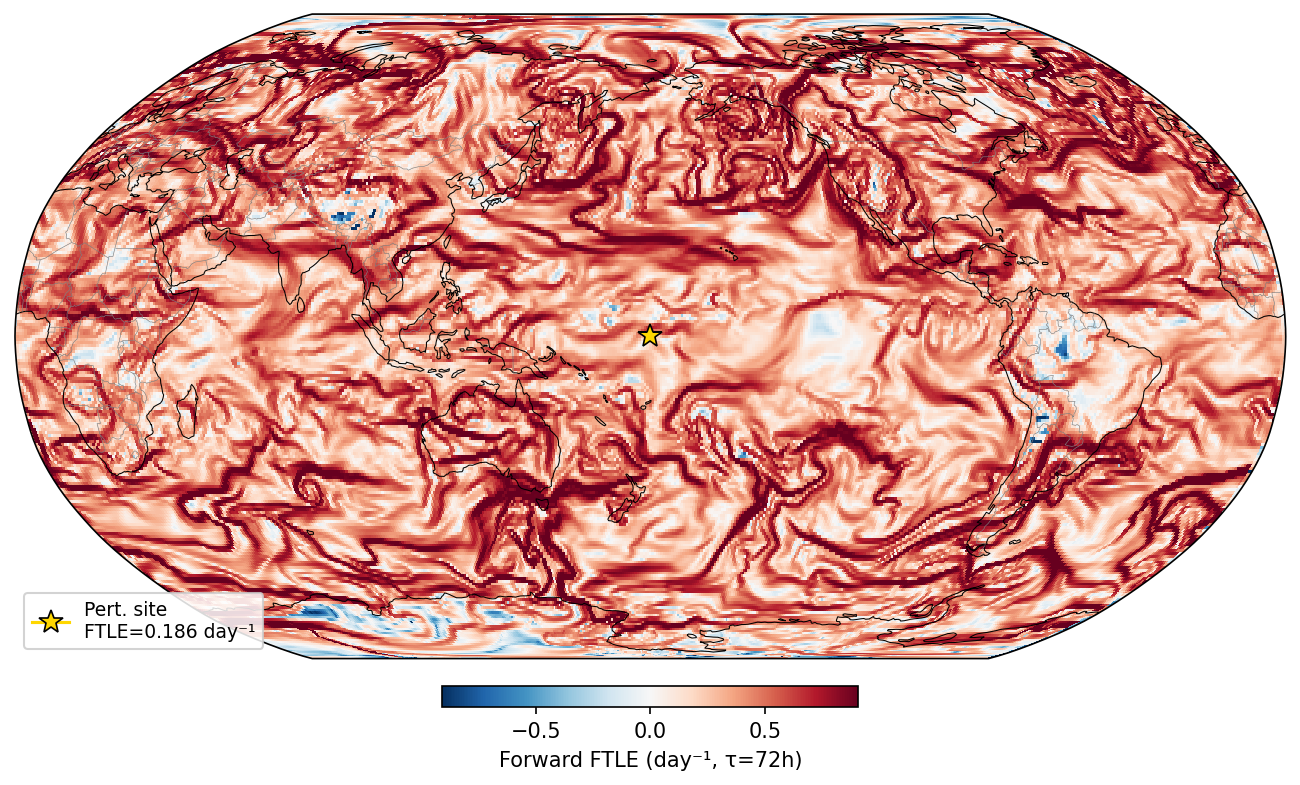}
  \caption{Forward FTLE field ($\tau=72$h, 700 hPa horizontal winds, 2023-01-01). The perturbation site (star) has FTLE $=0.186$ day$^{-1}$, positive but well below the field's maximum ($1.40$ day$^{-1}$); 92.6\% of the domain is positive over this window.}
  \label{fig:ftle}
\end{figure}

We compute the forward finite-time Lyapunov exponent (FTLE) at the site and date used for the perturbation experiments of \S\ref{subsec:drift comparison} (0.0\textdegree N, 180.0\textdegree E, 2023-01-01), from 700 hPa horizontal winds over a $\tau=72$h window (Figure~\ref{fig:ftle}). The local FTLE at this site is positive: $0.186$ day$^{-1}$. It is not the most unstable point available -- 92.6\% of the global domain is positive over this window, up to $1.40$ day$^{-1}$, and the perturbation site sits in a visibly calmer region of the field, consistent with the "calm winter day" character of 2023-01-01 established in \S\ref{subsec:Forecast Skill and Generalization}.

A positive local FTLE predicts a medium-horizon predictability window of the kind already captured by the AR(1) fit of \S\ref{subsec:Error Growth Dynamics}, while also indicating a degree of nonlinearity in Aurora's local dynamics beyond what a purely linear, AR(1)-type process would produce. One could argue this merely reflects the model shadowing some trajectory rather than exhibiting genuine sensitivity; however, 2023-01-01 is out-of-sample for Aurora's training window (\S\ref{subsec:Data and Model}), which weighs against that explanation.

Consistent with this positive FTLE, perturbing at this site produces trajectory divergence that increases as the time lag increases: chaotic drift grows from 6h to 12h for every variable examined (Figure~\ref{fig:drift_comparison}, \S\ref{subsec:drift comparison}), rather than remaining fixed or damping out. This is the behavior a positive local Lyapunov exponent predicts, and it is the behavior an AR(1)-type process -- which has no flow-dependent notion of local stability -- would not produce on its own.

This bears on a natural objection to using $\mathcal{R}_t$ as a proxy for local instability: $\gamma_t$ is an RMSE ratio for one fixed, untargeted perturbation shape at one horizon, not a tangent-space stretching rate along the most unstable direction, so it is not formally the same quantity as $\lambda(\mathbf{x}_t,T)$. But the two quantities are both measures of local sensitivity to perturbation, and we would expect $\mathcal{R}_t$'s magnitude to trend directionally with the local FTLE even without being numerically identical to it. The result above is a first, single-site instance of that directional correspondence: positive FTLE, growing chaotic drift. Extending this check across multiple sites and flow regimes, rather than the single date examined here, would establish the trend more rigorously and is a natural next step.
 
\subsection{Latent Space Analysis}\label{subsec:latent space analysis}
A control strategy must ultimately exploit the structure of the state space through which trajectories pass. If the weather foundation model's internal representations have no meaningful geometry relative to meteorological phenomena of interest, gradient-based steering becomes unreliable. We analyze Aurora's encoder and backbone representations along two axes: first, whether they capture large-scale thermodynamic structure at all; and second, whether they distinguish extreme atmospheric states from normal ones. The answers to these two questions have different implications for the design of perturbation strategies, as the following subsections \S\ref{subsec:seasonal structure} and \S\ref{subsec:extreme event encoding} show.

\subsection{Seasonal Structure}\label{subsec:seasonal structure}
\begin{figure}[h]
  \centering
  \begin{subfigure}[b]{0.48\textwidth}
    \centering
    \includegraphics[width=\textwidth]{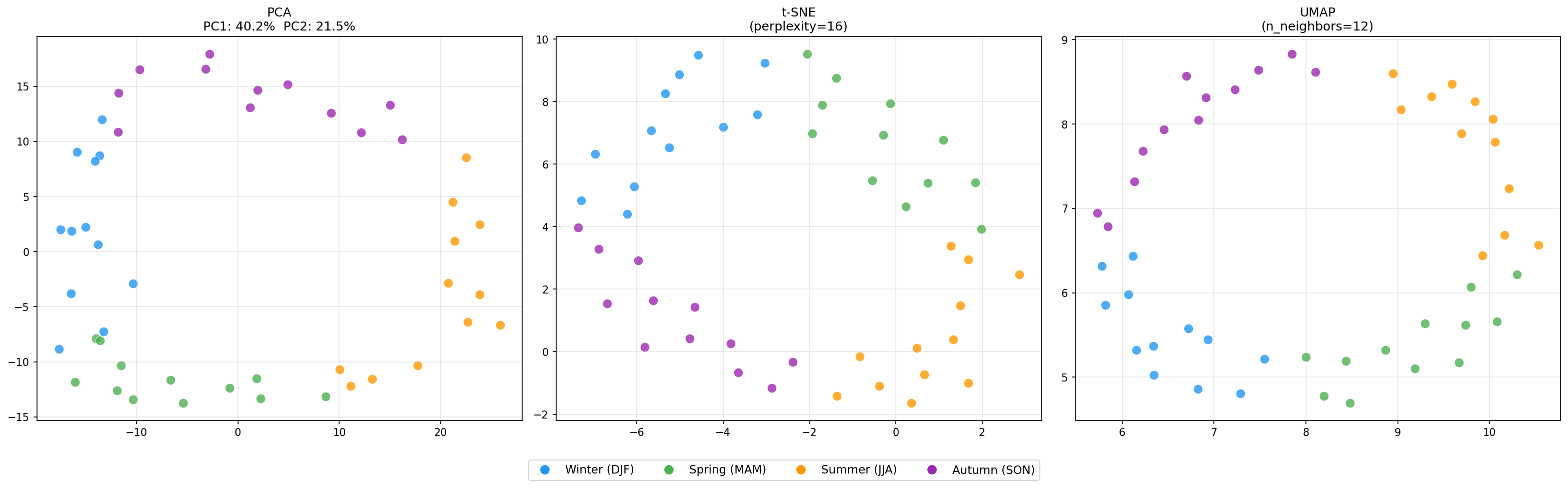}
    \caption{Encoder representations}
    \label{fig:clustering_encoder}
  \end{subfigure}
  \hfill
  \begin{subfigure}[b]{0.48\textwidth}
    \centering
    \includegraphics[width=\textwidth]{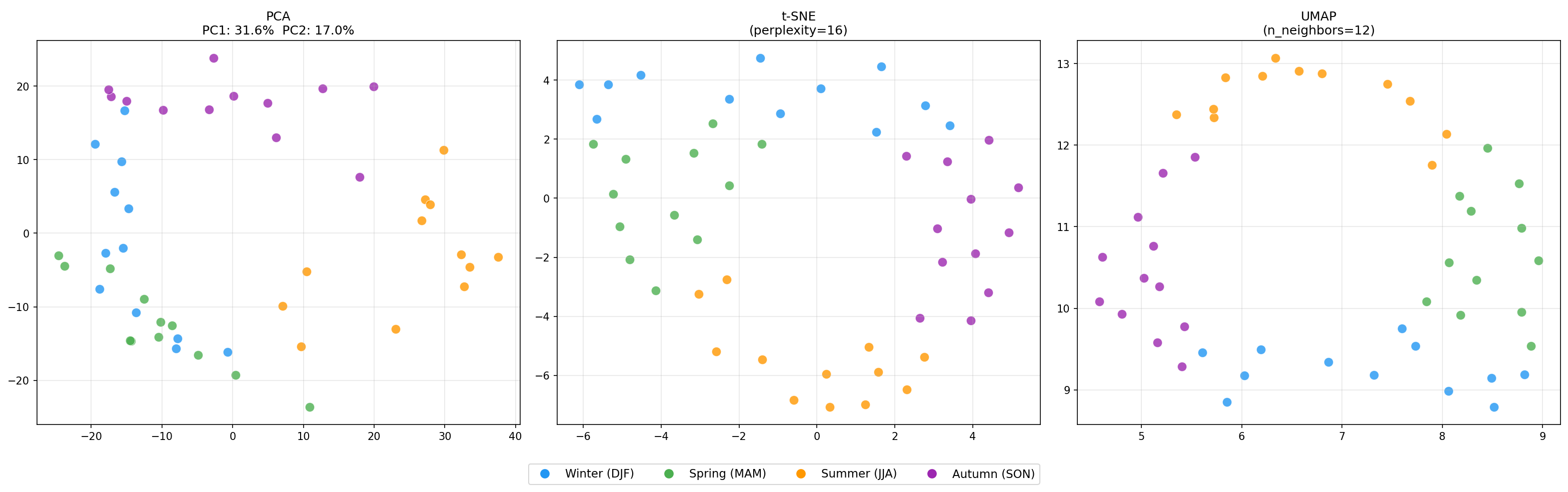}
    \caption{Backbone representations}
    \label{fig:clustering_backbone}
  \end{subfigure}
  \caption{PCA, t-SNE, and UMAP projections of Aurora's latent representations of 48 dates, colored by meteorological season (DJF, MAM, JJA, SON) for the encoder (left) and backbone (right). The encoder exhibits stronger seasonal clustering.}
  \label{fig:clustering_comparison}
\end{figure}

Aurora's encoder representations exhibit clear and consistent seasonal clustering across all three projection methods (Figure \ref{fig:clustering_comparison}). Winter (DJF), summer (JJA), spring (MAM), and autumn (SON) states occupy well-separated regions in both the t-SNE and UMAP embeddings, with the UMAP projection revealing a smooth arc structure that traces the annual cycle continuously rather than producing discrete jumps between seasons. The encoder captures substantially more variance in its top two principal components (PC1: 40.2\%, PC2: 21.5\%) than the backbone (PC1: 31.6\%, PC2: 17.0\%), suggesting that the patch-embedding stage retains more of the large-scale thermodynamic signal, while the backbone's deeper representations distribute that variance across a larger number of dimensions.

This is an important positive result. It confirms that Aurora's latent geometry is not arbitrary: the model internally organizes atmospheric states according to the annual cycle, and that organization is recoverable and coherent across dimensionality reduction methods. Gradient signals computed in this space carry real meteorological information (a prerequisite for any gradient-based perturbation strategy). Whether this geometric coherence extends beyond seasonality to the distinction between extreme and normal states is the question that is addressed next.

This confirms that Aurora's encoder captures a meaningful structure: it knows what time of year it is, and it organizes atmospheric states accordingly. This is a positive result, which means the latent space is not arbitrary, and it is at least partially geometrically structured.

\subsection{Extreme Event Encoding}\label{subsec:extreme event encoding}
We define extreme events as dates exceeding a threshold of 1.20 on a multivariate anomaly score computed from ERA5 reanalysis fields. For each date and variable, grid-point values are standardized by subtracting the within-sample mean and dividing by the within-sample standard deviation (both computed across the 48 analysis dates), yielding a field of dimensionless z-scores that capture relative extremes within the sample rather than relative to a long-term climatological baseline. The area-weighted root-mean-square of this field is then computed globally, yielding a single scalar per variable per date. Scores for mean sea-level pressure (msl) and 2-meter temperature (2t) are combined as $\max(\text{msl score},\, \text{2t score})$, so that a date is classified as extreme if either variable exhibits widespread anomalous conditions.

\begin{figure}[h]
  \centering
  \begin{subfigure}[b]{0.48\textwidth}
    \centering
    \includegraphics[width=\textwidth]{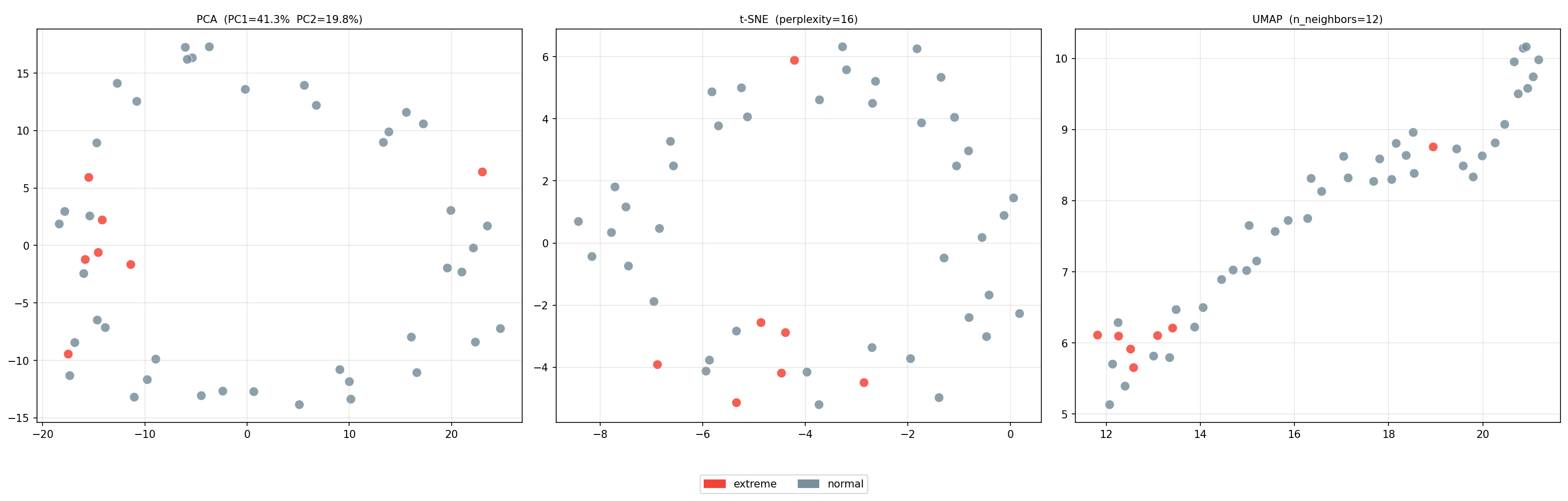}
    \caption{Encoder representations}
    \label{fig:extreme_label_encoder}
  \end{subfigure}
  \hfill
  \begin{subfigure}[b]{0.48\textwidth}
    \centering
    \includegraphics[width=\textwidth]{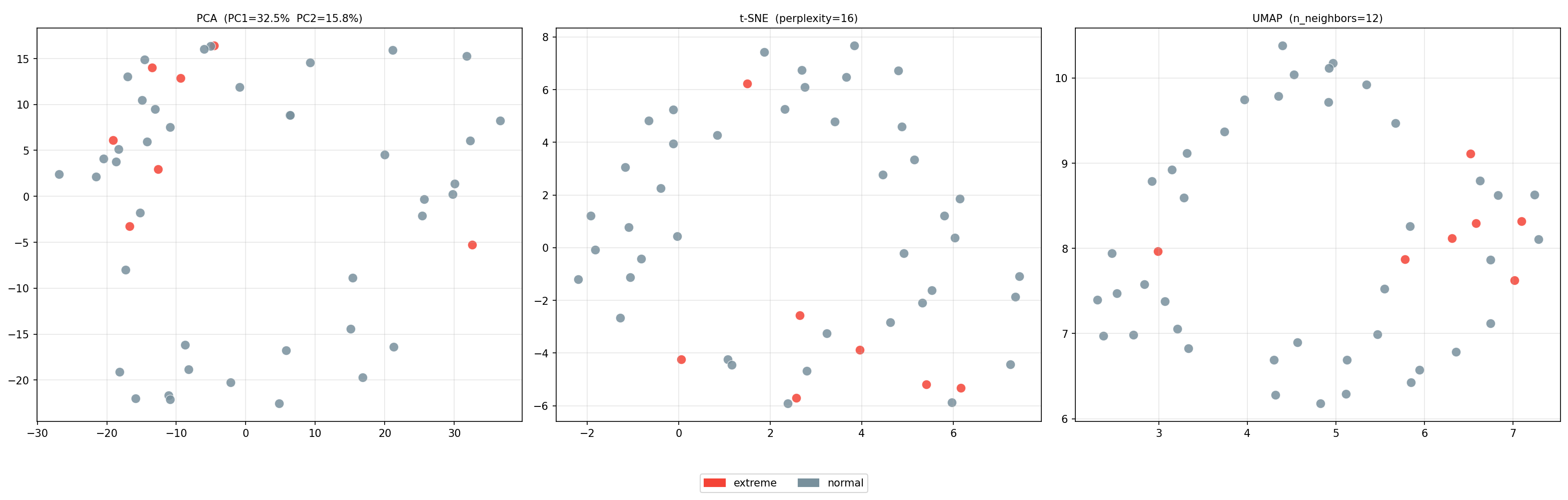}
    \caption{Backbone representations}
    \label{fig:extreme_label_backbone}
  \end{subfigure}
  \caption{PCA, t-SNE, and UMAP projections of Aurora's latent representations colored by extreme event intensity (normal vs.\ extreme) for the encoder (left) and backbone (right). Extreme and normal events are fully interleaved across all projections, with silhouette scores near zero, indicating no meaningful geometric separation.}
  \label{fig:extreme_label}
\end{figure}

We compare the latent representations of extreme and normal dates using the same projection methods as the preceding subsection. The result is unambiguous: extreme and normal events are completely interleaved across all projections and for both encoder and backbone representations (Figure \ref{fig:extreme_label}). Silhouette scores for this labeling are $0.005$ (encoder) and $-0.061$ (backbone). This shows that the backbone performs worse than random assignment. The joint season-by-intensity labeling yields an encoder silhouette of $0.089$, just below the meaningful threshold, confirming that adding extremity information to seasonal labels provides no additional geometric structure.

\begin{figure}[h]
  \centering
  \includegraphics[width=\textwidth]{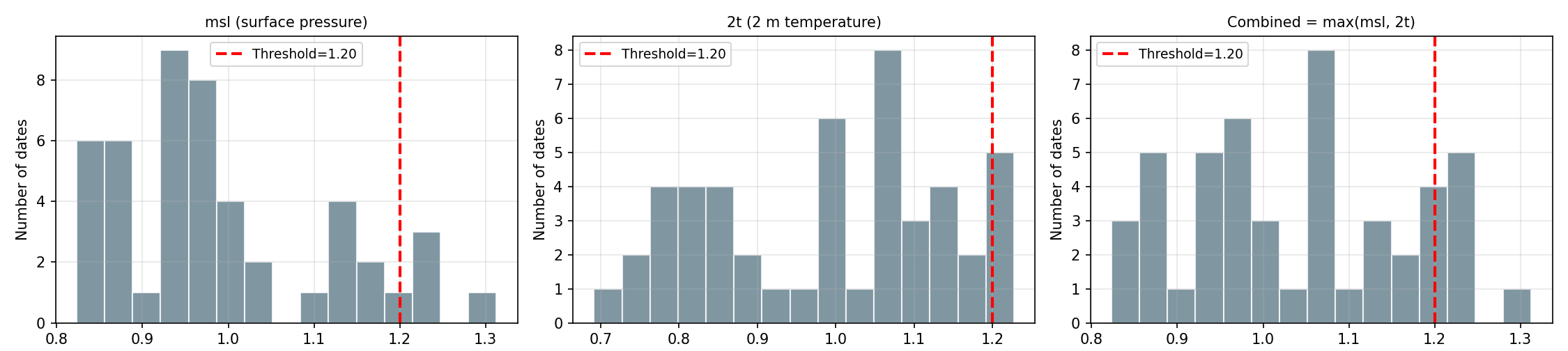}
  \caption{Distribution of anomaly scores across all samples with a threshold for 'extreme' marked by a red dashed line. The smooth, unimodal distribution confirms that extreme and normal atmospheric states do not occupy geometrically distinct regions of the latent space.}
  \label{fig:anomaly_score_distribution}
\end{figure}

\begin{figure}[h]
  \centering
  \includegraphics[width=0.49\textwidth]{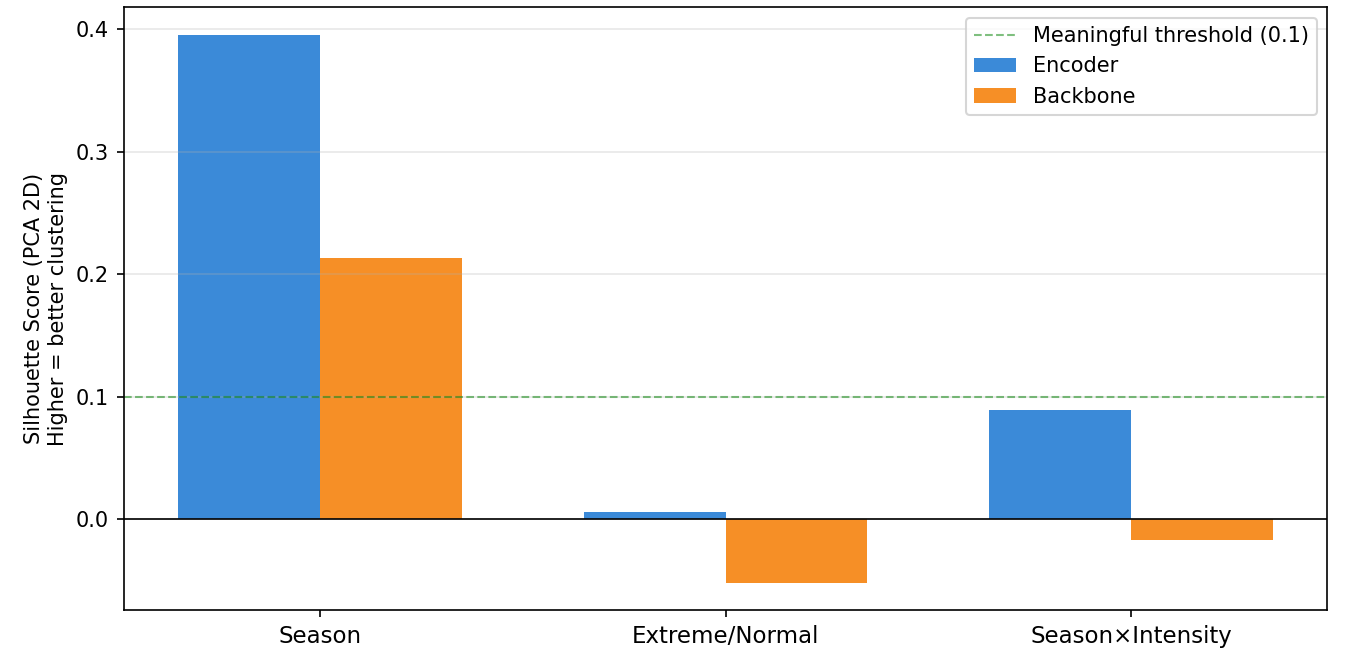}
  \caption{Silhouette scores for seasonal, extreme/normal, and joint season$\times$intensity labelings across Aurora's encoder and backbone representations (PCA 2D). Scores above 0.1 (dashed green line) indicate meaningful cluster separation. Season labeling yields strong separation in the encoder ($0.394$) and moderate separation in the backbone ($0.215$), while extreme/normal labeling produces scores near or below zero in both representations.}
  \label{fig:silhouette_comparison}
\end{figure}

The anomaly score distribution (Figure \ref{fig:anomaly_score_distribution}) reinforces this finding. Scores across the 48 dates are smooth and unimodal for all three panels, with no discontinuity separating extreme from normal states. Notably, the dates corresponding to documented extreme events, including major heat domes, cold waves, and tropical cyclones, score within the same range as matched normal days. This is a consequence of global area-weighting: a catastrophic regional event such as a heat dome or hurricane typically covers 2-5\% of the globe, and its signal is diluted when averaged over the remaining grid points. The 1.20 threshold reflects a human classification based on documented societal impact, not a physical boundary that manifests in globally-averaged metrics.
Figure \ref{fig:silhouette_comparison} summarizes all three labelings side by side, making the contrast between seasonal and extreme/normal geometry unambiguous.

This null result has a direct consequence for perturbation design, though its interpretation is not unambiguous. Because our extreme label is itself a spatially-aggregated, global summary, i.e., a single scalar per date, combining MSL and 2t anomalies averaged over the entire globe, it is not designed to isolate the regional or local character of an extreme event, the very spatial specificity that Weather Jiu-Jitsu's targeting strategy depends on. The latent-space geometry test above therefore asks whether Aurora's representations separate along a global label that does not itself preserve spatial structure, and a negative answer is at least as much a statement about the label's construction as about the latent space itself; a genuinely regional or event-local extremity criterion might reveal structure that this global criterion cannot. With that caveat, the practical implication stands: the latent space, as probed here, cannot serve as a detector or classifier of extreme atmospheric configurations, and perturbation targeting that relies on navigating toward an "extreme region" of the latent space has no geometric basis from this analysis. Instead, targeting must be grounded in physical diagnostics computed directly from the forecast fields.

\section{Beyond Global Clustering: Direction-based Analysis}\label{sec:direction-based}
The clustering analysis in \S\ref{subsec:extreme event encoding} found that extreme and normal atmospheric states are fully interleaved in Aurora's latent space, where the silhouette scores near zero (encoder 0.005, backbone -0.061). This section asks whether that null result is a limitation of the \textit{projection method} rather than of the latent space itself. The strategy shifts from asking "do extreme events form a cluster?" to "is there a latent direction along which extreme events consistently activate more strongly?"
\subsection{Dataset Expansion to 608 dates}
The 48-date sample in \S\ref{subsec:latent space analysis}-\ref{subsec:extreme event encoding} lacked statistical power to distinguish modest sensitivity ratios from noise. All analyses that follow use 608 ERA5 initialization dates spanning 1979-2021, retaining the same extreme classification criterion (area-weighted RMS standardized anomaly $\geq1.20$ on the combined MSL/2t score). This yields 139 extreme dates and 469 normal dates.

\subsection{Independent Component Analysis}
\label{subsec:ica}

A natural first attempt at a direction-based scan is ordinary PCA: rank the top 50 encoder principal components by an extreme-event sensitivity ratio and see whether any respond preferentially to extreme conditions. Appendix~\ref{subsec:appendix-higher-pcs} carries out this scan in full; briefly, it finds a sparse subset of five significant components (PC22, PC24, PC27, PC37, PC47, led by PC27 at $r=1.87$), but projecting onto that five-dimensional subspace fails to separate extreme from normal dates ($k$-means purity of 50\%, indistinguishable from chance). The sensitive PCs define a subspace in which extreme-relevant variance is elevated, but the discriminating direction within it is not aligned with any single PC axis. Recovering it requires a method that searches for structure within the subspace rather than treating each axis independently -- which motivates ICA, described next, and the archetype, PNS, and tail-dependence analyses of Appendix~\ref{sec:appendix-direction} that follow the same logic with different geometric assumptions.

While PCA maximises variance, ICA instead maximises statistical independence,
producing sparser and more physically interpretable directions. We apply FastICA
with 10 components to encoder and backbone representations across all 608 dates,
then rank components by extreme-event sensitivity ratio.

\begin{figure}[h]
\centering
\includegraphics[width=0.75\textwidth]{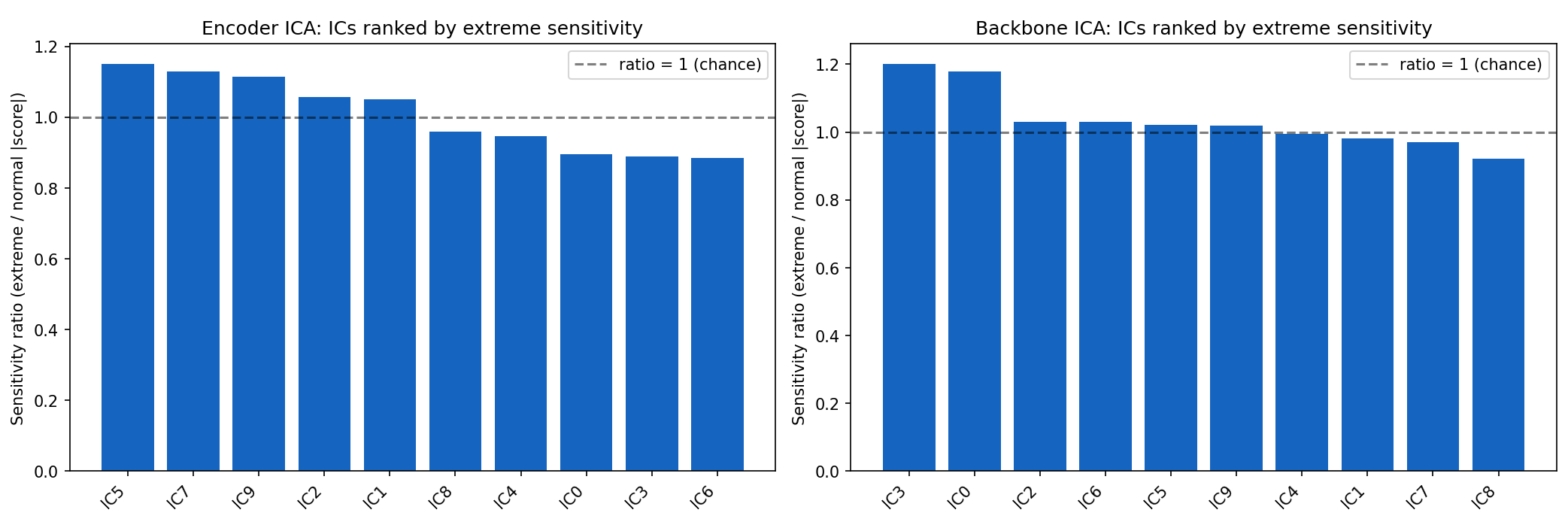}
\caption{Extreme-event sensitivity ratios for 10 ICs in the encoder (left) and backbone (right), ranked in descending order, in order to find the independent components that respond most to extreme events. The Dashed line marks a ratio = 1 (equal activation on extreme and normal dates). Encoder IC5 ($r = 1.15$) and backbone IC3 ($r = 1.20$) are the top-ranked directions in each layer; backbone IC3 is the highest ratio found across all methods in this study.}
\label{fig:ica-ratios}
\end{figure}

Figure~\ref{fig:ica-ratios} shows the ranked sensitivity ratios for both layers.
In the encoder, IC5 ($r = 1.15$) leads a gradual falloff across five components
that exceed ratio 1, suggesting diffuse sensitivity spread across multiple
directions. The backbone is sharper: IC3 ($r = 1.20$) and IC0 ($r = 1.18$)
stand clearly above the remaining components, which cluster near $r = 1$.
Backbone IC3 achieves the highest sensitivity ratio of any direction found
across all layers and methods in this study, and serves as the primary target
for gradient-based probing in \S\ref{subsec:jlens}.

\begin{figure}[h]
\centering
\includegraphics[width=0.75\textwidth]{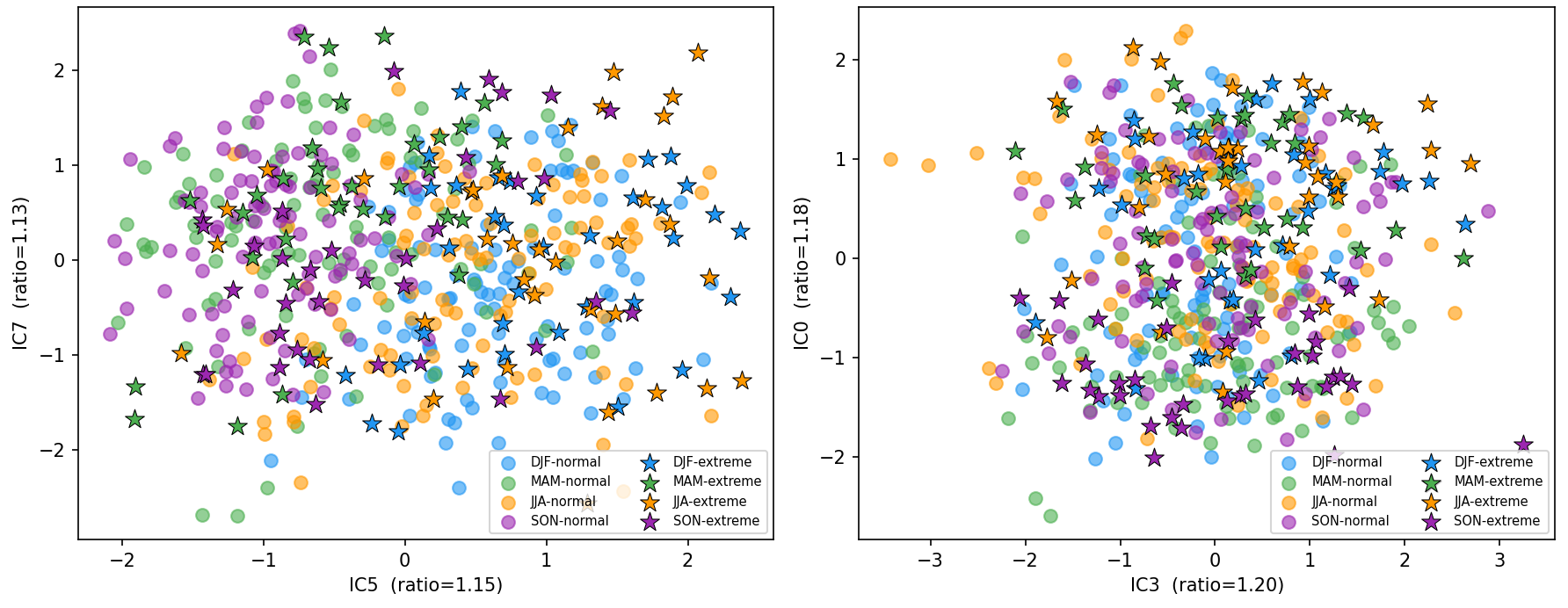}
\caption{IC scores for all 608 dates projected onto the top-two extreme-sensitive components: encoder IC5 vs IC7 (left) and backbone IC3 vs IC0 (right). Color encodes season (DJF: blue, MAM: green, JJA: orange, SON: purple); stars mark extreme dates, circles mark normal. Extreme dates span the entire IC plane but show a distributional shift to larger absolute scores along the top-ranked axis (IC5/IC3), visible as stars reaching further along the x-axis than the normal population.}
\label{fig:ica-scatter}
\end{figure}

An IC can be understood as a latent concept direction — a linear axis along
which Aurora's internal representation varies in a statistically independent
and non-Gaussian way. Figure~\ref{fig:ica-scatter} shows the top-two IC scores
for all 608 dates, colored by season and marked by extremity (stars = extreme,
circles = normal). Extreme dates are distributed across the full IC plane with
no spatial cluster, confirming the global null result of \S\ref{subsec:extreme event encoding}.
However, the marginal distribution along the top-ranked axis shifts: extreme
dates (stars) extend to larger absolute IC scores along IC5 (encoder) and IC3
(backbone) than the normal population. The directional sensitivity exists without
geometric separability — a distinction the gradient analyses of \S\ref{sec:jacobian}
are designed to probe.

Beyond ICA, we also examined archetype analysis, principal nested spheres, and tail-dependence PCA as alternative direction-finding methods over the same 608-date representations; these are reported in full in Appendix~\ref{sec:appendix-direction} rather than here. In brief, that appendix confirms the picture given by ICA -- the backbone is the more extreme-sensitive layer, with archetype analysis showing extreme dates spread across all four archetypal configurations rather than a single one (Appendix~\ref{subsec:appendix-archetypes}), principal nested spheres finding a comparable but slightly more diffuse signal to ICA in the encoder and an essentially linear counterpart to backbone IC3 (Appendix~\ref{subsec:appendix-pns}), and tail-dependence PCA isolating a sharply concentrated backbone direction, TPC2 ($r=1.28$), alongside IC3 and PNS7 (Appendix~\ref{subsec:appendix-tail}). Considered together, IC3, PNS7, and TPC2 confirm that the strongest co-extreme structure in Aurora's representations resides in the backbone, concentrated in a small number of directions rather than spread uniformly across the latent space.

\section{Gradient-Based Input Diagnostics}\label{sec:jacobian}
ICA identifies which latent directions are most sensitive to extreme events but not what in the ERA5 input fields drives that activation. Two gradient-based methods trace this information flow.

\subsection{Jacobian Saliency: Which Inputs drive IC5?}\label{subsec:jacobian}
Define the IC5 score for ERA5 initial condition $\text{x}$ as its projection onto the IC5 direction: $s(\text{x})=\langle E_\theta(\text{x}),\text{w}_{\text{IC5}}\rangle$. The Jacobian saliency is $\partial s/\partial\text{x}$, computed via backpropagation. To aggregate over 139 extreme dates while preserving sign information:
\begin{equation}
    J^-=\left(\frac{1}{N}\right)\sum_i \text{sgn}(s(\text{x}_i))\cdot\frac{\partial s}{\partial\text{x}}
\end{equation}
Sign-normalization ensures that extreme dates with positive and negative IC5 scores contribute additively rather than canceling.

\begin{table}[H]
	\caption{Jacobian saliency results by ERA5 input field.
	         Gradients computed as $\bar{\mathbf{J}} = \frac{1}{N}\sum_{i=1}^{N}
	         \mathrm{sign}(s(\mathbf{x}_i))\cdot\partial s/\partial\mathbf{x}_i$
	         averaged over $N=139$ extreme dates.}
	\centering
	\begin{tabular}{%
	  l
	  >{\raggedright\arraybackslash}p{4.2cm}
	  r
	  >{\raggedright\arraybackslash}p{4.0cm}
	}
		\toprule
		Field & Spatial pattern & Gradient scale & Role \\
		\midrule
		Z850
		  & Broad positive globally; mid-latitude peak
		  & ${\sim}10^{-6}\ \mathrm{m}^{-2}\,\mathrm{s}^{-2}$
		  & Primary driver — largest magnitude of any field \\
		Z500
		  & Broad positive, slightly weaker
		  & ${\sim}3\times10^{-7}$
		  & Strong secondary driver \\
		2m temperature
		  & Positive mid-latitudes, 30–70\textdegree{} N/S
		  & ${\sim}2\times10^{-7}\ \mathrm{K}^{-1}$
		  & Consistent thermodynamic signal \\
		MSL
		  & Positive NH mid-latitudes; patchy polar negative
		  & ${\sim}1.5\times10^{-7}$
		  & Seasonal mean dominates \\
		10m winds (U, V)
		  & Structured, mostly negative; storm-track hot spots
		  & ${\sim}10^{-8}$
		  & Secondary — $10\times$ smaller than Z850 \\
		\bottomrule
	\end{tabular}
	\label{tab:jacobian_saliency}
\end{table}

\begin{figure}[H]
  \centering
  \includegraphics[width=\textwidth]{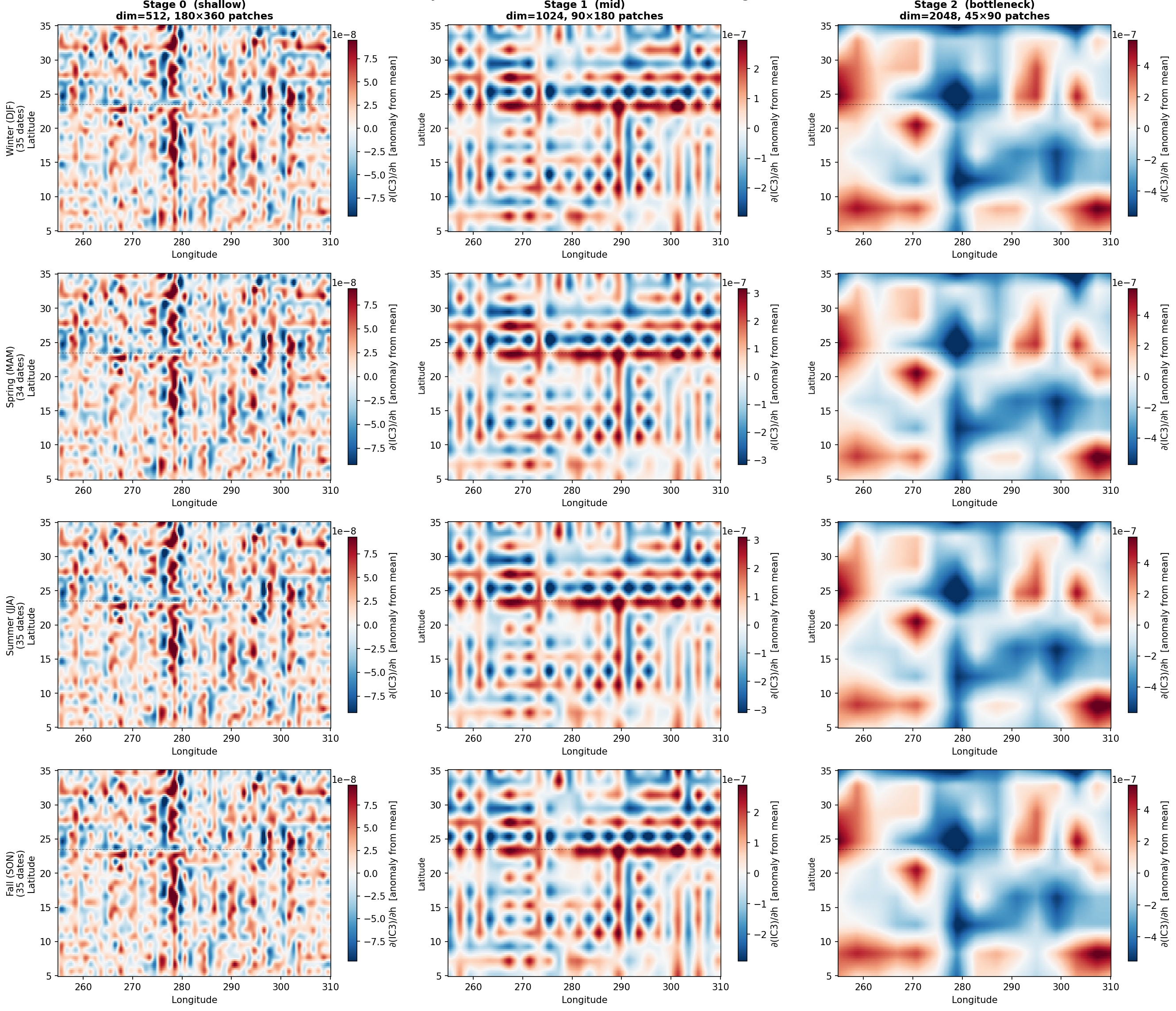}
  \caption{Mean sign-normalized Jacobian saliency $\bar{\mathbf{J}} = \frac{1}{N}\sum_i \mathrm{sign}(s(\mathbf{x}_i))\, \partial s_{\mathrm{IC5}}/\partial \mathbf{x}_i$ averaged over $N=139$ extreme dates (encoder IC5, sensitivity ratio $= 1.151$). Z850 (bottom right) carries the largest absolute gradient (${\sim}10^{-6}$), followed by Z500 and 2m temperature. Surface winds are an order of magnitude smaller.}
  \label{fig:saliency_raw}
\end{figure}
IC5 behaves as a global thermodynamic index, most strongly driven by warm temperature and elevated lower-tropospheric geopotential. This is consistent with synoptic-scale ridge patterns that precede and characterize heat extremes. The broad spatial footprint suggests IC5 encodes a hemispheric-scale thermodynamic state rather than a local feature.

\subsection{Anomaly-based Jacobian Saliency}\label{subsec:anom-jacobian}
The raw Jacobian in \S\ref{subsec:jacobian} reflects sensitivity to the full ERA5 state, including the seasonal mean. Because extreme events are non-uniformly distributed across seasons, averaged saliency maps may partly reflect the annual cycle. We recompute Jacobians with \textit{climatology-subtracted inputs}: for each date, its monthly ERA5 mean is subtracted before the forward pass, isolating the gradient with respect to the anomalous component.
\begin{table}
	\caption{Anomaly-based Jacobian saliency results by ERA5 field.
	         Monthly climatological means are subtracted from each input before
	         computing $\partial s_{\mathrm{IC5}}/\partial\tilde{\mathbf{x}}$,
	         isolating the gradient with respect to the anomalous component only.
	         Results averaged over $N=139$ extreme dates.}
	\centering
	\begin{tabular}{%
	  >{\raggedright\arraybackslash}p{2.6cm}
	  >{\raggedright\arraybackslash}p{4.4cm}
	  >{\raggedright\arraybackslash}p{4.2cm}
	}
		\toprule
		Field & Pattern after climatology removal & Interpretation \\
		\midrule
		MSL \& 10m winds
		  & Collapse to near-zero; no coherent spatial structure remains.
		  & Raw gradient dominated by seasonal mean. Event-specific anomaly carries little IC5 information. \\
		2m temperature
		  & Mid-latitude banding preserved; magnitude reduced but spatially coherent.
		  & Temperature \emph{anomalies} (not mean state) consistently displace IC5 during extreme events. \\
		Z850
		  & Clearest organised pattern: coherent wave-like banding across NH mid-latitudes (${\sim}30$--$60^\circ$\,N).
		  & \textbf{Primary target.} Upper-level ridge amplification signature consistent with blocking highs; survives seasonal removal. \\
		\bottomrule
	\end{tabular}
	\label{tab:anom_saliency}
\end{table}

\begin{figure}[h]
  \centering
  \includegraphics[width=\textwidth]{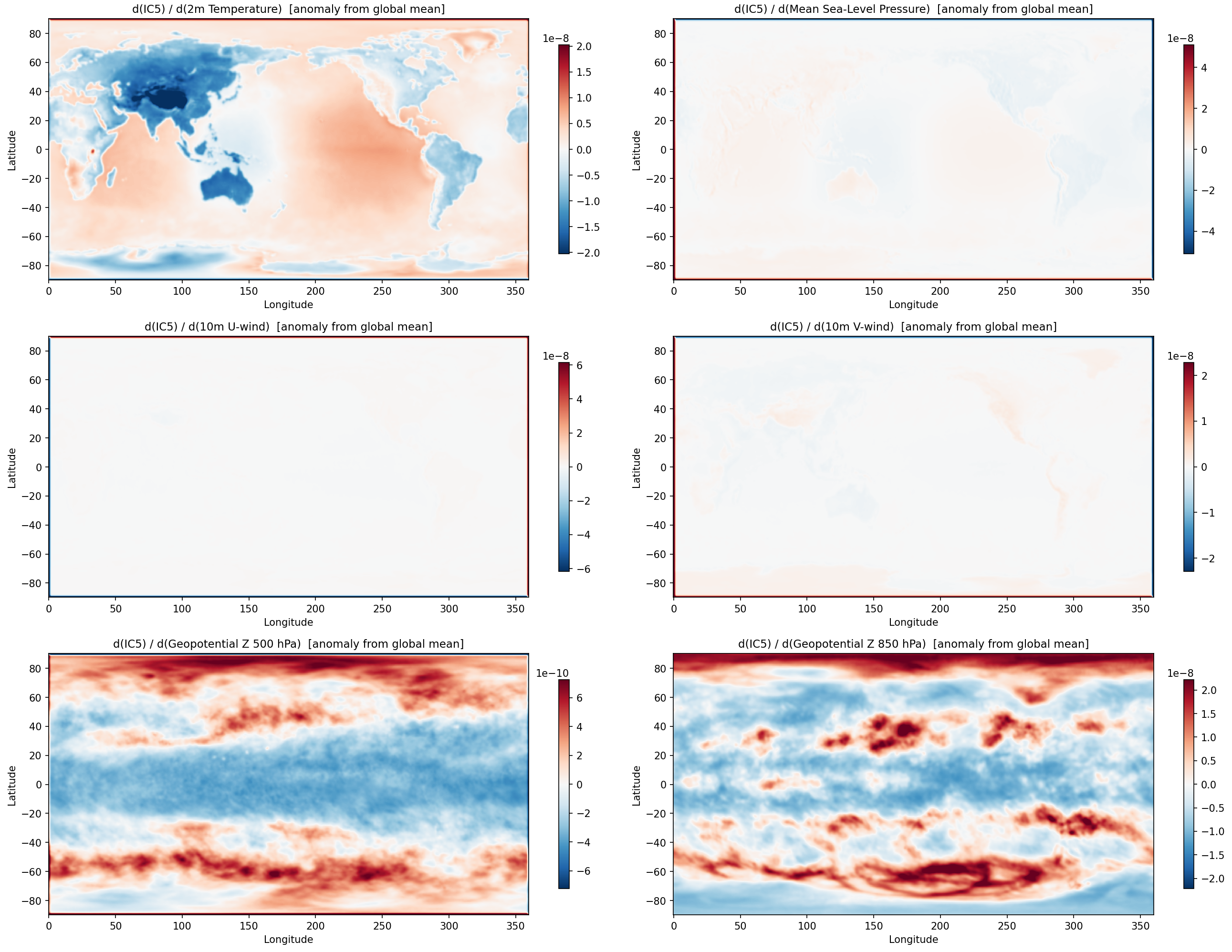}
  \caption{Anomaly-based Jacobian saliency on Aurora's encoder: same computation as Figure~\ref{fig:saliency_raw} but with monthly ERA5 climatological means subtracted from each input before the forward pass. MSL and 10\,m winds collapse to near-zero, revealing that their raw gradients were dominated by the seasonal mean. Z850 (bottom right) retains the clearest organized signal: coherent wave-like banding across NH mid-latitudes (${\sim}30$--$60^\circ$\,N), consistent with upper-level ridge amplification during blocking events.}
  \label{fig:saliency_anom}
\end{figure}
Z850 is the most informative single field for characterizing the ERA5 input structure driving Aurora's extreme-sensitive latent direction. Its anomaly saliency survives seasonal removal and shows a physically interpretable blocking/ridge pattern. Perturbation targeting should focus on lower-tropospheric geopotential anomalies over surface temperature or pressure.

\subsection{Jacobian Lens: Layer-wise Backbone Analysis} \label{subsec:jlens}

\begin{figure}[h]
\centering
\includegraphics[width=\textwidth]{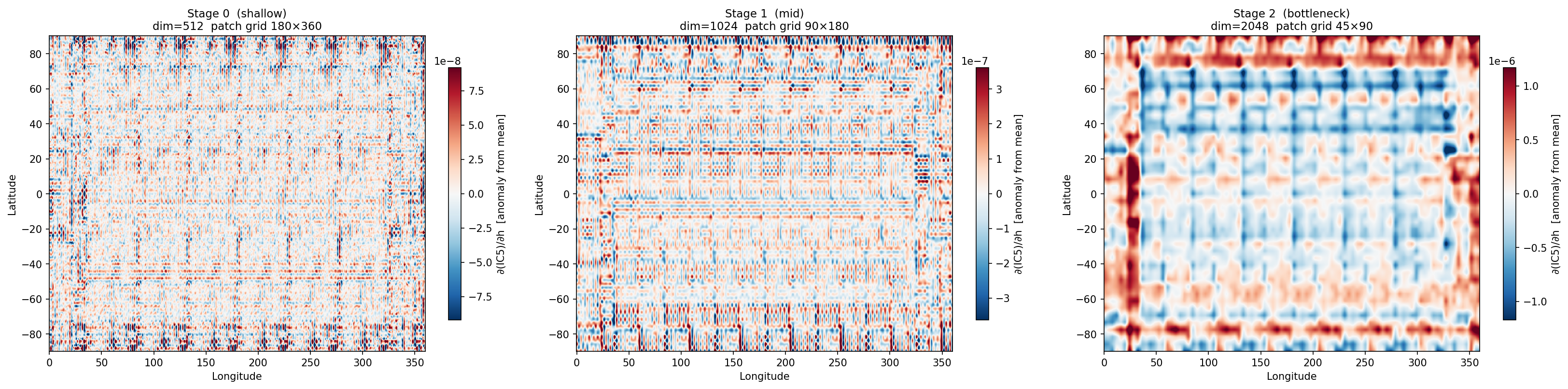}
\caption{J-lens signed gradient maps $\partial s_{\mathrm{IC3}} / \partial \mathbf{h}^{(k)}$ averaged over $N = 139$ extreme dates, shown as anomalies from each stage's own spatial mean. \textbf{Stage~0} (shallow; $180\!\times\!360$): near-uniform, no organized spatial structure. \textbf{Stage~1} (mid; $90\!\times\!180$): emerging large-scale contrast. \textbf{Stage~2} (bottleneck; $45\!\times\!90$): structured directionality — positive anomalies at $\sim\!30^\circ$--$60^\circ$\,N/S indicate that increasing mid-latitude hidden states pushes the representation toward higher IC3 scores; negative anomalies at tropical and polar latitudes indicate suppression. Note that these signed anomaly maps characterize gradient \emph{direction}; gradient \emph{magnitude} is analyzed separately in \S\ref{subsec:jlens-spatial}.}
\label{fig:jlens}
\end{figure}

The Jacobian saliency of \S\ref{subsec:jacobian}--\ref{subsec:anom-jacobian}
operates in input space. A complementary question is: \emph{at which stage of
Aurora's backbone does the IC3 extreme-sensitive information become concentrated?}
Following \citet{gurnee2026verbalizablerepresentationsformglobal}, who define a layer-wise gradient lens
$J_l = \mathbb{E}[\partial h_\text{final}/\partial h_l]$, we compute
\begin{equation}
    \mathbf{J}^{(k)}(\mathbf{x}) = \frac{\partial\, s_{\mathrm{IC3}}}{\partial\, \mathbf{h}^{(k)}}
\end{equation}
where $s_{\mathrm{IC3}}$ is the IC3 score at backbone output and $\mathbf{h}^{(k)}$
is the activation tensor at stage $k$ of Aurora's Swin3D backbone. Aurora's
backbone has three encoder stages of increasing depth and decreasing spatial
resolution (Table~\ref{tab:jlens}).
Although $\mathbf{J}^{(k)}$ is a first derivative, it is not the gradient of a linear readout of Aurora's state. The target score $s_{\mathrm{IC3}}$ is itself the projection onto a direction that ICA (\S\ref{subsec:ica}) identified as the axis of maximal statistical independence and non-Gaussianity in the backbone representation, a direction defined by higher-order statistical structure, not by variance alone. Backpropagating to this target therefore traces how the network's internal geometry responds to that non-Gaussian, extreme-sensitive structure, carrying information beyond what a purely linear sensitivity measure could capture. The strong spatial coherence of the Stage-2 gradient field reported below (Moran's I$=+0.570$, \S\ref{subsec:jlens-spatial}) is itself a property of this underlying geometry, and \S\ref{subsec:linearity of the j-lens} probes it more directly via the curvature of $s_{\mathrm{IC3}}$ along the Jacobian's leading singular directions. 
\paragraph{Gradient amplification.}
We compute $\mathbf{J}^{(k)}$ for each of the 139 extreme dates and report the spatial RMS of the gradient map per stage. Table~\ref{tab:jlens} shows a $5.3\times$ amplification from Stage~0 to Stage~2. Figure~\ref{fig:jlens} shows the average signed gradient maps, plotted as anomalies from each stage's own spatial mean. The Stage~2 map reveals structured directionality: positive anomalies at $\sim\!30^\circ$--$60^\circ$\,N/S and negative anomalies at tropical and polar latitudes. Interpreted geometrically, increasing mid-latitude hidden-state activations pushes the representation toward higher IC3 scores, while tropical and polar activations suppress it — a pattern consistent with upper-level ridge amplification associated with extreme precipitation. Stage~0 and Stage~1 maps are noisier and lack organized large-scale structure at the same scale. The $5.3\times$ amplification indicates that the IC3 direction is \emph{constructed} through Aurora's hierarchical transformer computation rather than present in any single early feature: the bottleneck (Stage~2) is where extreme atmospheric state information is most concentrated and most accessible to gradient-based targeting.
\begin{table}
	\caption{Jacobian lens results across Aurora's three Swin3D backbone encoder stages. Gradient RMS values $\|\mathbf{J}^{(k)}\|_{\mathrm{rms}}$ are averaged over $N = 139$ extreme dates (IC3, sensitivity ratio $= 1.201$). The bottleneck (Stage~2) accumulates a $5.3\times$ amplification relative to Stage~0.}
	\centering
	\begin{tabular}{c r r r c}
		\toprule
		Stage & Channels & Patch grid & {Mean gradient RMS} & Relative to Stage~0 \\
		\midrule
		0 & 512 & $180\times360$                  & 1.08e-7 & $1.0\times$ (baseline) \\
		1 & 1024 & $90\times180$                   & 2.51e-7 & $2.3\times$ \\
		2 & 2048 &       $45\times90$ \textit{(bottleneck)} & 5.72e-7 & $5.3\times$ \\
		\bottomrule
	\end{tabular}
	\label{tab:jlens}
\end{table}

\paragraph{IC variance contribution in activation space.}
The J-lens is a gradient-space probe. A complementary activation-space probe
asks: at each backbone patch, what fraction of the hidden-state variance is
attributable to the IC3 direction? We adapt Equation~3 of
Bodnar et al.~\citep{Bodnar2025}:

\begin{equation}
    v_c(t) = \frac{s_c(t)^2}{\|\mathbf{h}_\mathrm{std}(t)\|^2}
\end{equation}
where $s_c(t) = \mathbf{g} \cdot (\mathbf{h}(t) - \boldsymbol{\mu})$ is the IC
score at backbone patch $t$ and $\|\mathbf{h}_\mathrm{std}(t)\|^2$ is the total
standardized squared norm of the hidden state. Backbone activations are captured
via forward hook (shape $259{,}200 \times 1024 = 4 \times 180 \times 360$ patches
per date) and averaged over 50 extreme and 50 normal dates.\footnote{FastICA's
nondeterministic component labeling causes the extreme-sensitive backbone direction
to appear as IC3 in gradient analyses and IC7 in activation analyses; both refer
to the same physical direction (sensitivity ratio $\approx 1.20$).}

\begin{figure}[h]
\centering
\includegraphics[width=0.5\linewidth]{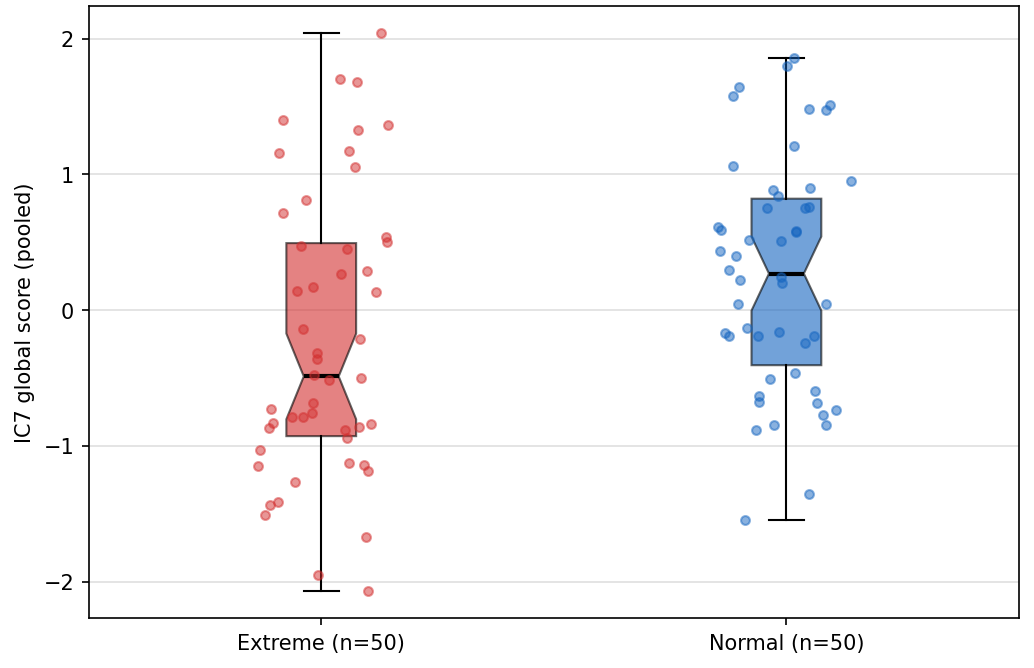}
\caption{Per-date IC7 global scores (mean-pooled over all backbone patches) for extreme ($n = 50$, red) and normal ($n = 50$, blue) dates. Extreme dates show a consistent negative shift (median $\approx -0.5$) relative to normal (median $\approx +0.25$); sensitivity ratio $= 1.194$.}
\label{fig:ic-boxplot}
\end{figure}

\begin{figure}[h]
\centering
\includegraphics[width=\linewidth]{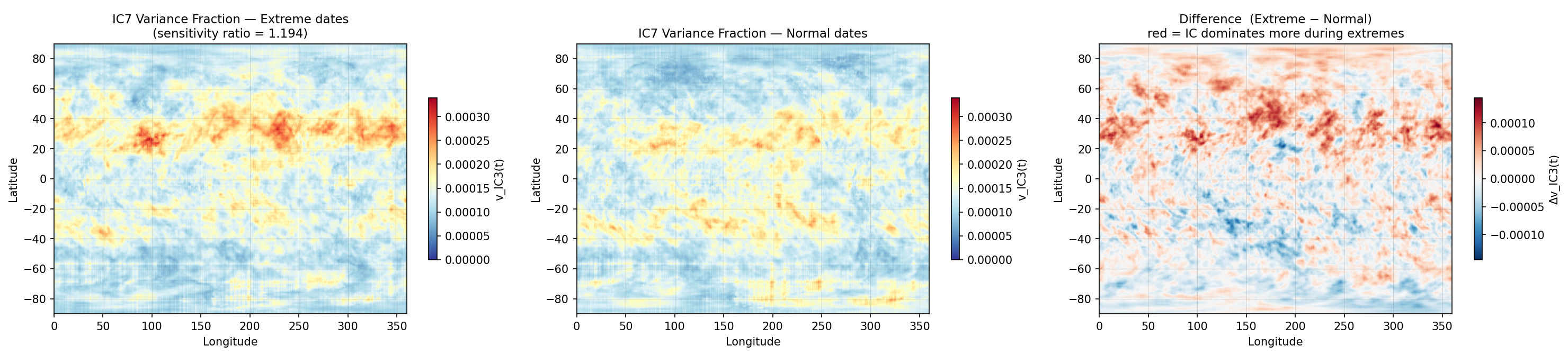}
\caption{Per-patch IC7 variance fraction $v_c(t)$ averaged over extreme dates
         (left) and normal dates (center), with difference map (right). Red
         in the difference map indicates regions where the IC direction accounts
         for a larger share of backbone variance during extreme events. The NH
         mid-latitude belt ($30$--$60^\circ$N) shows consistent enhancement,
         consistent with jet-stream and blocking-pattern activity.}
\label{fig:ic-varmap}
\end{figure}

Figure~\ref{fig:ic-boxplot} shows that extreme dates have systematically more
negative IC scores (median $\approx -0.5$) than normal dates
(median $\approx +0.25$). The geographic maps (Figure~\ref{fig:ic-varmap})
reveal that this shift has spatial structure: the Northern Hemisphere mid-latitude
belt ($\sim\!30^\circ$--$60^\circ$\,N) shows elevated IC variance fraction on
extreme dates relative to normal, as seen in the difference map. ~\citet{Bodnar2025} find that Aurora's latent space remains anchored to
meteorological structure even after atmospheric chemistry fine-tuning; our maps
confirm that, in the pretrained model, this anchoring has a specific geographic
footprint that shifts with meteorological extremity.

\subsection{Spatial Organization of J-Lens Gradients Across Backbone Stages}
\label{subsec:jlens-spatial}

Section~\ref{subsec:jlens} established that J-lens gradient RMS amplifies
$5.3\times$ from Stage~0 to Stage~2, and that the signed gradient map at
Stage~2 shows structured directionality. Here we characterize the
\emph{spatial organization} of the gradient \emph{magnitude} across stages using Moran's~I\footnote{A measure of spatial autocorrelation; Moran's I is large in regions of strong spatial correlation. \cite{af54b142-01be-3f96-af47-1730365d8376}}, LISA cluster maps, normalized gradient concentration, and
the block-average variance test. Together, these analyses reveal that Aurora's
backbone implements progressive spatial abstraction of the IC7 sensitivity signal.

\begin{figure}[h]
\centering
\includegraphics[width=\linewidth]{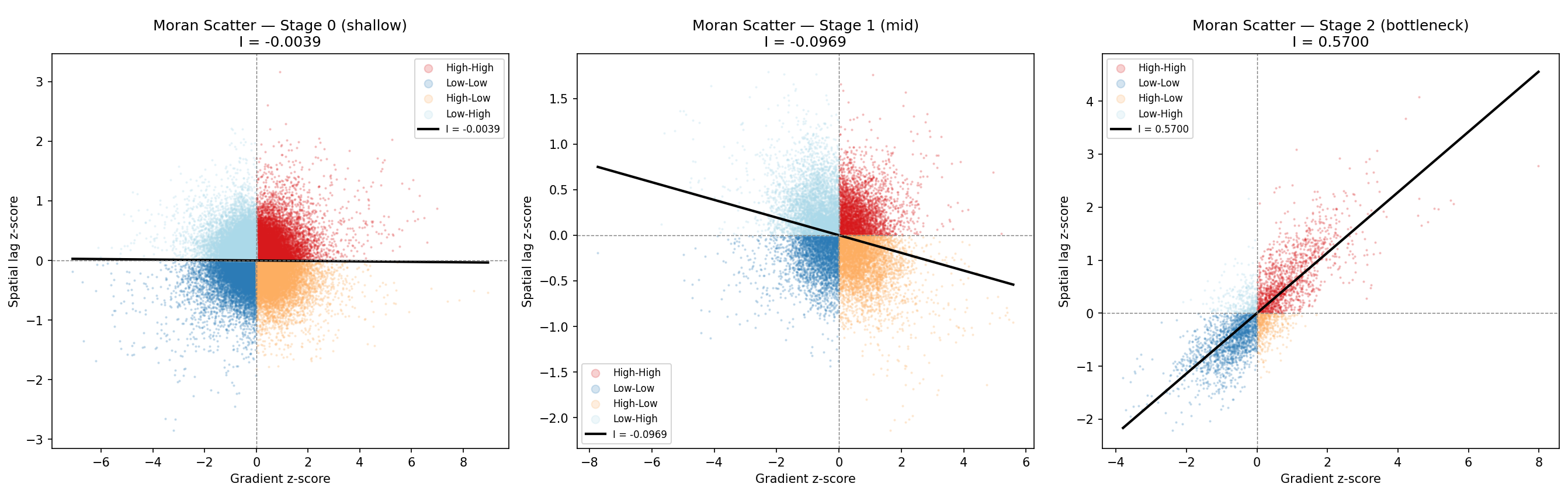}
\caption{Moran scatter plots ($z$ vs.\ spatial lag $Wz$) for J-lens gradient
         magnitude at Stage~0 (left), Stage~1 (center), and Stage~2 (right).
         The regression slope equals Moran's $I$. Stage~0: diffuse cloud, no
         spatial structure. Stage~1: negative slope (anti-clustering).
         Stage~2: pronounced positive slope ($I = +0.570$), indicating strong
         clustering of high-gradient patches.}
\label{fig:jlens-moran}
\end{figure}

\paragraph{Moran's I.}
We compute global Moran's~I for the J-lens gradient magnitude at each stage
using a row-standardized rook contiguity weight matrix with longitude wrapping.
Table~\ref{tab:jlens-moran} reveals a striking progression. Stage~0 is
spatially random ($I = -0.004$, not significant): gradient energy is uniformly
scattered at fine scales with no geographic preference. Stage~1 is weakly
anti-clustered ($I = -0.097$, $z = -12.3$): high-gradient patches are
surrounded by low-gradient neighbors. Stage~2 is strongly clustered
($I = +0.570$, $z = +36.3$) — the largest spatial autocorrelation measured
across all analyses in this study. Figure~\ref{fig:jlens-moran} confirms the
progression visually: Stage~0 is a diffuse cloud in the Moran scatter, Stage~1
has a negative slope, and Stage~2 has a sharp positive slope through the HH
and LL quadrants.

\begin{table}
	\caption{Global Moran's $I$ for J-lens gradient magnitude at each backbone
	         stage ($N = 139$ extreme dates, IC3, sensitivity ratio $= 1.201$).
	         Rook contiguity, row-standardized, longitude-periodic.}
	\centering
	\begin{tabular}{lcccc}
		\toprule
		Stage & Resolution & Cell size & Moran's $I$ & $z$-score \\
		\midrule
		0 (shallow)    & $180 \times 360$ & 1\textdegree/cell & $-0.004$ & $-1.0$ \\
		1 (mid)        & $90  \times 180$ & 2\textdegree/cell & $-0.097$ & $-12.3$ \\
		2 (bottleneck) & $45  \times 90$  & 4\textdegree/cell & $+0.570$ & $+36.3$ \\
		\bottomrule
	\end{tabular}
	\label{tab:jlens-moran}
\end{table}

\begin{figure}[h]
\centering
\includegraphics[width=0.75\linewidth]{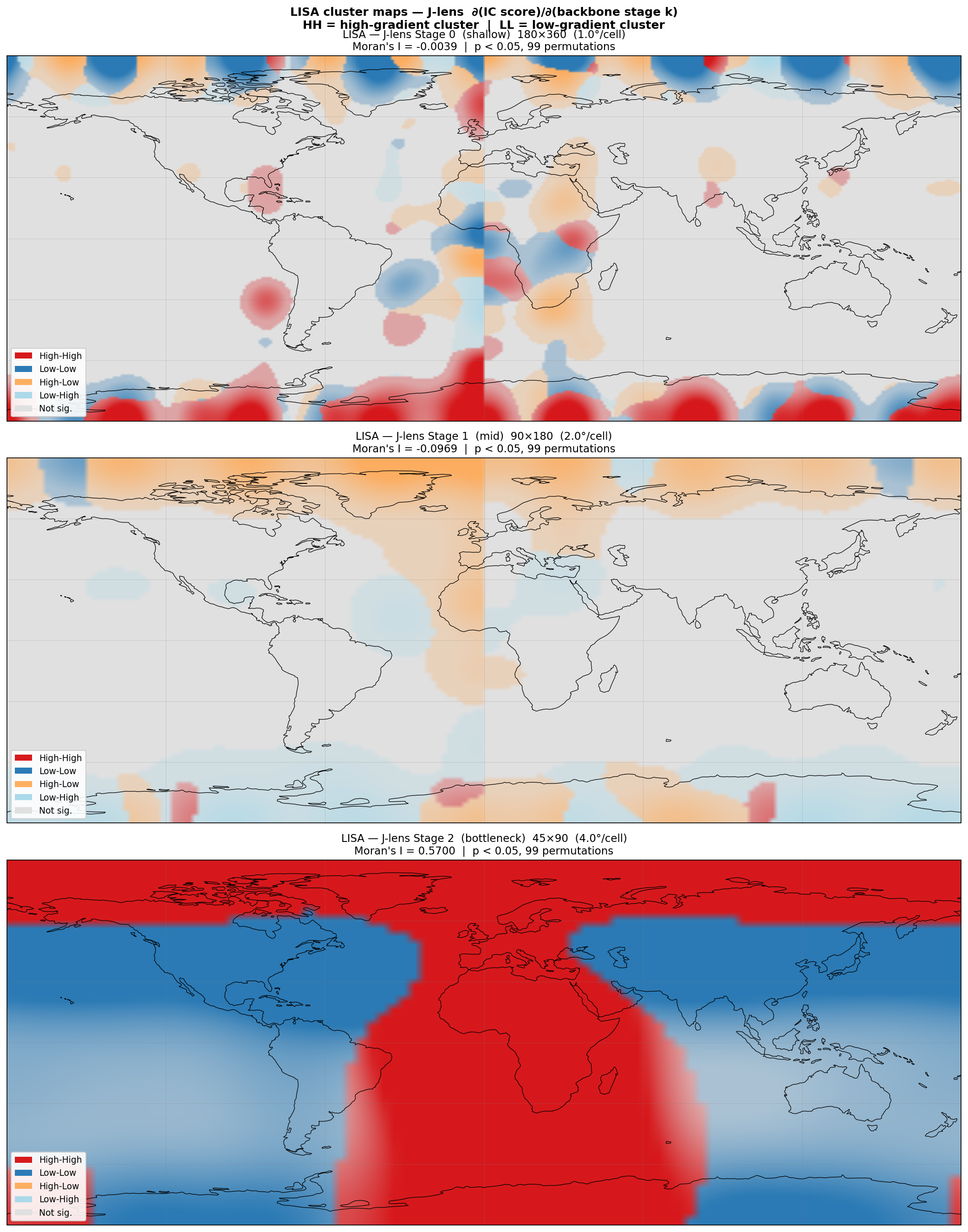}
\caption{LISA cluster maps for J-lens gradient magnitude at Stage~0 (top),
         Stage~1 (middle), and Stage~2 (bottom). Red = High-High (HH);
         blue = Low-Low (LL); orange = High-Low (HL); light blue = Low-High
         (LH); gray = not significant ($p \geq 0.05$, 99 permutations).
         Stage~2 resolves into two coherent regimes — tropical/polar HH and
         mid-latitude LL — with zero spatial outliers.}
\label{fig:jlens-lisa}
\end{figure}

\paragraph{LISA cluster maps.}
Figure~\ref{fig:jlens-lisa} shows the Local Indicators of Spatial Association
(LISA) maps for all three stages ($p < 0.05$, 99 permutations). Stage~0 displays
scattered HH, LL, HL, and LH clusters with no geographic coherence. Stage~1 is
dominated by High-Low (HL) spatial outliers, consistent with the anti-clustered
pattern: high-gradient patches are isolated spikes surrounded by low-gradient
neighbors. Stage~2 resolves into a clean two-regime partition covering the entire
globe: High-High (HH) clustering in the tropics and polar caps — with a prominent
concentration over the Indian Ocean and East Africa sector — and Low-Low (LL)
clustering in the mid-latitude storm tracks of both hemispheres. Stage~2 contains
\emph{zero} HL or LH cells: the bottleneck gradient is perfectly organized into
smooth contiguous geographic regions with no spatial outliers.

\begin{figure}[H]
\centering
\includegraphics[width=\linewidth]{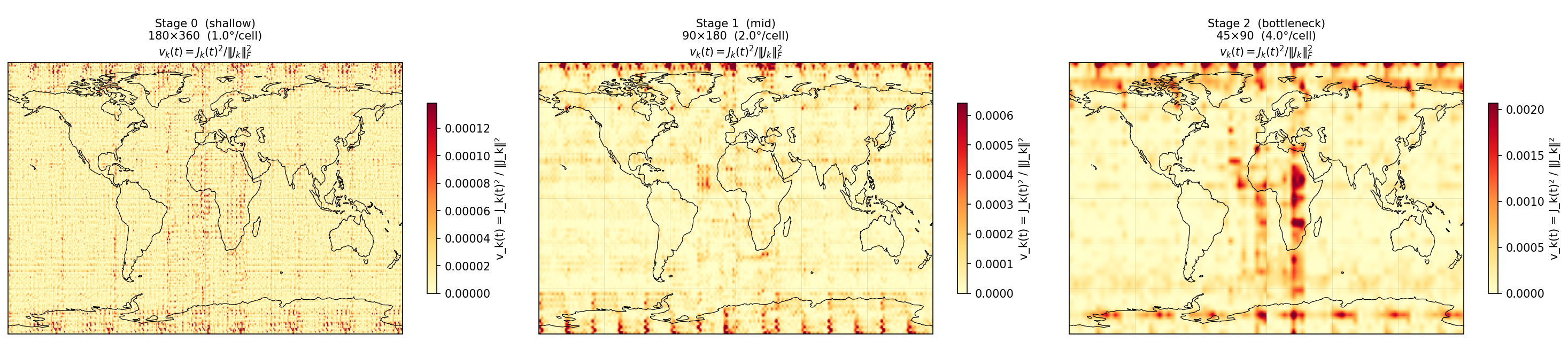}
\caption{Normalized gradient contribution $v_k(t) = J_k(t)^2 / \|\mathbf{J}_k\|_F^2$ for each backbone stage, upsampled to $180 \times 360$ for comparison. Stage~0 is near-uniform; Stage~2 concentrates IC3 gradient energy over the Indian Ocean and East Africa sector and the southern polar region, matching the Stage~2 LISA High-High cluster.}
\label{fig:jlens-eq3}
\end{figure}

\paragraph{Gradient concentration.}
We quantify how gradient energy concentrates across stages via the per-patch
normalized contribution:
\begin{equation}
    v_k(t) = \frac{J_k(t)^2}{\|\mathbf{J}_k\|_F^2}
\end{equation}
where $J_k(t)$ is the gradient magnitude at patch $t$ and
$\|\mathbf{J}_k\|_F^2$ is the total Frobenius norm at stage $k$.
Figure~\ref{fig:jlens-eq3} shows $v_k(t)$ upsampled to Stage~0 resolution
for direct comparison. At Stage~0 the distribution is near-uniform
(max $\sim 1.2 \times 10^{-4}$). By Stage~2, gradient energy concentrates
over the Indian Ocean and East Africa sector ($\sim 30$--$60^\circ$E,
tropical latitudes) and the southern polar region, directly matching the
Stage~2 LISA High-High cluster. This diagnostic is the gradient-space analog
of the activation-space metric $v_c(t)$ of \S\ref{subsec:jlens}:
both converge on the same geographic concentration, approached from
opposite directions.

\begin{figure}[H]
\centering
\includegraphics[width=0.75\linewidth]{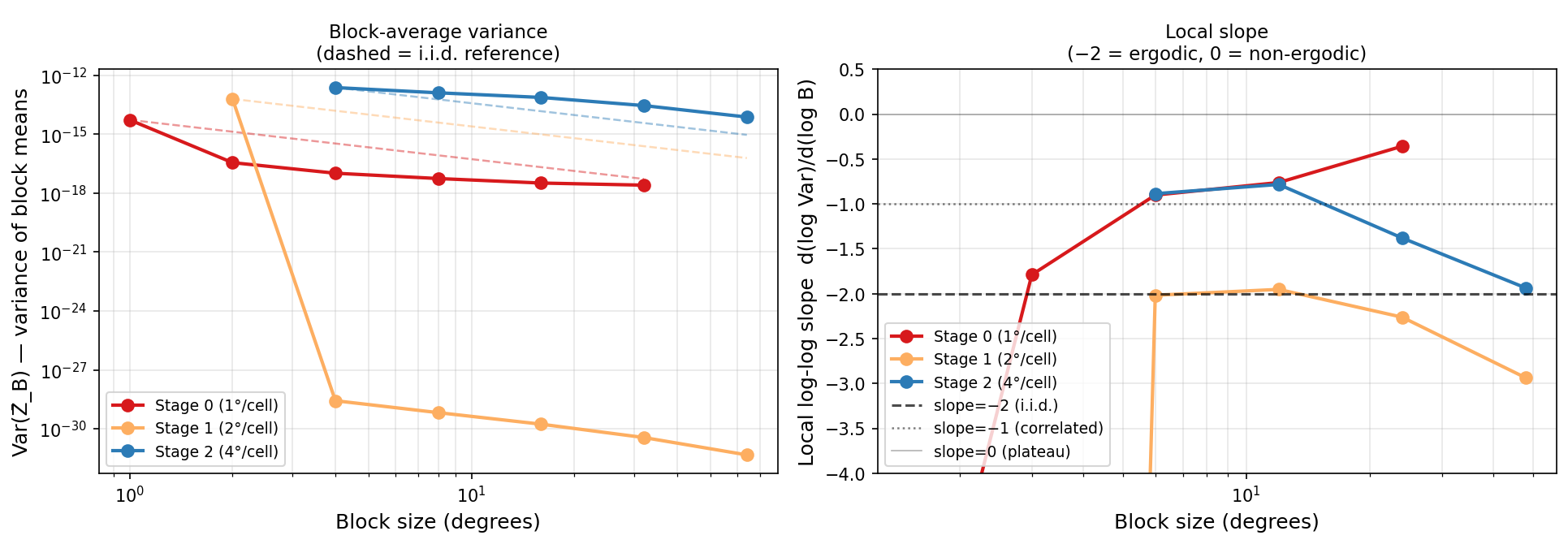}
\caption{Block-average variance test for J-lens gradient maps across backbone stages. Left: $\mathrm{Var}(\bar{Z}_B)$ vs.\ block size (log-log), where $Z_B$ is the block mean of the gradient magnitude field at block size $B$; dashed lines = i.i.d.\ reference ($\sigma^2/n_B$). Right: local log-log slope; dashed line at $-2$ marks the ergodic reference. Stage~1's slope below $-2$ reflects anti-clustering; Stage~2's near-ergodic slope reflects the equal-area balance of HH and LL regimes.}
\label{fig:jlens-block-var}
\end{figure}

\paragraph{Block-average variance and spatial ergodicity.}
Figure~\ref{fig:jlens-block-var} shows $\mathrm{Var}(\bar{Z}_B)$ — the
variance of non-overlapping block means — versus block size, with the
degree scale adjusted for each stage's cell size. Stage~1 produces a
log-log slope of $-2.94$, steeper than the i.i.d.\ reference of $-2$:
anti-correlated patches cancel variance more aggressively than white
noise when averaged, consistent with $I < 0$. Stage~2 produces a slope
of $-1.94$, near the ergodic reference despite its strong clustering
($I = +0.570$). This apparent paradox resolves because the HH (tropics
and poles) and LL (mid-latitudes) regions cover approximately equal
global area: sufficiently large blocks always sample both regimes and
converge to the global mean.

\paragraph{Summary}
These spatial analyses yield the same conclusion. 
\begin{itemize}
    \item \textbf{In gradient space} the IC3 sensitivity signal is diffuse and spatially random at Stage~0, then crystallizes at Stage~2 into a clean two-regime geographic partition — high sensitivity (HH) in the tropics and polar caps, low sensitivity (LL) in the mid-latitude storm tracks — with zero spatial outliers and gradient energy concentrated over the Indian Ocean and East Africa sector. 
    \item \textbf{In activation space}, the IC direction accounts for a disproportionately larger share of backbone variance during extreme dates in the NH mid-latitudes.
\end{itemize}

These probes are complementary: the gradient analysis identifies \textit{where sensitivity is encoded} at the bottleneck; the activation analysis identifies \textit{where that encoding manifests} during extremes. Aurora's extreme-sensitive representation is a bottleneck phenomenon — it does not exist in shallow layers and fully crystallizes only at the deepest stage of the encoder hierarchy.
\newpage
\subsection{Linearity of the J-Lens: Curvature at the Bottleneck}\label{subsec:linearity of the j-lens} 
The J-lens of \S\ref{subsec:jlens} is formally, a first-order local operator: $\mathbf{J}^{(k)}$ is the Jacobian of the (possibly highly nonlinear) map from stage-$k$ activations to the IC score $\text{s}_{\text{IC3}}$, evaluated at a single point. A large Jacobian entry identifies a direction of steep local ascent, not a nonlinear effect in its own right, and in the limit of infinitesimal perturbation, the leading Jacobian singular vectors coincide with the network's local finite-time Lyapunov vectors. What the Jacobian alone cannot say is how far a perturbation can travel along a given direction before this local linear picture breaks down. We answer that directly with the curvature of $\text{s}_{\text{IC3}}$.
For each of the $K=10$ leading right-singular directions $\mathbf{v}_k$ of the Stage-2 channel-gradient matrix $\bar{\mathbf{J}}^{(2)}$ (averaged over 138 extreme dates, reshaped to $\left(45\times90\right)\times2048$), we compute the scalar curvature of $\mathbf{s}_{\text{IC3}}$ along $\mathbf{v}_k$ via a Hessian-vector product, $\kappa_k=\mathbf{v}_k^\top\mathbf{H}\mathbf{v}_k$, where $\mathbf{H}=\partial^2\text{s}_{\text{IC3}}/\partial(\mathbf{h}^{(2)})^2$, obtained with one forward and two backward passes per (date, direction) pair, avoiding the intractable $8.3M\times8.3M$ full Hessian. We define a perturbation validity radius $\varepsilon^*_k=2\|\mathbf{J}\cdot\mathbf{v}_k\|/|\kappa_k|$: the amplitude along $\mathbf{v}_k$ at which the linear prediction and its quadratic correction become comparable. $\varepsilon^*_k\gg 1$ means the Jacobian is trustworthy well beyond realistic perturbation sizes; $\varepsilon^*_k<1$ means it is not.
\begin{figure}[h]
\centering
\includegraphics[width=\linewidth]{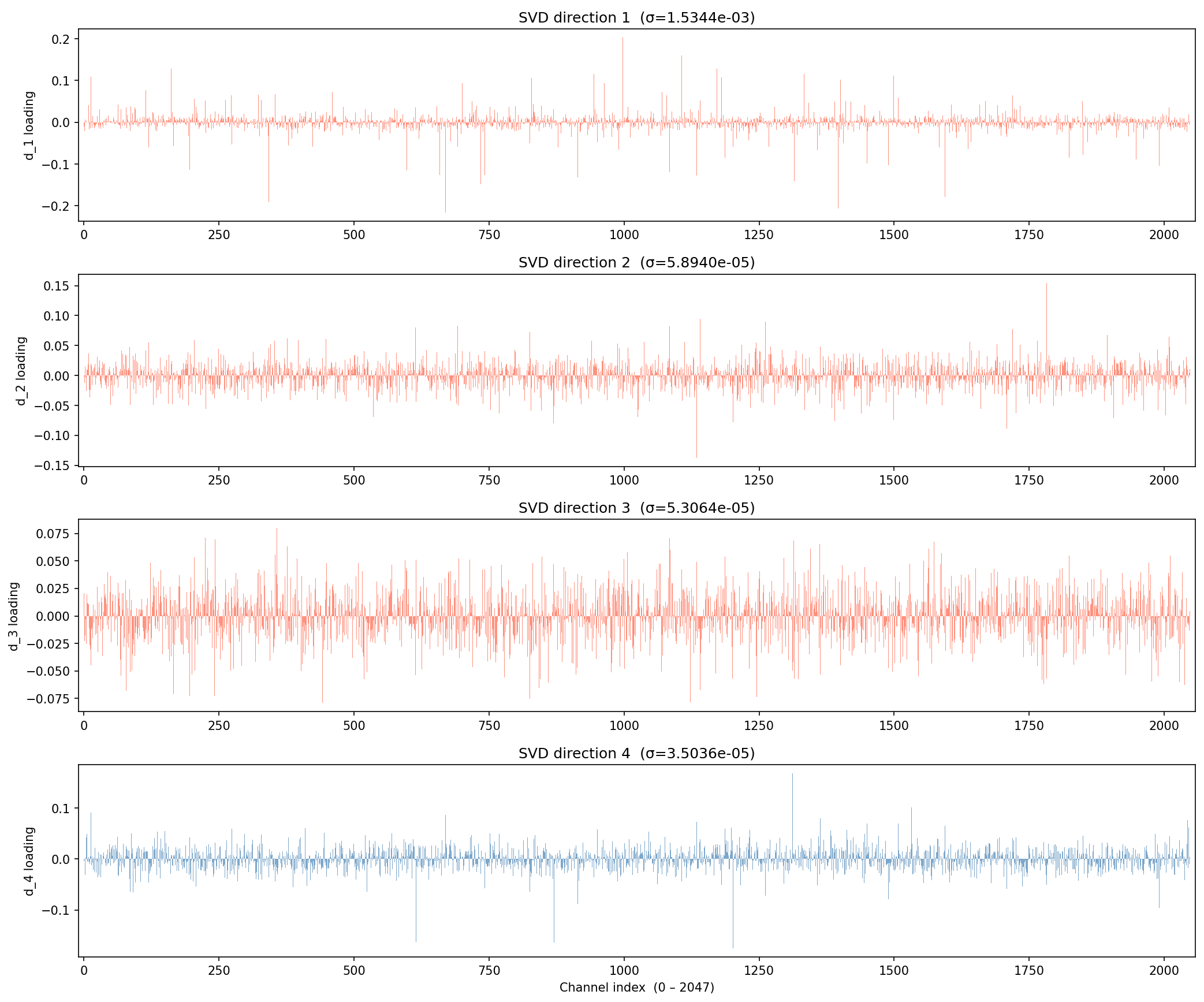}
\caption{Channel-space loadings of the four leading right-singular vectors of the Stage-2 channel-gradient matrix $\bar{\mathbf{J}}^{(2)}$ across all 2048 channels. Singular values decay sharply ($\sigma_1 = 1.53\times10^{-3}$, $\sigma_2 = 5.89\times10^{-5}$, $\sigma_3 = 5.31\times10^{-5}$, $\sigma_4 = 3.50\times10^{-5}$). None of the four directions is dominated by a small number of channels; each is a distributed combination across the full channel dimension.}
\label{fig:jlens-hessian-loadings}
\end{figure}

\begin{figure}[h]
\centering
\includegraphics[width=\linewidth]{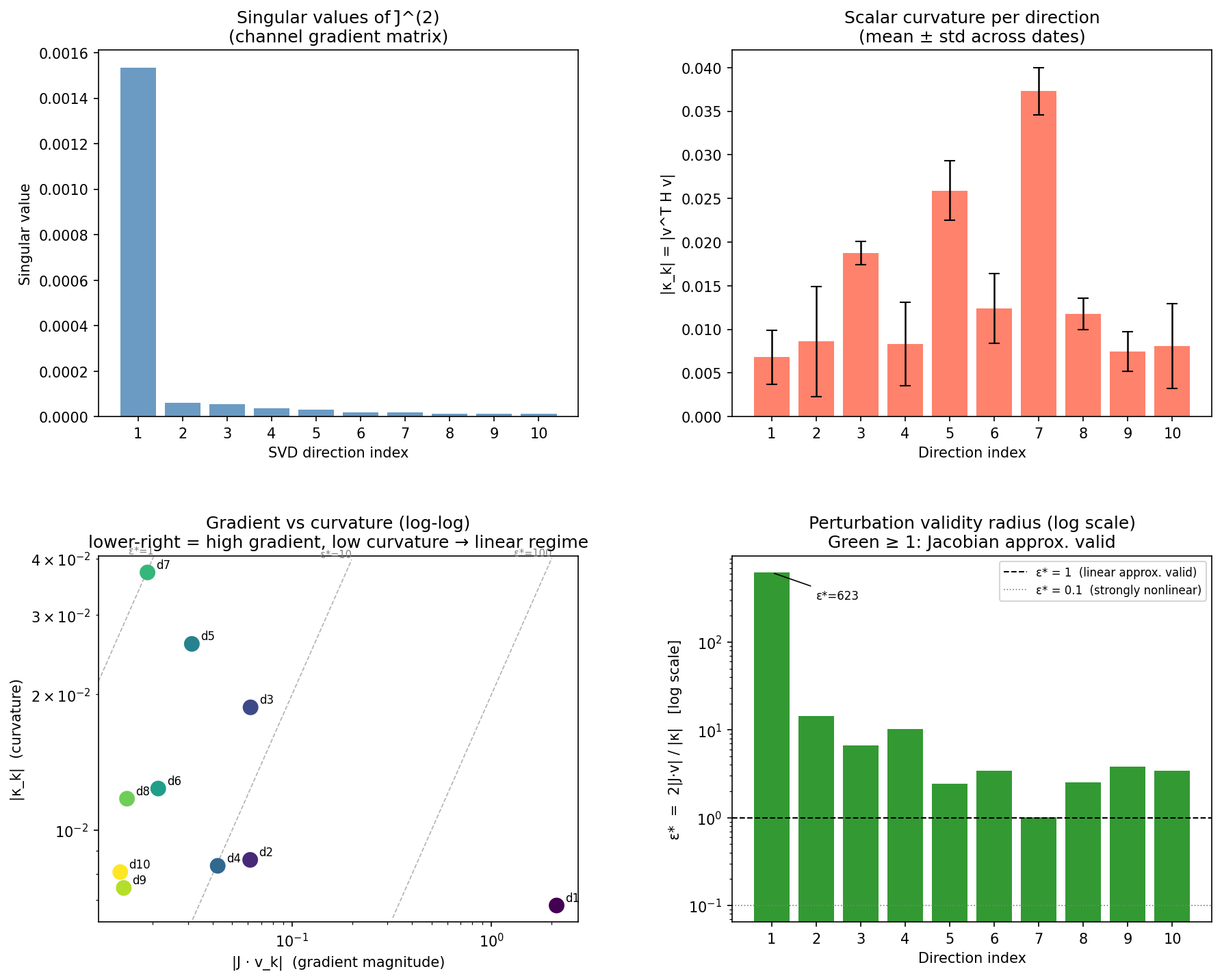}
\caption{Hessian curvature of $s_{\mathrm{IC3}}$ at the Stage-2 bottleneck, evaluated along the top-10 Jacobian SVD directions ($N=139$ extreme dates). Top left: singular values of $\bar{\mathbf{J}}^{(2)}$, showing the dominant causal direction is essentially one-dimensional. Top right: scalar curvature $|\kappa_k| = |\mathbf{v}_k^\top \mathbf{H} \mathbf{v}_k|$ per direction (mean $\pm$ std across dates); curvature peaks at direction 7, not direction 1. Bottom left: gradient magnitude vs.\ curvature (log-log); direction 1 sits in the high-gradient, low-curvature (linear) regime, while direction 7 sits in the low-gradient, high-curvature regime. Bottom right: perturbation validity radius $\varepsilon_k^{*} = 2\|\mathbf{J}\cdot\mathbf{v}_k\|/|\kappa_k|$ on a log scale; direction 1's validity radius ($\varepsilon_1^{*}\approx623$) is roughly two orders of magnitude above the $\varepsilon^{*}=1$ threshold, while direction 7 sits at the boundary of linear validity.}
\label{fig:jlens-hessian}
\end{figure}

Figure~\ref{fig:jlens-hessian} (top-left) shows the leading singular values of $\bar{\mathbf{J}}^{(2)}$ exceed the second by a factor of $\sim 26$ ($\sigma_1=1.53\times10^{-3}$ vs. $\sigma_2=5.89\times10^{-5}$), confirming that Stage-2's causal leverage over $\text{s}_{\text{IC3}}$ concentrates in an essentially one-dimensional subspace -- a sharper version of the small, disproportionately influential subspace \citet{gurnee2026verbalizablerepresentationsformglobal} report in language models. Curvature does not track gradient magnitude (Figure~\ref{fig:jlens-hessian}, top-right and bottom-left): direction 1, which carries nearly all of the linear leverage, has one of the \textit{smallest} curvatures among the ten and sits deep in the lower-right, linear regime corner of the gradient-curvature plane. Its validity radius $\varepsilon^*_1=623$ is roughly two orders of magnitude above threshold. Direction 7, by contrast, has the highest curvature of the ten despite a comparatively small gradient, and its validity radius sits right at $\varepsilon^*=1$ -- the one direction, of those examined, where a first-order treatment would not be safe. The channel-space loadings of these directions (Figure~\ref{fig:jlens-hessian-loadings}) show no single-channel dominance in any of the top four -- each is a distributed combination across the 2048 channels.
This is the direct answer to the concern that the Jacobian is unambiguously a first-order, local-tangent operator and says nothing about nonlinearity on its own. What the curvature analysis adds is not a claim that the Jacobian is nonlinear, but a measurement of \textit{how far} its local linear picture can be trusted. Because the dominant causal direction is also the most linear of the ten, the perturbation amplitudes used in the feasibility analysis of \S\ref{sec:jiu-jitsu feasibility analysis} fall safely within the Jacobian's regime of validity; a strategy that instead targeted a subdominant, high-curvature direction such as direction 7 would need the quadratic correction to remain accurate.

\section{Discussion and Conclusion}\label{sec:discussion-conclusion}
This work provides a systematic empirical characterization of Aurora as a dynamical system, motivated by the Weather Jiu-Jitsu framework of \citet{liu2026instabilityawaresteeringextremeatmospheric} and the question of whether small, targeted perturbations can leverage intrinsic atmospheric sensitivity to influence extreme weather evolution. Our results speak to four requirements of that framework.
First, Aurora performs genuine dynamic forecasting rather than sophisticated persistence: minimum RMSE against truth is achieved at zero time shift across all variables and lead times, and Aurora substantially outperforms persistence throughout the 7-day horizon, which is a prerequisite for containing the structured instability geometries that Weather Jiu-Jitsu exploits.
Second, Aurora's forecast uncertainty grows in a manner consistent with a chaotic dynamical system: R\textsuperscript{2} decay is well-approximated by an AR(1) process, and RMSE growth converges to a roughly constant effective Lyapunov exponent after an elevated-growth transient in the first 40 hours. This should be read against the recent debate over whether AI weather models reproduce a genuine butterfly effect \citep{selz2023,shen2025butterfly,hassanzadeh2026missingbutterfly}: we don't claim Aurora reproduces atmospheric chaos at PDE resolution, only that at its own 6-hour operating scale it exhibits local, spatially and temporally varying divergence consistent with a genuine, coarse-grained local Lyapunov exponent (the operationally relevant notion of chaos for Weather Jiu-Jitsu, and one consistent with the scale-dependent divergence reported in the Lorenz-model chaos-control work and the atmospheric-river steering paper). Our own latent-space diagnostics point the same way: the J-lens spatial analysis of \S\ref{subsec:jlens-spatial} finds that extreme-sensitive gradient energy is not uniform but organizes into coherent high- and low-sensitivity geographic regimes (Moran's I$=+0.570$ at the bottleneck), and the direction-based analysis of \S\ref{sec:direction-based} finds that sensitivity to extremity concentrates in specific latent directions rather than being spread uniformly across the representation -- both consistent with a local, spatially and directionally varying notion of sensitivity rather than a single global rate.
Third, Aurora's latent representations are organized by season but not by meteorological extremity. This null result is consistent with grounding perturbation targeting in physical diagnostics, i.e., FTLE fields, integrated vapor transport, jet-stream flank criteria, and relative humidity, rather than learned representations; global mean pooling appears to dilute local extreme signals. These findings rest on modest samples (48 and 608 dates against representations with thousands to hundreds of thousands of dimensions, over only two years of limited seasonal replication), so we treat them as exploratory rather than settled.
Fourth, a spatially fixed, dynamically uninformed perturbation (a Gaussian bump centered at the domain's midpoint) produces feasibility ratios well below 1 across all variables and magnitudes, motivating instability-aware targeting: the contrast with the FTLE-guided cloud-seeding operator, which achieves coherent downstream IVT reductions of up to 5.21\%, quantifies the leverage gained by targeting dynamically unstable regions rather than perturbing blindly.
Taken together, these results support the distinction between predictability limits and controllability potential: even where forecast uncertainty grows rapidly, Aurora's structured instability geometry creates windows in which small, targeted perturbations can influence trajectory evolution before nonlinear amplification fully develops, and our error growth, sensitivity map, and drift comparison diagnostics provide the empirical grounding to assess when and where such interventions are likely to succeed.
Several limitations warrant acknowledgment. All perturbation experiments use a single date (2023-01-01), a calm winter day without an active atmospheric river, limiting direct comparison with the atmospheric-river case study cited throughout; the feasibility ratio evaluates global RMSE rather than the regional IVT metric used there, making the analyses complementary rather than directly comparable; and extending the drift comparison and sensitivity analysis to active AR events and FTLE-identified perturbation sites remains for future work. Nonetheless, this analysis is consistent with Aurora containing the dynamical structure the WJJ framework requires, delineates where that structure is (and is not) accessible, and offers a diagnostic toolkit for evaluating perturbation feasibility across a broader range of atmospheric conditions and event types (diagnostics that motivate, but do not yet substitute for, a demonstrated control intervention). 

\appendix
\section{Supplementary Results: Direction-based Latent Analysis}\label{sec:appendix-direction}
\S\ref{sec:direction-based} presents independent component analysis (ICA) as the primary direction-finding method applied to Aurora's latent space. This appendix reports, in full, the other four direction-finding approaches explored over the same 608-date ERA5 sample (139 extreme, 469 normal dates) and extreme-event classification criterion described in \S\ref{sec:direction-based}: higher principal components (the motivating first pass that led to ICA), archetype analysis, principal nested spheres, and tail-dependence PCA.

\subsection{Higher Principal Components: A First Signal}\label{subsec:appendix-higher-pcs}
The latent space analysis of \S\ref{subsec:latent space analysis} projected encoder representations onto the top two principal components, which together capture the dominant seasonal variance (PC1: 41.3\%, PC2: 19.8\%). By construction, these components encode the annual cycle rather than meteorological extremity, explaining the null result of \S\ref{subsec:extreme event encoding}. Instead, we ask "\textit{across the full spectrum of principal directions, do any respond preferentially to extreme conditions?}"

\begin{figure}[H]
\centering
\includegraphics[width=\linewidth]{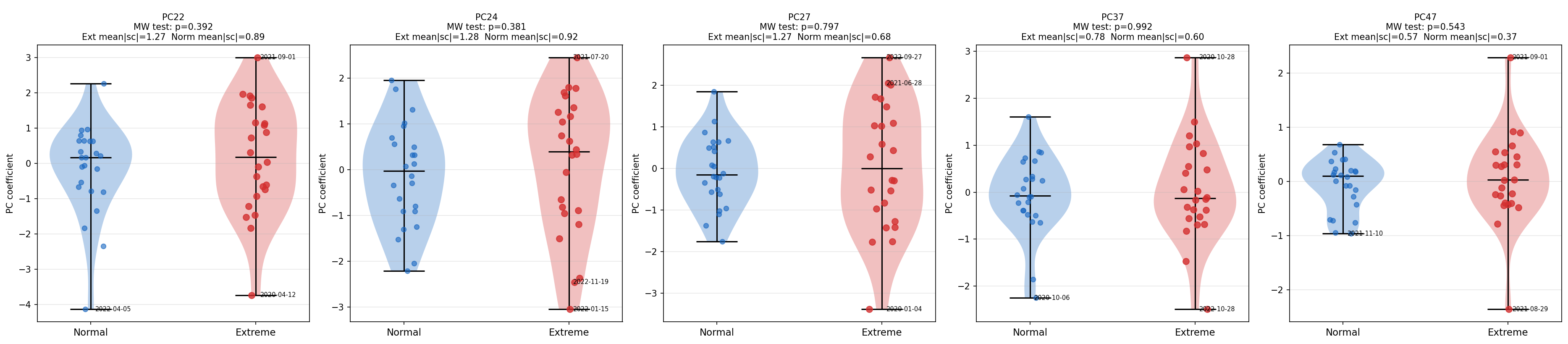}
\caption{Score distributions (Normal, blue; Extreme, red) for the five extreme-sensitive PCs in Aurora's encoder. The PCs are identified as both sparse and extreme-enriched. Extreme dates activate each component with larger absolute scores; separation is most pronounced for PC27 and PC47.}
\label{fig:pc-distributions}
\end{figure}

We perform a sensitivity scan by computing an extreme-event sensitivity ratio for each of the top 50 encoder PCs as,
\begin{equation}
    r_i=\frac{\mathbb{E}\left[|s_i|\mid\text{extreme}\right]}{\mathbb{E}\left[|s_i|\mid\text{normal}\right]}
\end{equation}
where $s_i$ is the PC $i$ score and expectation is taken over the respective date sets ($N=139$ extreme, matched normal sample). A ratio near 1 indicates equal activation and $r_i>1$ indicates preferential activation on extreme dates.
\begin{figure}[H]
\centering
\includegraphics[width=\linewidth]{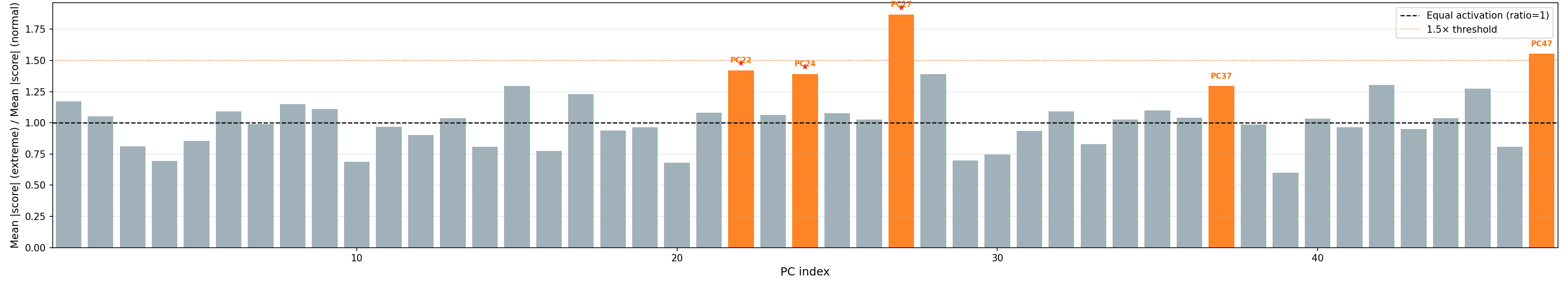}
\caption{Extreme-event sensitivity ratio $r_i$ for the top 50 encoder PCs. Gray bars: ratio near 1 (no extremity preference). Orange bars: components exceeding the $1.5\times$ threshold (Mann-Whitney U, $p < 0.05$). Five components — PC22, PC24, PC27, PC37, PC47 — form a sparse extreme-sensitive subset; PC27 achieves the highest ratio ($r = 1.87$). The dashed line marks equal activation ($r = 1$).}
\label{fig:pc-sensitivity}
\end{figure}
Figure~\ref{fig:pc-sensitivity} shows $r_i$ for all 50 components. The large majority cluster near $r=1$, confirming that most principal directions carry no extreme information. A sparse subset of five components -- PC22, PC24, PC27, PC37, and PC47 -- exceed the $1.5\times$ threshold with statistical significance (Mann-Whitney U, $p<0.05$). PC27 exhibits the highest ratio ($r=1.87$), followed by PC47 ($r=1.55$).
The score distributions for these five components, as shown in Figure~\ref{fig:pc-distributions}, confirm that extreme dates activate them with systematically larger magnitude, with the separation most pronounced for PC27.

\begin{figure}[H]
\centering
\includegraphics[width=\textwidth]{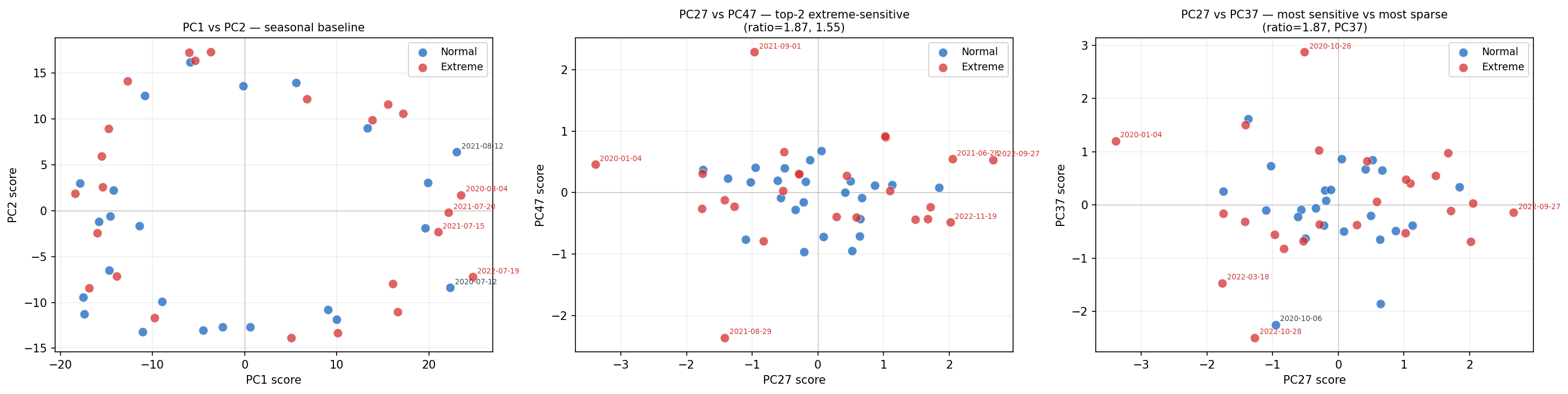}
\caption{Scatter plots of encoder PC scores for normal (blue) and extreme (red) dates. Left: PC1 vs PC2 (seasonal baseline) — no separation by design. Center: PC27 vs PC47 (top-2 extreme-sensitive components, $r = 1.87$ and $1.55$). Right: PC27 vs PC37. Despite elevated sensitivity ratios, extreme and normal dates remain fully interleaved in the sensitive PC subspace, with $k$-means achieving 50.0\% purity.}
\label{fig:pc-scatter}
\end{figure}

Figure~\ref{fig:pc-scatter}, however, shows that projecting onto this five-dimensional subspace does not recover a usable separation between extreme and normal dates. Scatter plots in the PC1--PC2 seasonal baseline (left) and in the PC27--PC47 extreme-sensitive plane (center and right) show complete interleaving of extreme (red) and normal (blue) dates. $k$-means clustering in the full five-dimensional sensitive subspace achieves a purity of 50\% (indistinguishable from chance), compared to a 54.2\% bound from the full PC space.

The failure of simple projection is informative. The sensitive PCs define a \emph{subspace} in which extreme-relevant variance is elevated, but the discriminating direction within that subspace is not aligned with any single PC axis. Recovering it requires a method that searches for structure within the subspace rather than treating each axis independently, which is what motivates the ICA analysis of \S\ref{subsec:ica}.

\subsection{Archetype Analysis}\label{subsec:appendix-archetypes}
\begin{figure}[H]
\centering
\includegraphics[width=0.75\textwidth]{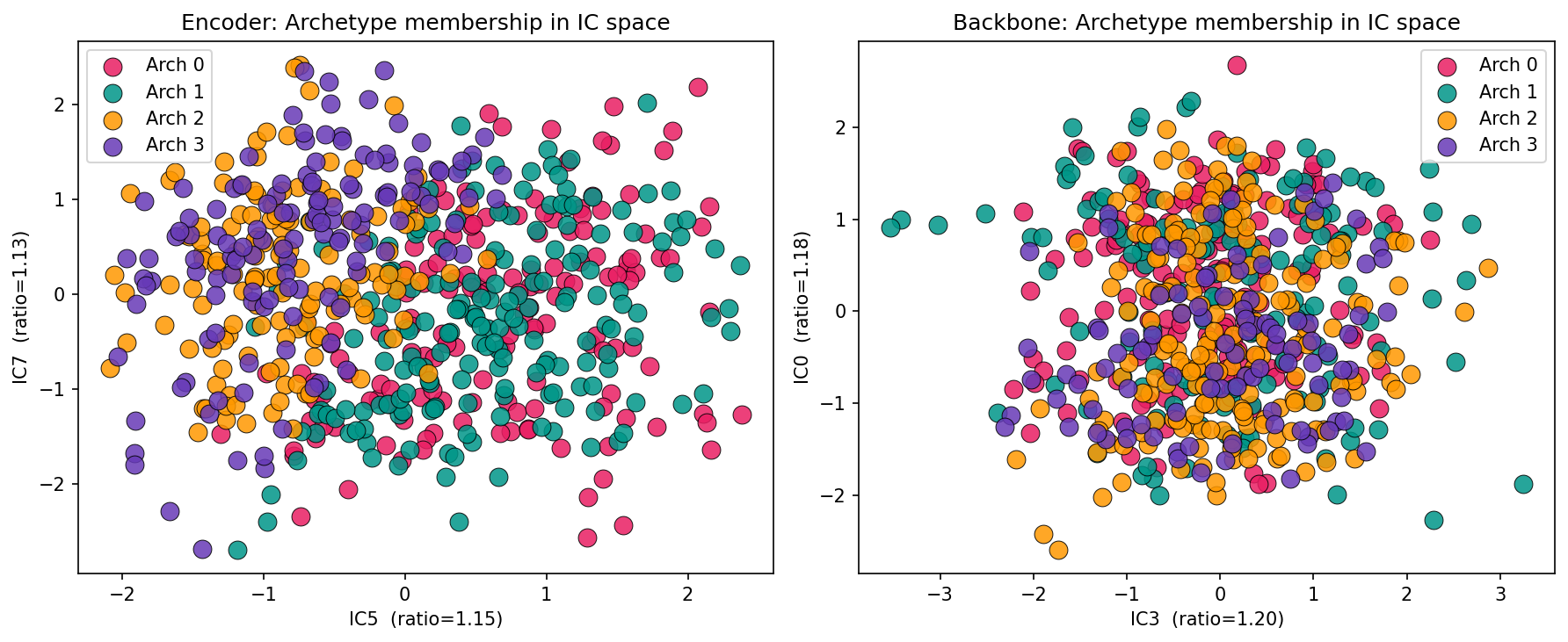}
\caption{Archetype membership in IC space for encoder (IC5 vs IC7, left) and backbone (IC3 vs IC0, right). Each point is one of the 608 dates; color indicates assignment to one of four archetypes. The archetypes occupy distinct corners of the IC plane, capturing maximally different atmospheric configurations. Extreme dates are distributed across all four archetypes with no dominant assignment.}
\label{fig:archetypes}
\end{figure}

Archetypal analysis finds the most extreme prototype points on the convex hull of the data distribution. Unlike $k$-means centroids, archetypes represent maximally distinct configurations and are guaranteed to lie at the boundary of the data. We fit four archetypes to each layer's IC-space
embedding. Figure~\ref{fig:archetypes} shows archetype membership in encoder IC5-IC7 space (left) and backbone IC3-IC0 space (right). The four archetypes partition the IC plane into distinct corners, capturing qualitatively different atmospheric configurations. Inspecting which archetype each extreme date is assigned to reveals that extreme dates are distributed across all four archetypes with no dominant assignment, implying that no archetype functions as a universal "extreme archetype." This null result is interpretable. Extreme precipitation events are meteorologically heterogeneous: a heat-dome-driven moisture event and a frontal-system event may produce comparable precipitation totals while occupying completely different regions of Aurora's latent space. Archetypal analysis makes this heterogeneity explicit: The four corners of the IC-space distribution correspond to distinct synoptic configurations, and extreme events are not concentrated in any one of them. Any control strategy that assumes a single target direction in latent space will therefore fail to generalize across event types.

\subsection{Principal Nested Spheres (PNS)}\label{subsec:appendix-pns}
PNS generalizes PCA to curved manifolds using a greedy backward-reduction sequence. It can capture intrinsically curved structure that linear methods miss. We apply FastPNS \citep{monem2025principalnestedsphereshighdimensional} with $n_{\text{pca}}=20$ pre-whitening components to both layers.
Encoder PNS ratios slightly exceed ICA ratios suggesting the extreme-sensitive structure has a mildly curved geometry in the encoder. For the backbone, PNS7 and IC3 are comparable, indicating a more linear structure there.
\begin{figure}[H]
  \centering
  \includegraphics[width=0.75\linewidth]{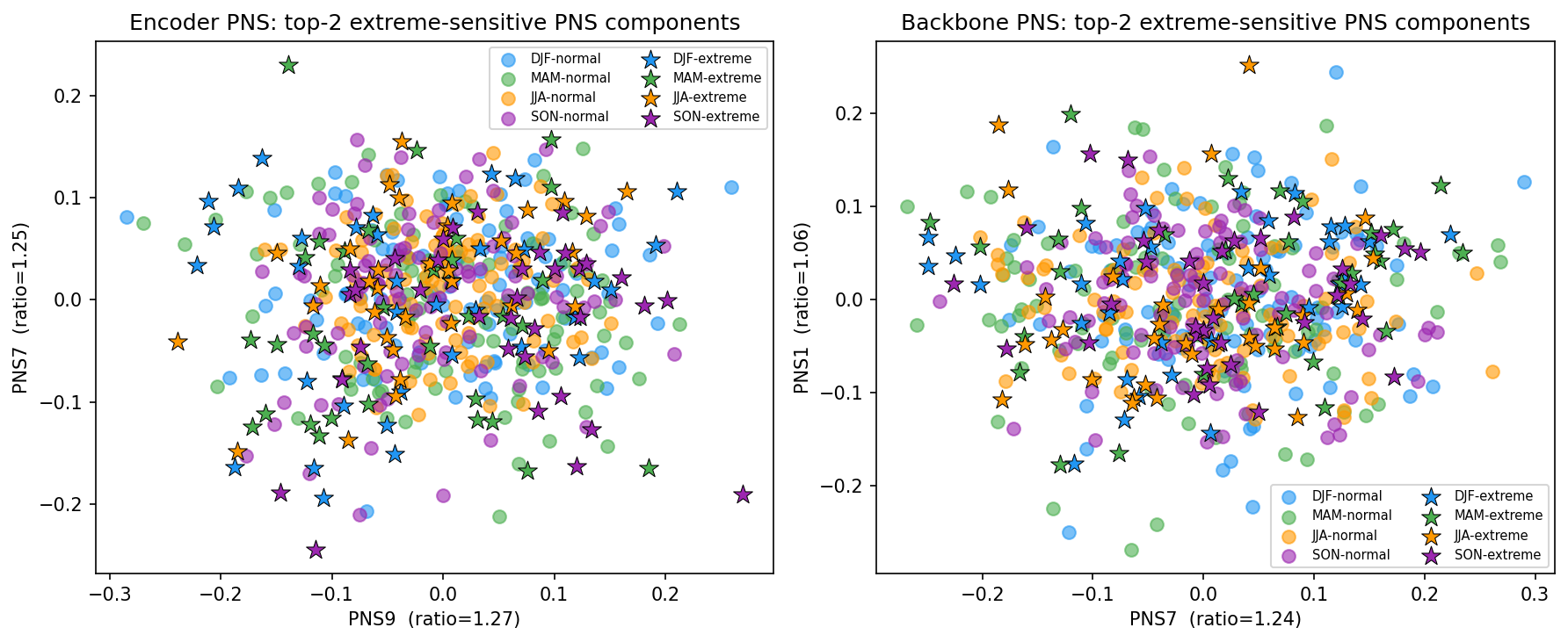}\\[4pt]
  \includegraphics[width=0.75\linewidth]{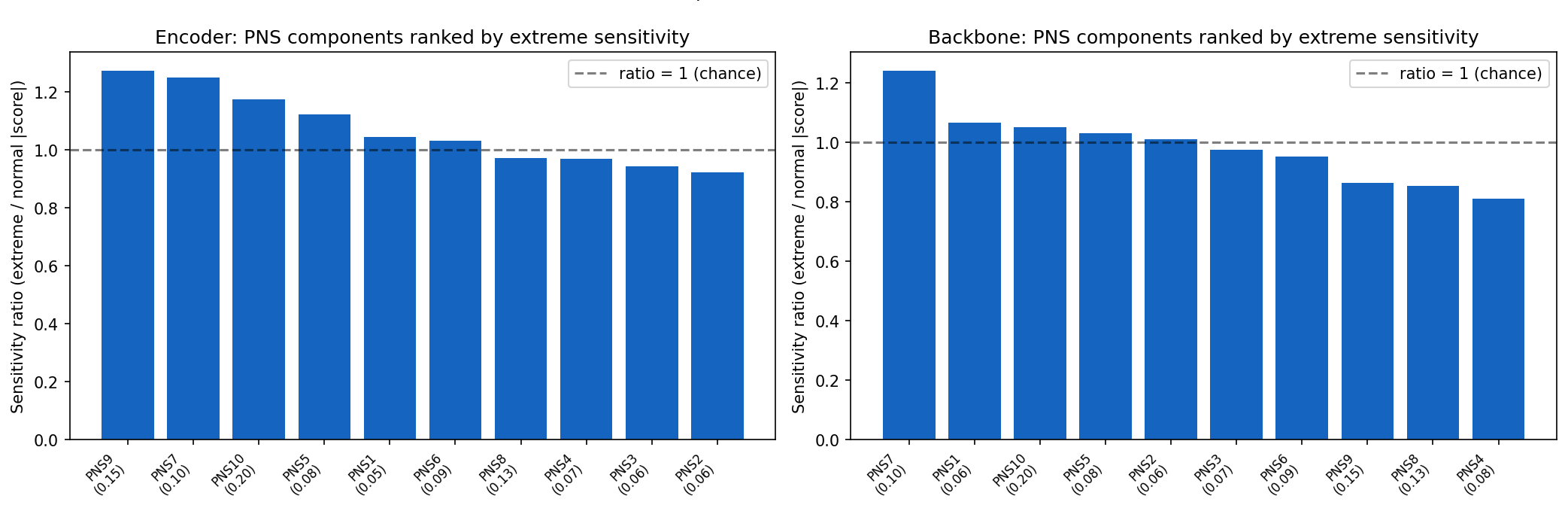}
  \caption{Principal Nested Spheres (FastPNS, $n_{\text{pca}}=20$) applied to Aurora encoder and backbone representations across 608 dates. \textit{Top:} Projection onto the top-2 extreme-sensitive PNS components for the encoder (left, PNS9/PNS7) and backbone (right, PNS7/PNS1). Circles denote normal dates; stars denote extreme-event dates; colors encode meteorological season. \textit{Bottom:} All PNS components ranked by their extreme-sensitivity ratio (mean $|$score$|$ for extreme dates divided by mean $|$score$|$ for normal dates); the dashed line marks ratio $= 1$ (chance). Variance explained by each component is shown in parentheses on the $x$-axis.}
  \label{fig:pns}
\end{figure}
The sensitivity ratio bar charts (Figure~\ref{fig:pns}, bottom) show that the encoder retains a broader diffuse signal: six of ten components exceed the chance ratio of 1, with the leading components PNS9 and PNS7 reaching ratios of $1.27$ and $1.25$ respectively. These results are slightly higher than the ICA extreme-sensitivity scores reported in \S\ref{subsec:ica}, which is consistent with a mild curvature in the encoder's extreme-event geometry. The backbone's signal is more concentrated, where PNS7 peaks at a ratio of $1.24$ but the remaining components rapidly fall toward or below chance, with only three components clearly exceeding ratio $=1$. PNS7 and IC3 attaining comparable ratios in the backbone suggests that the backbone's extreme-sensitive direction is essentially linear since FastPNS recovers no additional curvature beyond what ICA already captures. The scatter plots (Figure~\ref{fig:pns}, top) reflect these ratios, where extreme event dates (stars) are broadly interleaved with normal dates (circles) in both the encoder and backbone PNS spaces, with no visible cluster cleanly separating by intensity.

\subsection{Tail-Dependence PCA}\label{subsec:appendix-tail}
Rather than asking which directions activate more on extreme dates, tail-dependence PCA asks which directions capture correlated tail activation across dates, constructing a covariance matrix weighted towards the upper quantile (q=0.75) of joint activations.

The latent space contains no geometric cluster for extreme events, but it does contain specific directions that are preferentially activated. The strongest signals consistently reside in the backbone: IC3 (1.201), PNS7 (1.24), TPC2 (1.27). The backbone is a more extreme-sensitive layer than the encoder.
\begin{figure}[H]
  \centering
  \includegraphics[width=0.75\linewidth]{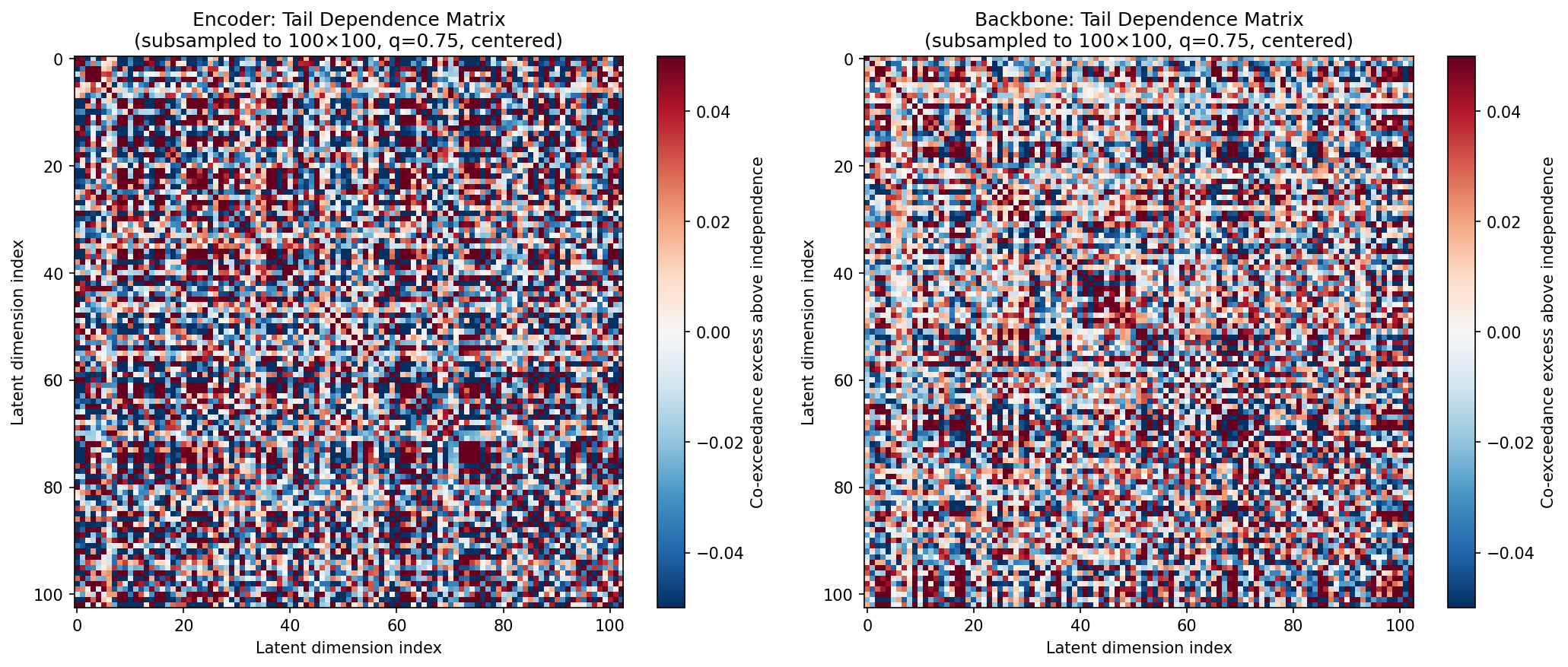}
  \caption{Tail-dependence co-exceedance matrices ($q=0.75$, subsampled to $100\times100$ latent dimensions, centered) for the encoder (left) and backbone (right). The encoder matrix is largely unstructured; the backbone exhibits pronounced positive co-exceedance blocks, indicating that groups of latent dimensions jointly activate in the tail regime, which is the structured dependence that motivates a tail-weighted PCA.}
  \label{fig:tdpca_matrix}
\end{figure}

\begin{figure}[H]
  \centering
  \includegraphics[width=0.75\linewidth]{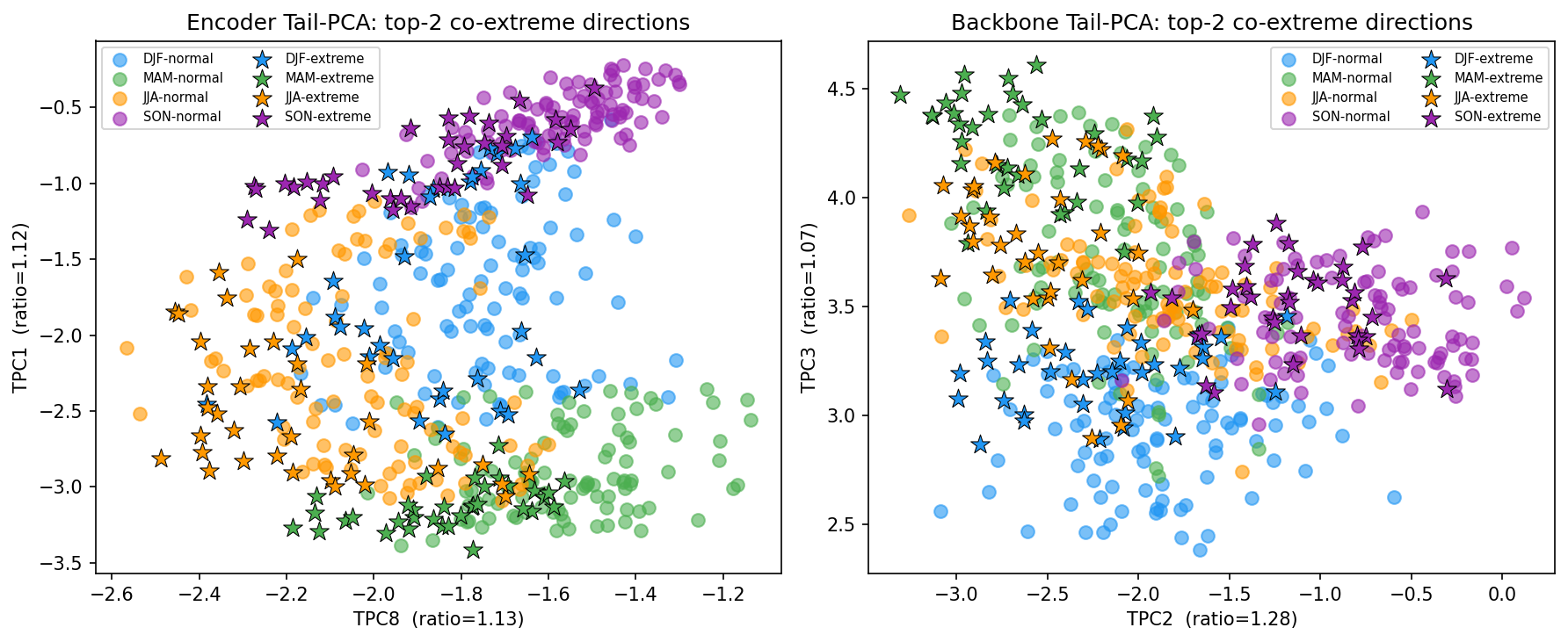}\\[4pt]
  \includegraphics[width=0.75\linewidth]{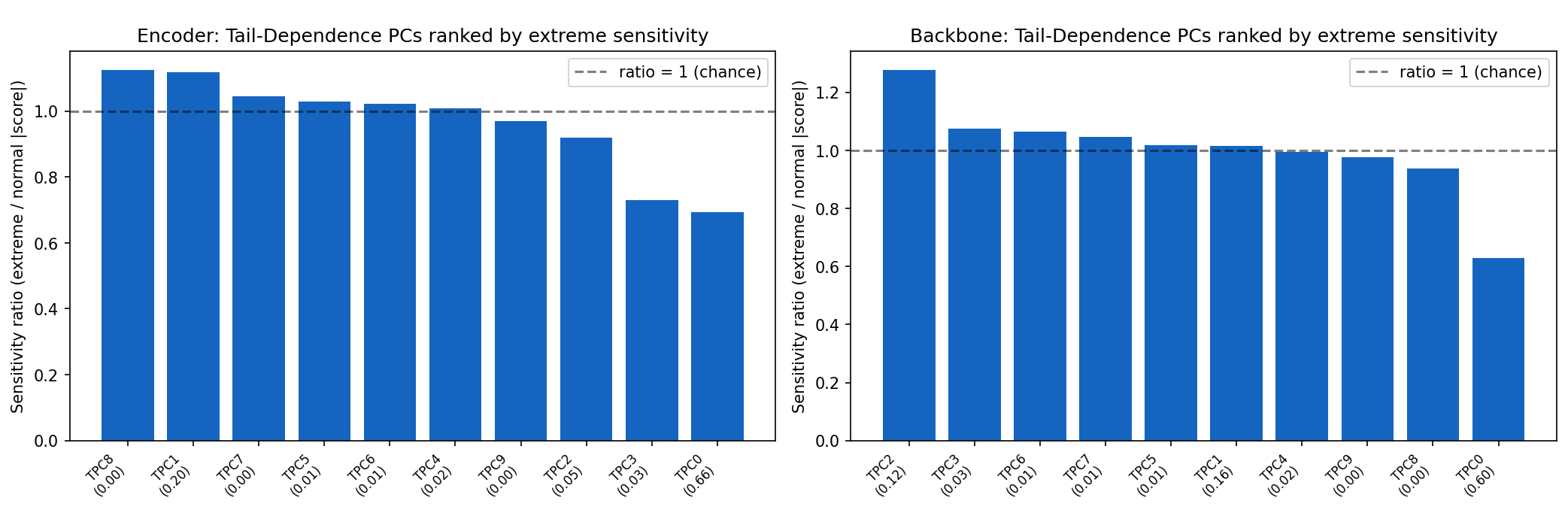}
  \caption{Tail-Dependence PCA ($q=0.75$) applied to Aurora encoder and backbone representations across 608 dates. \textit{Top:} Projection onto the top-2 extreme-sensitive tail-PCs for the encoder (left, TPC8/TPC1) and backbone (right, TPC2/TPC3). Circles denote normal dates; stars denote extreme-event dates; colours encode meteorological season. \textit{Bottom:} All tail-PCs ranked by extreme-sensitivity ratio; dashed line marks ratio $= 1$ (chance). Variance explained by each component is shown in parentheses.}
  \label{fig:tdpca}
\end{figure}
The tail-dependence matrices (Figure~\ref{fig:tdpca_matrix}) reveal why the two layers respond differently to tail weighting: the encoder co-exceedance matrix is largely unstructured, whereas the backbone exhibits pronounced positive blocks, indicating that backbone dimensions jointly activate in the tails across dates. Tail-PCA on the encoder distributes sensitivity diffusely across several components (TPC8, TPC1, TPC7) with no single direction standing out (Figure~\ref{fig:tdpca}, bottom left). The backbone concentrates its signal sharply in TPC2 (ratio $1.28$), which drops to 1.07 for TPC3 and near-chance for all remaining components (Figure~\ref{fig:tdpca}, bottom right). The backbone TPC2/TPC3 scatter (Figure~\ref{fig:tdpca}, top right) makes this concrete: TPC2 stratifies dates strongly by season, and within each season extreme events are systematically displaced toward the more negative tail of TPC2 relative to their normal counterparts. No analogous pattern emerges in the encoder projection, where seasons partially overlap and extreme dates are interleaved with normal ones.
\end{document}